\documentclass[showpacs,preprintnumbers,amsmath,amssymb]{revtex4}

\usepackage{epsfig}
\usepackage{graphicx}
\usepackage{dcolumn}
\usepackage{bm}

\begin{document}


\title{Description of the Energy Eigenstates of the 1D Hubbard Model\\
in Terms of Rotated-Electron Site Distribution Configurations}
\author{J. M. P. Carmelo}
\affiliation{GCEP-Center of Physics, University of Minho, Campus
Gualtar, P-4710-057 Braga, Portugal}
\date{22 November 2002}


\begin{abstract}
In this paper we describe the pseudoparticles, holons, and spinons
whose occupancy configurations describe the energy eigenstates of
the one-dimensional (1D) Hubbard model in terms of rotated
electrons. Rotated electrons are related to electrons by a mere
unitary transformation such that rotated electron double
occupation is a good quantum number and the {\it effective
electronic lattice} occupied by rotated electrons is identical to
the {\it real-space lattice} occupied by electrons. Moreover, we
find that the band-momentum pseudoparticle description associated
with the Bethe-ansatz Takahasi's thermodynamic equations is
related by Fourier transform to a {\it local pseudoparticle}
representation in terms of occupancy configurations of spatial
coordinates. Such spatial coordinates correspond to an {\it
effective pseudoparticle lattice}. There is an effective
pseudoparticle lattice for each pseudoparticle branch with finite
occupancy in a given state. This description introduces the local
pseudoparticles whose spatial coordinate is the conjugate of the
band momentum. We describe the energy eigenstates in terms of
local pseudoparticle site distribution configurations in such
lattices. Moreover, we relate both the local pseudoparticle
internal structure and the latter configurations to the
rotated-electron site distribution configurations which describe
the energy eigenstates. The electron - rotated-electron unitary
transformation is such that the latter configurations are
independent of the value of the on-site Coulombian repulsion. Our
findings provide useful information about the relation of the
exotic pseudoparticles, holons, and spinons that diagonalize the
non-perturbative many-electron problem to the original electrons.
We provide an example showing how the derived local pseudoparticle
representation can be used in the evaluation of finite-energy
few-electron spectral function expressions.
\end{abstract}


\maketitle
\section{INTRODUCTION}

Low-dimensional, correlated systems have attracted much attention
in recent years due to a large variety of unconventional
electronic properties directly connected with electronic
correlations. In this context one-dimensional (1D) systems are of
special interest, essentially stimulated by theoretical
predictions like charge-spin separation for the 1D Hubbard model
\cite{Woy,Ogata,Penc95,Penc96,Penc97}. While the 1D Hubbard model
\cite{Lieb,Takahashi,Martins98,Rasetti,Hubbard} is the prototype
of an exactly solvable model for correlated electrons, such a
charge-spin separation corresponds to the description of the
energy eigenstates in terms of occupancy configurations of holons
and spinons. Recently the charge-spin separation of that model was
shown to occur for all values of excitation energy, the
corresponding holon and spinon description being extended to the
complete set of energy eigenstates which span the Hilbert space
\cite{I,II,III}. On the other hand, quasi-1D materials are ideal
model systems which allow the study of basic physical concepts in
one dimension. In the last years there has been a renewed interest
on the unconventional spectral properties of these materials
\cite{Hussey,Menzel,Fuji02,Hasan,Ralph,Gweon,Monceau,Takenaka,Mizokawa,Moser,Mihaly,Vescoli00,Denlinger,Fuji,Kobayashi,Bourbonnais,Vescoli,Zwick,Neudert,Mori,Kim}.
Moreover, recent angle-resolved ultraviolet photoemission
spectroscopy revealed very similar spectral fingerprints from both
high-$T_c$ superconductors and quasi-1D compounds \cite{Menzel}.
The similarity of the ultraviolet data for these two different
systems could be evidence of the occurrence of the above mentioned
charge-spin separation associated with holons and spinons. The
anomalous temperature dependence of the spectral function could
also indicate a dimensional crossover
\cite{Menzel,Granath,Orgard,Carlson}. The results of Refs.
\cite{Zaanen,Antonio} also suggest that the unconventional
spectral properties observed in two-dimensional (2D) materials
could have a 1D origin. Thus the holons and spinons could play an
important role in spectral properties of both 1D and 2D
low-dimensional materials.

Recently it was found that for the 1D Hubbard model there are
exact selection rules which limit the number of holons and spinons
created by application of rotated electron operators onto any
eigenstate of the spin $\sigma$ electron number \cite{III}. The
concept of rotated electron is associated with a unitary
transformation introduced in Ref. \cite{Harris}. For such rotated
electrons double occupation is a good quantum number for all
values of the on-site Coulombian repulsion $U$. In order to
distinguish the rotated-electron description from the electron
representation, the concept of {\it effective electronic lattice}
was introduced \cite{III}. It is such that rotated electrons
occupy the effective electronic lattice, whereas electrons occupy
the {\it real-space lattice}. However, these lattices are
identical, having the same lattice constant $a$ and length
$L=a\,N_a$, where $N_a$ designates the number of sites. The
occurrence of rotated electrons leads to clear finger prints in
the few-electron spectral properties. Indeed, the few-electron
spectral functions include several upper Hubbard bands
corresponding to different and separated energy scales. The
$D_r^{th}$ band is spanned by excited states of rotated-electron
double occupation $D_r$. However, in general only the first few
upper bands have a significant amount of spectral weight. Although
the above-mentioned selection rules refer to rotated-electron
operators, their existence induces restrictions in the numbers of
holons, spinons, and pseudoparticles contained in few-electron
excitations. For instance, according to the results found in Ref.
\cite{III}, about 99\% of the upper Hubbard band spectral weight
of the one-electron addition spectral function corresponds to the
first upper Hubbard band. Interestingly, rotated electron double
occupation $D_r$ equals the number of holons of $\eta$ spin
projection $-1/2$ and charge $-2e$. Here $-e$ denotes the
electronic charge and the $\eta$ spin is a $SU(2)$ algebra
associated with the charge degrees of freedom \cite{HL,Yang89}.
The charge $-2e$ and $+2e$ holons refer to all values of the
on-site Coulomb repulsion $U$ and are related to the large $U$
dublons and holons respectively, studied in Ref. \cite{Tohyama}.

The related to the electron - rotated electron unitary
transformation is part of the non-perturbative diagonalization of
the quantum problem. Through out this paper the designation
rotated-electron site distribution configuration (and electron
site distribution configuration) refers to the effective
electronic lattice (and real-space lattice). A property of deep
physical meaning is that the energy eigenstates are described by
the the same rotated-electron site distribution configurations for
all values of $U/t$, where $t$ is the first-neighbor transfer
integral. This is in contrast to the complex electron site
distribution configurations of these states, which change upon
variations of the value of the on-site repulsion. Fortunately, the
electron - rotated electron unitary transformation becomes the
unit transformation as $U/t\rightarrow\infty$. Thus one can reach
the $U/t$ independent rotated-electron site distribution
configurations which describe the energy eigenstates by studying
the corresponding electron site distribution configurations for
the 1D Hubbard model in the limit $U/t\rightarrow\infty$. In the
present paper we study such $U/t$ independent rotated electron
configurations. In part, this is achieved by considering the
corresponding problem of the electron real-space lattice charge
(and spin) sequences which describe the energy eigenstates of the
1D Hubbard model in the limit $U/t\rightarrow\infty$. The problem
of the 1D Hubbard model in the limit of $U/t\rightarrow\infty$ has
been previously studied by many authors
\cite{Ogata,Penc95,Penc96,Penc97,Harris,Beni,Klein,Carmelo88,Mac,Parola,Ricardo,Eskes,Geb,RicCar}.
In such a limit there is a huge degeneracy of $\eta$-spin and spin
occupancy configurations. Thus there are several choices for
complete sets of energy eigenstates with the same energy and
momentum spectra. However, only one of these choices is associated
with the rotated-electron site distribution configurations reached
by the electron - rotated electron canonical unitary
transformation. It refers to the complete set of energy
eigenstates generated from the corresponding energy eigenstates of
the 1D Hubbard model for finite values of $U/t$ by turning off
adiabatically the parameter $t/U$. We call the obtained set of
states of the model {\it band-momentum energy eigenstates}. The
electron occupancy configurations which describe these states have
not been studied so far. In addition to study this problem, in
this paper we discuss the relation of the obtained states to an
alternative choice of energy eigenstates whose electron site
distribution configurations were studied in Ref. \cite{Geb}.

Concerning the open questions whose clarification our results are
useful for, one should mention the evaluation of few-electron
spectral functions for finite values of excitation energy. Except
in the limit $U/t\rightarrow\infty$ \cite{Penc95,Penc96,Penc97},
this is an important open problem of interest for the further
understanding of the unconventional spectral properties observed
in low-dimensional materials. The relationship of the
pseudoparticles, holons, and spinons to the electrons is a complex
problem of crucial importance for the evaluation of these spectral
functions. As discussed in later sections, the concepts of local
pseudoparticle and effective pseudoparticle lattices introduced in
this paper, as well as the relationship of the energy-eigenstate
pseudoparticle occupancy configurations in these effective
lattices to the rotated-electron occupancy configurations, provide
useful information about such a relationship. As open questions
are concerned, we emphasize that the studies of this paper are a
necessary step and a valuable contribution for the evaluation of
the overlap between few-electron excitations and energy
eigenstates, as further discussed in Sec. VI. Elsewhere the tools
and concepts introduced in this paper are used in the study of
few-electron spectral distributions for finite values of momentum
and excitation energy \cite{IIIb,V}. Importantly, predictions
obtained recently by application of a preliminary version of the
theory introduced in this paper and in Refs. \cite{IIIb,V} seem to
describe both qualitatively and quantitatively the one-electron
removal spectral lines observed in real quasi-1D materials by
photoemission experiments for finite values of the excitation
energy \cite{spectral0}.

The paper is organized as follows: In Sec. II we introduce the 1D
Hubbard model and summarize the concept of rotated electron as
well as the description of the model in terms of holons, spinons,
and pseudoparticles. In Sec. III we introduce the complete set of
energy eigenstates of the 1D Hubbard model in the limit
$U/t\rightarrow\infty$ which refers to the electron - rotated
electron canonical unitary transformation. This includes the
introduction of a set of basic properties which turn out to be
useful for finding the internal structure of the local
$\alpha,\nu$ pseudoparticle, which is one of the subjects of Sec.
IV. In Sec. IV we introduce a complete basis of local states which
we express in terms of rotated-electron local charge, spin, and
$c$ pseudoparticle sequences. In that section we also find the
rotated-electron site distribution configurations which describe
the internal structure of the local $\alpha,\nu$ pseudoparticle.
The local $c$ pseudoparticle and $\alpha ,\nu$ pseudoparticle
effective lattices are introduced in Sec. V. In that section we
also express the energy eigenstates in terms of the
Fourier-transform superpositions of local charge, spin, and $c$
pseudoparticle sequences introduced in the previous section. A
pratical application of the derived local pseudoparticle
representation is given in Sec. VI. Finally, in Sec. VII we
present the concluding remarks.

\section{THE 1D HUBBARD MODEL, ROTATED ELECTRONS, AND SUMMARY OF THE
PSEUDOPARTICLE, HOLON, AND SPINON DESCRIPTION}

In this section we introduce the model used in our studies. In
addition, we discuss the concept of rotated electron and summarize
some basic information about the pseudoparticle, holon, and spinon
description which is useful for the studies of this paper.

\subsection{THE MODEL}

In a chemical potential $\mu $ and magnetic field $H$ the 1D
Hubbard Hamiltonian can be written as,

\begin{equation}
\hat{H} = {\hat{H}}_{SO(4)} + \sum_{\alpha}\mu_{\alpha
}\,{\hat{S}}^{\alpha}_z \label{H}
\end{equation}
where the Hamiltonian,

\begin{equation}
{\hat{H}}_{SO(4)} = {\hat{H}}_H  - (U/2)\,\hat{N} + (U/4)\,N_a \,
, \label{HSO4}
\end{equation}
has $SO(4)$ symmetry \cite{HL,Yang89,Essler} and

\begin{equation}
{\hat{H}}_H = \hat{T} + U\,\hat{D} \, , \label{HH}
\end{equation}
is the simple Hubbard model. The operators

\begin{equation}
\hat{T} = -t\sum_{j =1}^{N_a}\sum_{\sigma =\uparrow
,\downarrow}\sum_{\delta =-1,+1}
c_{j,\,\sigma}^{\dag}\,c_{j+\delta,\,\sigma} \, , \label{Top}
\end{equation}
and

\begin{equation}
\hat{D} = \sum_{j =1}^{N_a}
\hat{n}_{j,\,\uparrow}\,\hat{n}_{j,\,\downarrow} \, , \label{Dop}
\end{equation}
on the right-hand side of Eq. (\ref{HH}) are the kinetic energy
operator and the electron double occupation operator respectively.
The operator $\hat{n}_{j,\,\sigma}= c_{j,\,\sigma }^{\dagger
}\,c_{j,\,\sigma }$ on the right-hand side of Eq. (\ref{Dop})
counts the number on spin $\sigma$ electrons at real-space lattice
site $j=1,2,3,...,N_a$, where the number of lattice sites $N_a$ is
even and large, $N_a/2$ is odd, and we consider periodic boundary
conditions. The associated spin $\sigma$ electron number operator
reads,

\begin{equation}
{\hat{N}}_{\sigma}=\sum_{j} \hat{n}_{j,\,\sigma} \, . \label{Nsi}
\end{equation}
The operators $c_{j,\,\sigma }^{\dagger }$ and $c_{j,\,\sigma}$
which appear in the above equations are the spin $\sigma $
electron creation and annihilation operators at site $j$
respectively. Moreover, on the right-hand side of Eq. (\ref{H}),
$\mu_c=2\mu$, $\mu_s=2\mu_0 H$, $\mu_0$ is the Bohr magneton, and
${\hat{S }}^c_z= -{1\over 2}[N_a-\hat{N}]$ and ${\hat{S }}^s_z=
-{1\over 2}[{\hat{N}}_{\uparrow}-{\hat{N}}_{\downarrow}]$ are the
diagonal generators of the $SU(2)$ $\eta$-spin $S^c$ and spin
$S^s$ algebras \cite{HL,Yang89} respectively. In the latter
expressions ${\hat{N}}=\sum_{\sigma} \hat{N}_{\sigma}$ is the
electron number operator and the operator ${\hat{N}}_{\sigma}$
counts the number of spin $\sigma$ electrons and is given in Eq.
(\ref{Nsi}). We denote by $N_{\uparrow}$ and $N_{\downarrow}$ the
number of spin-up electrons and spin-down electrons respectively,
and by $N=N_{\uparrow}+N_{\downarrow}$ the number of electrons.
The lattice constant is denoted by $a$ and thus the length of the
system is $L=N_a\,a$. We consider electronic densities and the
spin densities given by $n=n_{\uparrow }+n_{\downarrow}$ and
$m=n_{\uparrow}-n_{\downarrow}$ respectively, where
$n_{\sigma}=N_{\sigma}/L$ and $n=N/L$. These densities and spin
densities belong to the domains defined by the following
inequalities $0\leq n \leq 1/a$\, ; $1/a\leq n \leq 2/a$ and
$-n\leq m \leq n$\, ; $-(2/a-n)\leq m \leq (2/a-n)$ respectively.

The Hamiltonian $\hat{H}_{SO(4)}$ defined in Eq. (\ref{HSO4})
commutes with the six generators of the $\eta$-spin $S_c$ and spin
$S_s$ algebras \cite{HL,Yang89,Essler}. While the expressions of
the diagonal generators were provided above, the off-diagonal
generators of these two algebras read,

\begin{equation}
{\hat{S}}^c_{+}=\sum_{j}(-1)^j\, c_{j,\,\downarrow}^{\dag}
c_{j,\,\uparrow}^{\dag} \, ; \hspace{1cm}
{\hat{S}}^c_{-}=\sum_{j}(-1)^j\,
c_{j,\,\uparrow}c_{j,\,\downarrow} \, , \label{Sc}
\end{equation}
and

\begin{equation}
{\hat{S}}^s_{+}= \sum_{j}c_{j,\,\downarrow}^{\dag}c_{j,\,\uparrow}
\, ; \hspace{1cm} {\hat{S}}^s_{-}=\sum_{j}c_{j,\,\uparrow}^{\dag}
c_{j,\,\downarrow} \, , \label{Ss}
\end{equation}
respectively. We note that the Bethe-ansatz solution of the 1D
Hubbard model refers to the Hilbert subspace spanned by the
lowest-weight states (LWSs) of the $\eta$-spin and spin algebras,
{\it i.e.} such that $S^{\alpha}= -S^{\alpha}_z$ \cite{Essler}.

Finally, the momentum operator is given by,

\begin{equation}
\hat{P} = - {i\over 2}
\sum_{\sigma}\sum_{j=1}^{N_a}\Bigl[\,c^{\dag
}_{j,\,\sigma}\,c_{j+1,\,\sigma} - c^{\dag
}_{j+1,\,\sigma}\,c_{j,\,\sigma}\Bigr] \, , \label{Popel}
\end{equation}
and commutes with the Hamiltonians introduced in Eqs. (\ref{H})
and (\ref{HSO4}).

\subsection{ROTATED ELECTRON OPERATORS}

Rotated electrons are associated with a unitary transformation
first introduced by Harris and Lange \cite{Harris,Mac}. Such a
transformation plays a key role in the expression of the holon and
spinon number operators in terms of electronic operators for the
whole parameter space of the model \cite{III}. The electron
operators that occur in the 1D Hubbard model (\ref{H}) and
(\ref{HH}) are defined by $c_{j,\,\sigma}^{\dag}$, while the
rotated electron operator ${\tilde{c}}_{j,\,\sigma}^{\dag}$ is
given by,

\begin{equation}
{\tilde{c}}_{j,\,\sigma}^{\dag} =
{\hat{V}}^{\dag}(U/t)\,c_{j,\,\sigma}^{\dag}\,{\hat{V}}(U/t) \, ,
\label{c+til}
\end{equation}
where ${\hat{V}}(U/t)$ is the Harris and Lange unitary operator.
As mentioned in Sec. I, such rotated electrons conserve double
occupation for all finite values of $U/t$. The operators
${\hat{V}}^{\dag}(U/t)$ and ${\hat{V}}(U/t)$ can be written as,

\begin{equation}
{\hat{V}}^{\dag}(U/t) = e^{-\hat{S}} \, ; \hspace{0.5cm}
{\hat{V}}(U/t) = e^{\hat{S}} \, . \label{SV}
\end{equation}
These operators are uniquely defined by the following two
equations,

\begin{equation}
{\tilde{H}}_H = {\hat{V}}^{\dag}(U/t)\,{\hat{H}}_H\,{\hat{V}}(U/t)
= {\hat{H}}_H + [{\hat{H}}_H,\,{\hat{S}}\,] + {1\over
2}\,[[{\hat{H}}_H,\,{\hat{S}}\,],\,{\hat{S}}\,] + ... \,
,\label{HHtil}
\end{equation}
and

\begin{equation}
[{\hat{H}}_H,\,{\hat{V}}^{\dag}(U/t)\,\hat{D}\,{\hat{V}}(U/t)] =
[{\hat{H}}_H,\,\tilde{D}] = 0 \, . \label{HHDtil}
\end{equation}
In these equations the Hamiltonian ${\hat{H}}_H$ is given in Eq.
(\ref{HH}) and and the rotated-electron double occupation operator
$\tilde{D}$ reads,

\begin{equation}
\tilde{D} \equiv {\hat{V}}^{\dag}(U/t)\,\hat{D}\,{\hat{V}}(U/t) =
\sum_{j}\, {\tilde{c}}_{j,\,\uparrow }^{\dagger
}\,{\tilde{c}}_{j,\,\uparrow }\, {\tilde{c}}_{j,\,\downarrow
}^{\dagger }\,{\tilde{c}}_{j,\,\downarrow } \, , \label{Doptil}
\end{equation}
where $\hat{D}$ is the electron double occupation operator given
in Eq. (\ref{Dop}). Note that $c_{j,\,\sigma}^{\dag}$ and
${\tilde{c}}_{j,\,\sigma}^{\dag}$ are only identical in the
$U/t\rightarrow\infty$ limit where electron double occupation
becomes a good quantum number. The expression of any rotated
operator $\tilde{O}$ such that,

\begin{equation}
\tilde{O} = {\hat{V}}^{\dag}(U/t)\,\hat{O}\,{\hat{V}}(U/t) \, ,
\label{Otil}
\end{equation}
in terms of the rotated electron operators
${\tilde{c}}_{j,\,\sigma}^{\dag}$  and ${\tilde{c}}_{j,\,\sigma}$
is the same as the expression of the corresponding general
operator $\hat{O}$ in terms of the electron operators
$c_{j,\,\sigma}^{\dag}$ and $c_{j,\,\sigma}$ respectively.
Equations (\ref{HHtil}) and (\ref{HHDtil}) can be used to derive
an expression for the unitary operator ${\hat{V}}(U/t)$ in terms
of electronic elementary operators order by order in $t/U$. The
authors of Ref. \cite{Mac} carried out this expansion up to eighth
order (see foot note [12] of that reference).

Operators which commute with the Harris and Langue unitary
operator have the same expressions in terms of both electronic and
rotated-electronic operators. Examples of such operators are the
six generators of the $SU(2)$ $\eta$-spin and spin algebras. Thus
it follows that the commutators $
[\hat{V}(U/t),{\vec{\hat{S}}}_{\alpha}.{\vec{\hat{S}}}_{\alpha}]$
and $[\hat{V}(U/t),{\hat{N}}_{\sigma}]$ vanish, where
${\vec{\hat{S}}}_c.{\vec{\hat{S}}}_c\equiv{\vec{\hat{\eta}}}.{\vec{\hat{\eta}}}$
and ${\vec{\hat{S}}}_s.{\vec{\hat{S}}}_s\equiv
{\vec{\hat{S}}}.{\vec{\hat{S}}}$ are the square $\eta$-spin and
spin operators respectively, and ${\hat{N}}_{\sigma}$ is the spin
$\sigma$ electron number operator (\ref{Nsi}). The Harris and
Langue unitary operator also commutes with the momentum operator
(\ref{Popel}).

\subsection{SUMMARY OF
THE PSEUDOPARTICLE, HOLON, AND SPINON DESCRIPTION}

Let us summarized the aspects of the holon, spinon, and
band-momentum pseudoparticle description introduced in Ref.
\cite{I} and further investigated in Refs. \cite{II,III} which are
useful for the studies of this paper. Such a description is
extracted directly from combination of the Bethe-ansatz
Takahashi's thermodynamic equations with the $SO(4)$ symmetry of
the model and the symmetries associated with the concept of
rotated electron. Indeed, holons, spinons, and band-momentum
pseudoparticles emerge naturally from the non-perturbative
diagonalization of the model. The first step of such a
diagonalization is precisely the Hilbert-space unitary rotation
which maps electrons onto rotated electrons and is such that
rotated-electron double occupation is a good quantum number. Let
us denote by $N_c$ the number of rotated-electron singly occupied
sites. Thus the numbers $N_c$, $[N-N_c]/2$, and $[N^h-N_c]/2$
equal the number of rotated-electron singly occupied sites, doubly
occupied sites, and empty sites respectively. (Here
$N^h=[2N_a-N]$.) Importantly, once rotated electron double
occupation is a good quantum number also all these numbers are
good quantum numbers for all values of $U/t$. There are precisely
$N_c$ $c$ pseudoparticles and $N_c$ spinons. These $c$
pseudoparticles (and spinons) describe the charge part (and spin
part) of the $N_c$ rotated electrons which singly occupy lattice
sites. Thus the charge and spin degrees of freedom of these $N_c$
rotated electrons separate. For each $c$ pseudoparticle there is a
chargeon and a rotated-electronic hole. The chargeon corresponds
to the charge part of the rotated electron which singly occupies
the site. The non-perturbative organization of the electronic
degrees of freedom associates the $[N-N_c]$ rotated electrons
which doubly occupy $[N-N_c]/2$ sites with $[N-N_c]/2$ holons of
$\eta$-spin projection $-1/2$. Indeed each rotated-electron doubly
occupied site corresponds to a spin singlet rotated electron pair
which is nothing but a $-1/2$ holon. Finally, the $[N^h-N_c]/2$
rotated-electron empty sites are nothing but the $[N^h-N_c]/2$
holons of $\eta$-spin projection $+1/2$. Such an analysis confirms
the importance played by the unitary operator $\hat{V}(U/t)$ in
the description of the quantum objects whose occupancy
configurations describe all energy eigenstates of the model.

Let us summarize some of the properties of the quantum objects
which we have briefly related to rotated-electrons above. One
should distinguish the total $\eta$ spin (and spin) of the system,
which we denote by $S_c$ (and $S_s$) and the corresponding
$\eta$-spin (and spin) projection, which we denote by $S_c^z$ (and
$S_s^z$) from the $\eta$ spin (and spin) carried by the elementary
quantum objects. We call $s_c$ (and $s_s$) the $\eta$ spin (and
spin) carried by the holons, spinons, and pseudoparticles and
$\sigma_c$ (and $\sigma_s$) their $\eta$-spin (and spin)
projection. The holons are such that $s_c=1/2$, $s_s=0$, and
$\sigma_c =\pm 1/2$ whereas the spinons have $s_s=1/2$ and
$\sigma_s=\pm 1/2$. Through out this paper we denote the holons
and spinons according to their $\pm 1/2$ $\eta$-spin and spin
projections respectively. The $-1/2$ and $+1/2$ holons have charge
$-2e$ and $+2e$ respectively. It is shown in later sections that
the band-momentum $c$ pseudoparticle is associated with the local
$c$ pseudoparticle. The latter local quantum object describes the
chargeon and the corresponding rotated electronic hole of each
rotated-electron singly occupied site. There are also $2\nu$-holon
(and $2\nu$-spinon) composite $c,\nu$ pseudoparticles (and $s,\nu$
pseudoparticles). The $c,\nu$ pseudoparticle has $s_c=0$ and
$s_s=0$ (and the $s,\nu$ pseudoparticle has $s_s=0$ and no charge
degrees of freedom) and contains an equal number $\nu$ of $-1/2$
and $+1/2$ $\eta$-spin projection holons (and spin projection
spinons). The $c$ pseudoparticle has $s_c=0$ and no spin degrees
of freedom. The $\pm 1/2$ holons (and $\pm 1/2$ spinons) which are
not part of $2\nu$-holon composite $c,\nu$ pseudoparticles (and
$2\nu$-spinon composite $s,\nu$ pseudoparticles) are called $\pm
1/2$ Yang holons (and $\pm 1/2$ HL spinons). In the designations
{\it HL spinon} and {\it Yang holon}, HL stands for Heilmann and
Lieb and Yang refers to C. N. Yang respectively, who are the
authors of Refs. \cite{HL,Yang89}.

All energy eigenstates can be specified by the $c$ pseudoparticle
band-momentum distribution function $N_c(q)$, by the set of
$\alpha ,\nu$ pseudoparticle band-momentum distribution functions
$\{N_{\alpha,\,\nu}(q)\}$ where $\alpha =c,s$ and $\nu =1,2,...$,
and by the set of numbers numbers $L_{\alpha ,\,-1/2}$ of
$s_c=1/2$, $s_s=0$, and $-1/2$ Yang holons ($\alpha =c$) and
$s_s=1/2$ and $-1/2$ HL spinons ($\alpha =s$). An important
concept is that of CPHS ensemble subspace where $CPHS$ stands for
{\it $c$ pseudoparticle, holon, and spinon}. This is a Hilbert
subspace spanned by all states with fixed values for the $-1/2$
Yang holon number $L_{c,\,-1/2}$, $-1/2$ HL spinon number
$L_{s,\,-1/2}$, $c$ pseudoparticle number $N_c$, and for the sets
of $\alpha ,\nu$ pseudoparticle numbers $\{N_{c,\,\nu}\}$ and
$\{N_{s,\,\nu}\}$ corresponding to the $\nu=1,2,3,...$ branches.

The variable $q$ appearing in the above distribution functions
$N_c(q)$ and $\{N_{\alpha,\,\nu}(q)\}$ where $\alpha =c,s$ and
$\nu =1,2,...$ is the continuum band momentum associated with the
discrete band-momentum values $q_j$ which are of the following
form,

\begin{eqnarray}
q_j & = & {2\pi\over L}\, I_j^c \,  ; \hspace{1cm}
j=1,...,N_a \, , \nonumber \\
q_j & = & {2\pi\over L}\, I_j^{\alpha,\,\nu} \,  ; \hspace{1cm}
j=1,...,N_{\alpha,\,\nu} \,  ; \hspace{0.5cm} \alpha = c,\,s \, ,
\hspace{0.5cm} \nu= 1,2,... \,  . \label{qj}
\end{eqnarray}
The $I_j^c$ numbers are integers (half-odd integers), if
${N_a\over 2} - \sum_{\alpha =c,s}\sum_{\nu = 1}^{\infty}N_{\alpha
,\,\nu}$ is odd (even) and the $I_j^{\alpha,\,\nu}$ numbers are
integers (half-odd integers), if $N^*_{\alpha ,\,\nu}$ is odd
(even). The discrete index $j$ can have the following values
$j=1,2,3,...,N_a$ and $j=1,2,3,...,N^*_{\alpha ,\,\nu}$ for the
$c$ and $\alpha,\nu$ pseudoparticles respectively, where the
number of lattice sites $N_a$ and the number

\begin{equation}
N^*_{\alpha,\,\nu} = N_{\alpha,\,\nu} + N^h_{\alpha,\,\nu} \, ,
\label{N*sum}
\end{equation}
where

\begin{equation}
N^h_{\alpha,\,\nu} = L_{\alpha} + 2\sum_{\nu'=\nu +1}^{\infty}
(\nu' -\nu) N_{\alpha,\,\nu'} \, , \label{Nhag}
\end{equation}
give the total number of corresponding discrete band-momentum
values $q_j$. On the right-hand side of Eq. (\ref{N*sum})
$N_{\alpha ,\,\nu}$ and $N^h_{\alpha ,\,\nu}$ are the number of
$\alpha ,\nu$ pseudoparticles and $\alpha ,\nu$ pseudoparticle
holes respectively. On the right-hand side of Eq. (\ref{Nhag})
$L_{\alpha}$ is the number of Yang holons ($\alpha =c$) and HL
spinons ($\alpha =s$) such that $L_{\alpha}=2S_{\alpha}$ where
$S_{\alpha}$ is the $\eta$-spin ($\alpha =c$) and spin ($\alpha
=s$) respectively \cite{I}. On the other hand, the number of $c$
pseudoparticle holes is $N^h_c=[N_a-N_c]$, as given below in Eq.
(\ref{Mcs}). The pseudoparticles obey a Pauli principle respecting
the band-momentum occupancies, {\it i.e.} a discrete band-momentum
value $q_j$ can either be unoccupied or singly occupied by a
pseudoparticle.

It follows from Eq. (\ref{qj}) that the discrete band-momentum
values $q_j$ are such that,

\begin{equation}
q_{j+1}-q_j = {2\pi\over L} \,  . \label{qj1j}
\end{equation}
The band-momentum distribution functions $N_c (q)$ and
$\{N_{\alpha,\,\nu}(q)\}$ where $\alpha =c,s$ and $\nu =1,2,...$
read $N_c (q_j)=1$ and $N_{\alpha ,\,\nu} (q_j)= 1$ for occupied
values of the discrete band momentum $q_j$ and $N_c (q_j)=0$ and
$N_{\alpha ,\,\nu} (q_j)= 0$ for unoccupied values of that band
momentum. The limiting band-momentum value which defines the
continuum band-momentum $q$ range $-q_{\alpha ,\,\nu}\leq q \leq
q_{\alpha ,\,\nu}$ of the $\alpha ,\nu$ pseudoparticle branch is
given by,

\begin{equation}
q_{\alpha ,\,\nu} = {\pi\over L}[N^*_{\alpha ,\,\nu}-1] \approx
{\pi N^*_{\alpha ,\,\nu}\over L} \, . \label{qag}
\end{equation}
The $c$-pseudoparticle continuum band-momentum limiting values
read

\begin{equation}
q_c^{\pm} = \pm q_c \, , \label{qcev}
\end{equation}
for ${N_a\over 2} - \sum_{\alpha =c,s}\sum_{\nu =
1}^{\infty}N_{\alpha ,\,\nu}$ even and

\begin{equation}
q_c^{+} = q_c + {\pi\over L}\, ; \hspace{1cm} q_c^{-} = - q_c +
{\pi\over L} \, , \label{qcodd}
\end{equation}
for ${N_a\over 2} - \sum_{\alpha =c,s}\sum_{\nu =
1}^{\infty}N_{\alpha ,\,\nu}$ odd. Here

\begin{equation}
q_c = {\pi\over a}\bigl[1-{1\over N_a}\bigr] \approx {\pi\over a}
\, . \label{qc}
\end{equation}

The numbers $L_{\alpha ,\,{\sigma}_{\alpha}}$ of $\sigma_c$ Yang
holons ($\alpha =c$) and of $\sigma_s$ HL spinons ($\alpha =s$)
such that $L_{\alpha}=\sum_{{\sigma}_{\alpha}=\pm 1/2}L_{\alpha
,\,{\sigma}_{\alpha}}$ are given by,

\begin{equation}
L_{\alpha ,\,{\sigma}_{\alpha}} = S_{\alpha} -2{\sigma}_{\alpha}\,
S_{\alpha}^z = {L_{\alpha}\over 2} -2{\sigma}_{\alpha}\,
S_{\alpha}^z\, ; \hspace{1cm} \alpha = c,s \, , \label{Las}
\end{equation}
where the values $S_{\alpha}$ of $\eta$ spin ($\alpha =c$) or spin
($\alpha =s$) of the system can be written as,

\begin{equation}
S_{\alpha} = {1\over 2}[L_{\alpha,\,+1/2}+ L_{\alpha,\,-1/2}] =
{1\over 2}M_{\alpha} - \sum_{\nu =1}^{\infty}\nu\,N_{\alpha
,\,\nu} \, ; \hspace{1cm} \alpha = c,s \, . \label{Scs}
\end{equation}
Here $M_{\alpha}$ denotes the total number of holons $M_c$ and of
spinons $M_s$ and $N_{\alpha,\,\nu}$ gives the number of
$\alpha,\nu$ pseudoparticles. The $\eta$-spin ($\alpha =c$) and
spin ($\alpha =s$) projections read,

\begin{eqnarray}
S_c^z & = & -{1\over 2}[L_{c,\,+1/2}- L_{c,\,-1/2}] = -{1\over
2}[M_{c,\,+1/2}- M_{c,\,-1/2}] \, ; \nonumber \\ S_s^z & = &
-{1\over 2}[L_{s,\,+1/2}- L_{s,\,-1/2}] = -{1\over
2}[M_{s,\,+1/2}- M_{s,\,-1/2}] \, , \label{Szhs}
\end{eqnarray}
respectively, where $M_{c,\,\sigma_c}$ and $M_{s,\,\sigma_s}$ give
the total number of $\sigma_c$ holons and of $\sigma_s$ spinons
respectively. These numbers can be written as follows,

\begin{equation}
M_{\alpha,\,\sigma_{\alpha}} = L_{\alpha ,\,\sigma_{\alpha}}+
\sum_{\nu =1}^{\infty} \nu \, N_{\alpha,\,\nu} \, ; \hspace{1cm}
\alpha = c,s \, . \label{Mas}
\end{equation}
The holon and spinon numbers $M_c$ and $M_s$ respectively, can be
expressed as,

\begin{equation}
M_c = \sum_{\sigma_c =\pm 1/2}\,M_{c,\,\sigma_c} = N^h_c = [N_a -
N_c] \, ; \hspace{1cm} M_s = \sum_{\sigma_s =\pm
1/2}\,M_{s,\,\sigma_s} = N_c \, , \label{Mcs}
\end{equation}
where $N_c$ and $N^h_c$ are the numbers of $c$ pseudoparticles and
$c$ pseudoparticle holes respectively.

The above equations are valid for all energy eigenstates. Let us
now consider the particular case of a ground state. In the case of
electronic densities $n$ and spin densities $m$ such that $0\leq
n\leq 1/a$ and $0\leq m\leq n$ respectively, a ground state
belongs to a CPHS ensemble subspace with the values for the
pseudoparticle, $\pm 1/2$ holon, and $\pm 1/2$ spinon numbers
given by $L_{c,\,-1/2} =M_{c,\,-1/2}=0$, $M_{c,\,+1/2} =
L_{c,\,+1/2} = (N_a - N^0)$, $N_c = N^0$, and $N_{c,\,\nu} = 0$ in
the charge sector. In the spin sector the numbers are
$M_{s,\,-1/2} = N^0_{\downarrow}$, $M_{s,\,+1/2} =
N^0_{\uparrow}$, $L_{s,\,+1/2}=(N^0_{\uparrow}-N^0_{\downarrow})$,
$N_{s,\,1} = N^0_{\downarrow}$, and $L_{s,\,-1/2}= N_{s,\,\nu} =
0$. In these expressions the electron numbers $N^0_{\downarrow}$,
$ N^0_{\downarrow}$, and $N^0=N^0_{\downarrow}+N^0_{\downarrow}$
refer to the ground state. In the case of the ground-state CPHS
ensemble subspace the expression of the number
$N^*_{\alpha,\,\nu}$ given in Eqs. (\ref{N*sum}) and (\ref{Nhag})
also simplifies \cite{II} and reads,

\begin{equation}
N^*_{c,\,\nu}=(N_a -N^0) \, ; \hspace{1cm}
N^*_{s,\,1}=N^0_{\uparrow} \, ; \hspace{1cm}
N^*_{s,\,\nu}=(N^0_{\uparrow} -N^0_{\downarrow}) \, ,
\hspace{0.3cm} \nu > 1 \, , \label{N*csnu}
\end{equation}
whereas $N^*_c$ is given by $N^*_c=N_a$ for all energy
eigenstates. It follows that also the expressions of the limiting
band-momentum values (\ref{qag})-(\ref{qc}) simplify in the case
of the ground-state CPHS ensemble subspace. Disregarding
corrections of the order of $1/L$, in the case of the ground state
these limiting band-momentum values can be written as $\pm q^0_c$
and $\pm q^0_{\alpha,\,\nu}$ where,

\begin{eqnarray}
q^0_c & = & \pi/a \, ; \hspace{1cm} q^0_{s,\,1} =
k_{F\uparrow} \, ;  \nonumber \\
q^0_{c,\,\nu} & = & [\pi/a -2k_F]  \, ; \hspace{1cm} q^0_{s,\,\nu}
= [k_{F\uparrow}-k_{F\downarrow}] \, , \hspace{0.25cm} \nu >1 \, .
\label{qcanGS}
\end{eqnarray}
On the right-hand side of Eq. (\ref{qcanGS}) the Fermi momenta are
such that $2k_F=k_{F\uparrow}-k_{F\downarrow}$,
$k_{F\uparrow}=\pi\,n_{\uparrow}$, and
$k_{F\downarrow}=\pi\,n_{\downarrow}$. In most situations one can
disregard the $1/L$ corrections and use the band-momentum limiting
values given in Eq. (\ref{qcanGS}).

\section{ENERGY EIGENSTATES AND BASIC PROPERTIES ASSOCIATED WITH THE
PSEUDOPARTICLE INTERNAL STRUCTURE AND {\it EMPTY SITES}}

As mentioned in Secs. I and II, the energy eigenstates of the 1D
Hubbard model are described by the the same rotated-electron site
distribution configurations for all values of $U/t$. Fortunately,
since the electron - rotated electron unitary transformation
becomes the unit transformation as $U/t\rightarrow\infty$, one can
reach the $U/t$ independent rotated-electron site distribution
configurations which describe the energy eigenstates by studying
the corresponding electron site distribution configurations for
the 1D Hubbard model in the limit $U/t\rightarrow\infty$. Thus the
main goal of this section is to extract from the physics
associated with the $U/t\rightarrow\infty$ limit of the model
useful information for the construction of the rotated-electron
occupancy configurations that describe the energy eigenstates for
finite values of $U/t$.

In the limit $U/t\rightarrow\infty$ there is a huge degeneracy of
$\eta$-spin and spin occupancy configurations. Such a degeneracy
results from the simple form that the kinetic energy $T$ and
potential energy $V$ spectra associated with the operators
(\ref{Top}) and (\ref{Dop}) respectively, have in the limit
$U/t\rightarrow\infty$,

\begin{equation}
E_H = T + V \, ; \hspace{1cm} T = 2t \sum_{j=1}^{N_a}N_c
(q_j)[-2t\cos q_j] \, ; \hspace{1cm} V/U = D\, , \label{EHUinf}
\end{equation}
where $E_H$ is the energy spectrum of the Hamiltonian (\ref{HH})
and the electron double occupation $D$ is a good quantum number
which equals rotated-electron double occupation in that limit. The
kinetic energy equals that of a system of free spin-less fermions
\cite{Ogata,Penc95,Penc96,Penc97}. The corresponding momentum
spectrum is given in Eq. (\ref{PpUinf}) of Appendix A. A short
discussion of the relation of our results to some of the well
known concepts of the $U/t\rightarrow\infty$ physics is presented
in that Appendix. Since the energy spectrum (\ref{EHUinf}) is
independent of the $\eta$-spin and spin occupancy configurations,
there are several choices for complete sets of energy eigenstates
with the same energy and momentum spectra. However, only one of
these choices is associated with the rotated-electron site
distribution configurations which describe the energy eigenstates
for all values of $U/t$. The discussion presented in Appendix A
and in this section about the relation of the energy-eigenstate
rotated-electron description introduced in this paper to other
choices of energy eigenstates and to the $t/U<<1$ physics
associated with the leading-order $t/U$ corrections to the energy
spectrum (\ref{EHUinf}) contextualizes the quantum problem studied
in this paper and contributes to its deeper understanding.

The pseudoparticle band momentum obeys well defined boundary
conditions which are a necessary condition for the fulfilment of
the periodic boundary conditions for the original electrons.
However, such a pseudoparticle band-momentum boundary conditions
are not a sufficient condition to ensure the electronic periodic
boundary conditions. A second condition imposes that the internal
structure of the local $\alpha,\nu$ pseudoparticles introduced in
later sections must be of a specific form. In this section we
introduce a set of properties which are rather useful both for the
construction of the specific rotated-electron site distribution
configurations which describe the internal structure of such local
$\alpha,\nu$ pseudoparticles and the definition of the
corresponding pseudoparticle {\it empty sites} in terms of
occupancy configurations of rotated-electron sites. As further
discussed in Sec. VI, the construction of the rotated-electron
site distribution configurations which describe the energy
eigenstates and the introduction of the associated concepts of
local pseudoparticle and effective pseudoparticle lattice are
necessary steps for the evaluation of few-electron spectral
functions at finite values of excitation energy.

Our choice of energy eigenstates corresponds to the band-momentum
energy eigenstates associated with the thermodynamic Bethe-ansatz
equations introduced by Takahashi \cite{Takahashi}. As discussed
in Appendix A, for $U/t\rightarrow\infty$ there are other choices
for complete sets of energy eigenstates. One of the alternative
complete set of energy and momentum eigenstates for the model in
the limit $U/t\rightarrow\infty$ can be explicitly constructed for
the Harris-Lange model mentioned in Appendix A by the method used
in Ref. \cite{Geb}. Both the band-momentum energy eigenstates and
the {\it symmetrized energy eigenstates} used in that reference
are superpositions of charge (and spin) sequences formed by local
electron distribution configurations of doubly occupied and empty
sites (and spin-down and spin-up singly occupied sites). The
expression of both these two sets of energy eigenstates ensures
the periodic boundary conditions for the original electronic
problem.

Through out this paper we denote the rotated-electron doubly
occupied and empty sites by $\bullet$ and $\circ$ (and the
spin-down and spin-up rotated-electron singly occupied sites by
$\downarrow$ and $\uparrow$) respectively. In the present
$U/t\rightarrow\infty$ limit such a concept also refers to
electrons. Often we add an index to these symbols which defines
the position of the doubly occupied site or empty site (and
spin-down singly occupied site or spin-up singly occupied site).
In the case of the {\it symmetrized} energy eigenstates the charge
and spin sequences are properly symmetrized owing to the periodic
boundary conditions of the original problem. (This justifies the
designation of {\it symmetrized energy eigenstates}.) The
procedure used in such a symmetrization involves powers of
suitable charge and spin operators ${\hat{\cal{T}}}_C$ and
${\hat{\cal{T}}}_S$ respectively, for cyclic permutation of the
electron site distribution configurations of the local charge and
spin sequences \cite{Geb}. For instance, let

\begin{equation}
(\bullet,\,\bullet,\,\circ,\,\bullet,\,\circ,...,\circ,\,\bullet,\,\bullet)
\, , \label{cseq}
\end{equation}
and

\begin{equation}
(\downarrow,\,\uparrow,\,\downarrow,\,\uparrow,\,\uparrow,...,
\downarrow,\,\downarrow,\,\uparrow) \, , \label{sseq}
\end{equation}
be a charge and a spin sequence respectively. Then the operators
${\hat{\cal{T}}}_C$ and ${\hat{\cal{T}}}_S$ are such that,

\begin{equation}
{\hat{\cal{T}}}_C\,(\bullet,\,\bullet,\,\circ,\,\bullet,\,\circ,...,\circ,
\,\bullet,\,\circ) =
(\circ,\,\bullet,\,\bullet,\,\circ,\,\bullet,\,\circ,...,\circ,\,\bullet)\,
, \label{Tcseq}
\end{equation}
and

\begin{equation}
{\hat{\cal{T}}}_S\,(\downarrow,\,\uparrow,\,\downarrow,\,\uparrow,\,
\uparrow,...,\downarrow,\,\downarrow,\,\uparrow) =
(\uparrow,\,\downarrow,\,\uparrow,\,\downarrow,\,\uparrow,\,\uparrow,...,
\downarrow,\,\downarrow)\, , \label{Tsseq}
\end{equation}
respectively. The powers $[{\hat{\cal{T}}}_C]^{K_C}$ and
$[{\hat{\cal{T}}}_S]^{K_S}$ were called in Ref. \cite{Geb} $K_C$
and $K_S$ respectively. This introduces the {\it charge momentum}
$k_C$ and {\it spin momentum} $k_S$ respectively, such that,

\begin{eqnarray}
k_C & = & {2\pi\over K_C\,a}\,m_C \, ; \hspace{1cm} m_C =
0,1,...,K_C-1 \, , \nonumber \\
k_S & = & {2\pi\over K_S\,a}\,m_S \, ; \hspace{1cm} m_S =
0,1,...,K_S-1 \, . \label{kCkS}
\end{eqnarray}

The symmetrized energy eigenstates are classified according to
their charge and spin sequence, their charge momentum $k_C$ and
spin momentum $k_S$, and a number $N_C$ of momenta which in the
notation of Ref. \cite{Geb} equals the number of charges. The
latter discrete momenta are closely related to the discrete
band-momentum values occupied by $c$ pseudoparticle holes in the
pseudoparticle representation of the band-momentum energy
eigenstates studied in later sections. Moreover, the number $N_C$
of charges equals both the number $N^h_c$ of $c$ pseudoparticle
holes and the number $M_c$ of holons. Thus this is a good quantum
number for both the band-momentum and symmetrized energy
eigenstates. However, the charge momentum $k_C$ and spin momentum
$k_S$ are eigenvalues of the charge momentum operator
${\hat{k}}_C$ and spin momentum operator ${\hat{k}}_S$
respectively, which in general do not commute with the set of
operators $\{{\hat{N}}_{\alpha,\,\nu}(q_j)\}$ of the band-momentum
basis. Thus in general the band-momentum energy eigenstates are
not eigenstates of the charge momentum operator ${\hat{k}}_C$ and
spin momentum operator ${\hat{k}}_S$. Exceptions are the
band-momentum energy eigenstates with occupancy of a single
$c,\nu$ pseudoparticle (and a single $s,\nu$ pseudoparticle) and
with no finite occupancy of Yang holons (and HL spinons) and of
$c,\nu'$ pseudoparticles (and $s,\nu'$ pseudoparticles) belonging
to other branches such that $\nu'\neq\nu$. We find below that in
this case the corresponding band-momentum energy eigenstate is
also an eigenstate of the charge (and spin) momentum operator of
eigenvalue $k_C =\pi /a$ (and $k_S =\pi /a$).

According to Eq. (\ref{qj1j}) the discrete values of the band
momentum $q_j$ are such that $q_{j+1}-q_j = 2\pi/L$ and
$q_j=[2\pi/L]\,I^c_j$ or $q_j=[2\pi/L]\,I^{\alpha,\,\nu}_j$ where
the numbers $I^c_j$ and $I^{\alpha,\,\nu}_j$ with $j=1,2,...,N_a$
and $j=1,2,...,N^*_{\alpha,\,\nu}$ respectively, are integers or
half-odd integers as a result of the following boundary
conditions,

\begin{equation}
e^{iq_j\,L}=(e^{i\pi})^{\textstyle [
\sum_{\alpha=c,\,s}\sum_{\nu=1}^{\infty}N_{\alpha,\,\nu}]} \, ,
\label{pbccp}
\end{equation}
in the case of the $c$ pseudoparticle branch and,

\begin{equation}
e^{iq_j\,L}=(e^{i\pi})^{[1+N^*_{\alpha,\,\nu}]}=(e^{i\pi})^{[1+L_{\alpha}+
N_{\alpha,\,\nu}]}=(e^{i\pi})^{[1+N_c + N_{\alpha,\,\nu}]} \, ;
\hspace{1cm} \alpha =c,s \, , \hspace{0.5cm} \nu =1,2,... \, ,
\label{pbcanp}
\end{equation}
in the case of the $\alpha,\nu$ pseudoparticle branches. Here
$N_{\alpha,\,\nu}$ and $L_{\alpha}$ denote the number of
$\alpha,\,\nu$ pseudoparticles and of Yang holons ($\alpha =c$) or
HL spinons ($\alpha =s$) respectively, and the number
$N^*_{\alpha,\,\nu}$ is given in Eq. (\ref{N*sum}). According to
Eq. (\ref{qag}), in the case of the $\alpha,\nu$ pseudoparticle
branches the $j=1$ minimum and $j=N^*_{\alpha,\,\nu}$ maximum
index of the band-momentum values $q_j$ are such that,

\begin{equation}
-q_1 = q_{N^*_{\alpha,\,\nu}} = q_{\alpha ,\,\nu} = {\pi\over
L}[N^*_{\alpha ,\,\nu}-1] \, . \label{limits}
\end{equation}
In the case of the $c$ pseudoparticles the limiting values are
$q_1=q_c^{-}$ and $q_{N_a}=q_c^{+}$ where the band momenta
$q_c^{\pm}$ are defined in Eqs. (\ref{qcev})-(\ref{qc}).

There is a holon, spinon, $c$ pseudoparticle separation for the
whole parameter space of the 1D Hubbard model \cite{I}. In the
case of the $t/U\rightarrow 0$ limit the description of the $c$
pseudoparticle excitation sector is very similar for both the
representations in terms of band-momentum energy eigenstates and
of symmetrized energy eigenstates. The separation of the charge
and spin excitation sectors occurs for these two alternative
representations. The construction of the band-momentum energy
eigenstates also involves superpositions of charge (and spin)
sequences associated with rotated-electron distribution
configurations of doubly occupied and empty sites (and spin-down
and spin-up singly occupied sites). Such superpositions are also
due to the periodic boundary conditions but are not in general
generated by the above symmetrization procedure. In spite of these
general similarities, the symmetrized and band-momentum energy
eigenstates of the 1D Hubbard model in the limit of
$t/U\rightarrow 0$ are different states. The difference between
the symmetrized and band-momentum energy eigenstates refers in
general both to the form of the local electron distribution
configurations of doubly occupied and empty sites (and spin-down
and spin-up singly occupied sites) which describe the charge (and
spin) sequences and to the form of the superposition of these
local sequences which describes the energy eigenstates.

Let us introduce eight properties which correspond to a first step
of the introduction of the concepts of local pseudoparticle and
effective pseudoparticle lattices. However, the precise definition
of these concepts involves clarification of several issues and is
only fulfilled in Sec. V. These properties are used in the ensuing
section in finding the electron site distribution configurations
of the local charge and spin sequences whose Fourier-transform
superpositions describe the band-momentum energy eigenstates of
the model in the limit $t/U\rightarrow 0$. These properties follow
in part from symmetries and features of the pseudoparticle, holon,
and spin description and related rotated electron representation
studied in Refs. \cite{I,II,III} and from well known properties
associated with the $t/U\rightarrow 0$ physics. We emphasize that
the properties given below also apply to finite values of $t/U$
provided that electrons are replaced by rotated electrons. These
useful properties read:\vspace{0.5cm}

1-III The numbers of electron doubly occupied sites, empty sites,
spin-down singly occupied sites, and spin-up singly occupied sites
are good quantum numbers whose values are equal to the total
numbers of $-1/2$ holons, $+1/2$ holons, $-1/2$ spinons, and
$+1/2$ spinons respectively, of the band-momentum energy
eigenstates.\vspace{0.5cm}

2-III The kinetic energy $T$ given in Eq. (\ref{EHUinf}) arises
from the movements of the singly occupied sites relative to the
doubly occupied and empty sides which do not change double
occupation. These movements are fully described by the $c$
pseudoparticles which are associated with the charge degrees of
freedom of these sites. In the limit $t/U\rightarrow 0$ these
quantum objects acquire a spin-less fermion spectrum. On the other
hand, the electron distribution configurations of doubly occupied
and empty sites and of spin-down and spin-up singly occupied sites
do not contribute to the kinetic energy $T$. Moreover these
electron site distribution configurations must remain unchanged in
spite of the movements of the $c$ pseudoparticles. Alternatively,
we can consider that the electron doubly occupied and empty sites
move relative to the singly occupied sites. In this case one
describes the movements of the $c$ pseudoparticles in terms of the
movements of the corresponding $c$ pseudoparticle holes. This is
the choice of Ref. \cite{Geb} for the case of the symmetrized
energy eigenstates. We recall that the numbers $N_c$ of $c$
pseudoparticles, $N^h_c$ of $c$ pseudoparticle holes, $M_s$ of
spinons, and $M_c$ of holons are such that $N_c+N^h_c=N_a$, $N_c
=M_s$, and $N^h_c=M_c$ and thus these two alternative descriptions
are fully equivalent. \vspace{0.5cm}

3-III We call {\it local charge sequences} and {\it local spin
sequences} the occupancy configurations of the $\pm 1/2$ holons
(and corresponding electron distribution configurations of doubly
occupied and empty sites) and the occupancy configurations of the
$\pm 1/2$ spinons (and corresponding electron distribution
configurations of singly occupied sites of spin projection $\pm
1/2$) respectively. These local charge (and spin) sequences can
also be expressed in terms of occupancy configurations of $\pm
1/2$ Yang holons and local $c,\nu$ pseudoparticles (and $\pm 1/2$
HL spinons and local $s,\nu$ pseudoparticles) belonging to
$\nu=1,2,3...$ branches. From the general properties introduced in
Ref. \cite{I} one finds that the electron description of these
quantum objects in terms of distribution configurations of
electron doubly occupied and empty sites (and spin-down and
spin-up singly occupied sites) is as follows: A $-1/2$ Yang holon
(and $-1/2$ HL spinon) is described by a doubly occupied site (and
a spin-down singly occupied site); A $+1/2$ Yang holon (and $+1/2$
HL spinon) is described by an empty site (and a spin-up singly
occupied site); A local $c,\nu$ pseudoparticle (and a local
$s,\nu$ pseudoparticle) is described by a number $\nu$ of doubly
occupied sites and a number $\nu$ of empty sites (and a number
$\nu$ of spin-down singly occupied sites and a number $\nu$ of
spin-up singly occupied sites). The electron distribution
configurations of doubly occupied and empty sites (and of
spin-down and spin-up singly occupied sites) of any local charge
(and spin) sequence can be expressed in terms of a corresponding
occupancy configuration of $\pm 1/2$ Yang holons and local $c,\nu$
pseudoparticles (and $\pm 1/2$ HL spinons and local $s,\nu$
pseudoparticles) belonging to the $\nu=1,2,3...$ branches. The
form of Eqs. (\ref{Pp}) and (\ref{PpUinf}) of Appendix A reveals
that a $-1/2$ holon carries momentum $\pi/a$ and that in the limit
$t/U\rightarrow 0$ each doubly occupied site must also be
associated with a momentum $\pi/a$ respectively. Therefore, the
expressions of the local charge sequences must include a phase
factor operator $\exp (i\pi\sum_{j}j\,{\hat{D}}_j)$ such that the
$j$ summation runs over the sites doubly occupied and empty and
the local double occupation operator has eigenvalues $1$ and $0$
if the site $j$ is doubly occupied and empty
respectively.\vspace{0.5cm}

4-III The number of different possible discrete locations of the
electron distribution configurations of doubly occupied and empty
sites ($\alpha =c$) or of spin-down and spin-up singly occupied
sites ($\alpha =s)$ describing a local $\alpha ,\nu$
pseudoparticle equals the number $N^*_{\alpha,\,\nu}$ given in Eq.
(\ref{N*sum}). Each of these possible positions corresponds to a
different local charge ($\alpha =c$) or spin ($\alpha =s$)
sequence. The number $N^*_{\alpha,\,\nu}$ is directly provided by
the Bethe-ansatz solution \cite{I} since it also equals the number
of different discrete band momentum values $q_j$, where
$j=1,2,...,N^*_{\alpha,\,\nu}$, of the band-momentum $\alpha ,\nu$
pseudoparticle band.\vspace{0.5cm}

5-III In case of energy eigenstates with finite occupancy of
$-1/2$ and $+1/2$ Yang holons (and $-1/2$ and $+1/2$ HL spinons)
the electron distribution configurations of doubly occupied and
empty sites (and spin-down and spin-up singly occupied sites)
describing the local $c,\nu$ pseudoparticle (and local $s,\nu$
pseudoparticle) must remain unchanged under the application of the
off-diagonal generators of the $SU(2)$ $\eta$-spin algebra given
in Eq. (\ref{Sc}) (and off-diagonal generators of the $SU(2)$ spin
algebra given in Eq. (\ref{Ss})). Moreover, in case of states with
no $+1/2$ or $-1/2$ Yang holons (and no $+1/2$ or $-1/2$ HL
spinons) application of the operators
${\hat{S}}^c_{+}=\sum_{j}(-1)^j\, c_{j,\,\downarrow}^{\dag}
c_{j,\,\uparrow}^{\dag}$ or ${\hat{S}}^c_{-}=\sum_{j}(-1)^j\,
c_{j,\,\uparrow}c_{j,\,\downarrow}$ (and ${\hat{S}}^s_{+}=
\sum_{j}c_{j,\,\downarrow}^{\dag}c_{j,\,\uparrow}$ or
${\hat{S}}^s_{-}=\sum_{j}c_{j,\,\uparrow}^{\dag}
c_{j,\,\downarrow}$) onto these states must give zero. These
requirements result from the value of the $\eta$ spin of the
$c,\nu$ pseudoparticles (and spin of the $s,\nu$ pseudoparticles)
which is given by $s_c=0$ (and $s_s=0$) \cite{I}. On the other
hand, the transformations generated by application of these
off-diagonal generators onto the electron distribution
configurations of doubly occupied and empty sites (and spin-down
and spin-up singly occupied sites) describing the $\pm 1/2$ Yang
holons (and $\pm 1/2$ HL spinons) must be the ones defined by the
$\eta$-spin (and spin) algebra \cite{I}. \vspace{0.5cm}

6-III Let us consider a local charge (and spin) sequence with no
Yang holons (and no HL spinons) and consisting of $\nu$ electron
doubly occupied sites and $\nu$ electron empty sites (and $\nu$
electron spin-down singly occupied sites and $\nu$ electron
spin-up singly occupied sites). If such a sequence describes a
single local $c,\nu$ pseudoparticle (and local $s,\nu$
pseudoparticle) it is properly symmetrized in such way that the
distribution configurations associated with the internal structure
of that quantum object remain unchanged under cyclic permutations.
Furthermore, property 5-III imposes that a band-momentum energy
eigenstate with the above Yang holon and $c,\nu$ pseudoparticle
(and HL spinon and $s,\nu$ pseudoparticle) numbers is an
eigenstate of the charge (and spin) momentum operator
${\hat{k}}_C$ (and ${\hat{k}}_S$) of eigenvalue $k_C =\pi /a$ (and
$k_S =\pi /a$). \vspace{0.5cm}

7-III The electron distribution configurations of the $\nu$ doubly
occupied sites and $\nu$ empty sites (and $\nu$ spin-down singly
occupied sites and $\nu$ spin-up singly occupied sites) which
describe the internal structure a local $c,\nu$ pseudoparticle
(and $s,\nu$ pseudoparticle) should be the same for all local
charge ($\alpha =c$) or spin ($\alpha =s$) sequences involved in
the description of the $4^{N_a}$ band-momentum energy eigenstates.
This is a necessary condition for the indiscernible character of
local $\alpha,\nu$ pseudoparticles with the same lattice position
but involved in occupancy configurations describing different
local charge ($\alpha =c$) or spin ($\alpha =s$) sequences. This
indiscernible character of the local $\alpha,\nu$ pseudoparticles
follows from the corresponding indiscernible character of the
$\alpha,\nu$ pseudoparticles of band-momentum $q_j$, which are
indistinguishable quantum objects.\vspace{0.5cm}

8-III Since in the limit $t/U\rightarrow 0$ there is
nearest-neighbor hopping only and it does not change double
occupation, the charge, spin, $c$ pseudoparticle separation
studied in Ref. \cite{I} implies that both the local charge and
spin sequences of the energy eigenstates must be separately
conserved. Moreover, the periodic boundary conditions of the
original electronic problem are ensured if both the pseudoparticle
band-momentum discrete values obey Eqs. (\ref{pbccp}) and
(\ref{pbcanp}) and the requirements of the basic properties 6-III
and 7-III are fulfilled. The band-momentum energy eigenstates are
Fourier transform superpositions of local charge sequences, spin
sequences, and $c$ pseudoparticle sequences corresponding to
Slater determinants involving the pseudoparticle band momentum
$q_j$ and the spatial coordinate $x_j$ of these quantum objects.
Both in the case of the $c$ pseudoparticle branch and of the
$\alpha,\nu$ pseudoparticle branches with finite occupancy in a
given state, such spatial coordinate is the conjugate of the band
momentum $q_j$ of the above Fourier transforms. The spatial
coordinate $x_j$ of the $c$ pseudoparticles (and $\alpha,\nu$
pseudoparticles) is associated with an effective pseudoparticle
lattice of length $L$, number of lattice sites $N_a$ (and
$N^*_{\alpha,\,\nu}$ given in Eq. (\ref{N*sum})) and lattice
constant $a$ (and $a_{\alpha,\,\nu}=L/N^*_{\alpha,\,\nu}$). The
possible values of these spatial coordinates are $x_j=a\,j$ where
$j=1,2,3,...,N_a$ (and $x_j=a_{\alpha,\,\nu}\,j$ where
$j=1,2,3,...,N^*_{\alpha,\,\nu}$). The effective pseudoparticle
lattices arise because the occupancy configurations of the
different pseudoparticle branches are separately
conserved.\vspace{0.5cm}

These eight basic properties play an important role in the
mechanisms which in the $t/U\rightarrow 0$ limit determine the
choice of the electron site distribution configurations of the
local charge, spin, and $c$ pseudoparticle sequences whose Fourier
transform superpositions describe the band-momentum energy
eigenstates. Property 1-III follows from well known properties of
the 1D Hubbard model in the limit $t/U\rightarrow 0$. Property
2-III is consistent with the finding of Ref. \cite{II} that in the
limit $t/U\rightarrow 0$ only the $c$ pseudoparticles move and
{\it carry} kinetic energy, whereas the $\pm 1/2$ holons and $\pm
1/2$ spinons correspond to unchanged occupancy configurations in
that limit. Such a property results from well established features
of the Bethe-ansatz solution \cite{Ogata,Penc97,II}. Property
3-III is a consequence of the combination of property 1-III with
the relation of Yang holons, HL spinons, and $\alpha ,\nu$
pseudoparticles to $\pm 1/2$ holons and $\pm 1/2$ spinons. In
property 4-III the number $N^*_{\alpha,\,\nu}$ given in Eq.
(\ref{N*sum}) plays a central role. The expression of that number
is valid for all values of $U/t$ and is provided by the Bethe
ansatz solution. This is consistent with the electron site
distribution configurations which describe the local $\alpha ,\nu$
pseudoparticles in the limit of $t/U\rightarrow 0$ being the same
as the rotated-electron configurations which describe these local
$\alpha ,\nu$ pseudoparticles for finite values of $t/U$. Property
5-III follows from general symmetries of the 1D Hubbard model
which are also valid for all values of $U/t$. This basic property
is also consistent with the above equivalence of the electron site
distribution configurations in the limit of $t/U\rightarrow 0$ and
of the rotated-electron configurations for finite values of $t/U$.
Property 6-III is a consequence of the periodic boundary
conditions in the particular case when a local charge or spin
sequence corresponds to a single $c,\nu$ or $s,\nu$ pseudoparticle
respectively. Property 7-III results from the indiscernible
character of the $\alpha,\nu$ pseudoparticles. In the case of
band-momentum pseudoparticles such an indiscernible character is
implicit in the description of the band-momentum energy
eigenstates in terms of pseudoparticle occupancy configurations
\cite{I,II}. Finally, property 8-III is related to the periodic
boundary conditions of the original electronic problem and to the
existence in this limit of nearest-neighbor hopping only which
does not change double occupation. Such a property is also related
to the charge, spin, $c$ pseudoparticle separation studied in Ref.
\cite{I} for all energy scales and for the whole parameter space
of the 1D Hubbard model. In addition, in the case of the
$\alpha,\nu$ pseudoparticle branches, we find below that the
occurrence of the effective pseudoparticle lattices mentioned in
this property is related to the separation of the internal and
translational degrees of freedom of the local $\alpha,\nu$
pseudoparticles.

Since in the $t/U\rightarrow 0$ limit the rotated electrons become
the electrons, often below we refer to the rotated-electron site
distribution configurations only. However, whenever referring to
rotated electrons we mean implicitly that in the limit
$t/U\rightarrow 0$ the electron site distribution configurations
which describe the band-momentum energy eigenstates are the same.

\section{THE PSEUDOPARTICLE INTERNAL STRUCTURE AND COMPLETE SET OF
LOCAL STATES IN TERMS OF CHARGE, SPIN, AND $c$ PSEUDOPARTICLE
SEQUENCES}

In this section we find the rotated-electron site distribution
configurations which describe the local charge, spin, and $c$
pseudoparticle sequences. This requires the study of the internal
structure of the local $\alpha,\nu$ pseudoparticles. In addition,
such an analysis requires the definition of the positions in the
effective electronic lattice of the rotated-electron site
distribution configurations which describe both the local $c$ and
$\alpha,\nu$ pseudoparticles and their {\it empty sites}. We find
that each specific site distribution configurations of a local
charge, spin, and $c$ pseudoparticle sequence defines a local
state. The set of all different possible such states constitutes a
complete set of local states. In the ensuing section we express
the energy eigenstates of the model as a Fourier-transform
superposition of these local states. Such a study reveals the
connection of the energy eigenstates to the rotated electron
distribution configurations of doubly occupied sites, empty sites,
spin-down singly occupied sites, and spin-up singly occupied sites
which describe the local states.

\subsection{THE LOCAL $\alpha,\nu$ PSEUDOPARTICLE INTERNAL STRUCTURE}

In the following we use in part the basic properties 1-III to
8-III to study the rotated-electron site distribution
configurations which describe the local pseudoparticles introduced
in this paper. We start by restricting our study to
rotated-electron site distribution configurations which describe
the lowest-weight states of both the $SU(2)$ $\eta$-spin and spin
algebras. These states have no $-1/2$ Yang holons and no $-1/2$ HL
spinons. Thus in this case the number of Yang holons (and of HL
spinons) is such that $L_c=L_{c,+1/2}$ (and $L_s=L_{s,+1/2}$).

In the limit $U/t\rightarrow\infty$ the $N_c$ $c$ pseudoparticles
behave as spin-less fermions in a lattice of
$N_a=N_c+N^h_c=M_s+M_c$ sites. Such a lattice is nothing but the
effective $c$ pseudoparticle lattice mentioned in property 8-III.
As in the case of the band-momentum $q_j$, such an effective $c$
pseudoparticle lattice and its site coordinates $x_j$ remain the
same for the whole parameter space of the model. Also the $c$
pseudoparticle occupancy configurations of such a lattice
describing a given energy eigenstate are the same for all values
of $U/t$, density $n$, and spin density $m$. At a fixed value
$N_c=M_s$ of the numbers $N_c$ of $c$ pseudoparticles and $M_s$ of
spinons the number of occupancy configurations of the local $c$
pseudoparticles in such an effective lattice is given by,

\begin{equation}
{N_a\choose N_{c}} = {N_a!\over N_c!\,N^h_c!} = {N_a!\over
M_s!\,M_c!}\, . \label{NcNhc}
\end{equation}
Each of these $c$ pseudoparticle occupancy configurations of the
effective $c$ pseudoparticle lattice defines a {\it local $c$
pseudoparticle sequence}.

Let $x_{j_l}=a\,j_l$ where $l=1,2,..., N_c$ be the actual set of
occupied coordinates of the effective $c$ pseudoparticle lattice
out of the available $x_j=a\,j$ coordinate sites where $j=1,2,...,
N_a$. We note that these $N_a$ sites have the same coordinates and
lattice constant $a$ as the $N_a$ sites of the effective
electronic lattice. Moreover, the set of coordinates
$x_{j_l}=a\,j_l$ where $l=1,2,..., N_c$ corresponding to the
occupied sites of the effective $c$ pseudoparticle lattice are
precisely the same as the coordinates of the rotated electron
singly occupied sites. Thus we can define each of the occupancy
configurations whose total number is given in Eq. (\ref{NcNhc}) by
the coordinates in units of the lattice constant $a$ of the
rotated-electron singly occupied sites,

\begin{equation}
(j_{1},\,j_2,...,j_{N_c}) \, . \label{lcp}
\end{equation}
The positions occupied by the rotated-electron doubly occupied and
empty sites are then these left over by the rotated-electron
singly occupied sites. Since the sequence (\ref{lcp}) also gives
the positions of the local $c$ pseudoparticles in their effective
lattice, the sites left over by these pseudoparticles define the
positions of the local $c$ pseudoparticle holes. Thus the relation
of the numbers $N_c$ and $N_c^h$ of local $c$ pseudoparticles and
local $c$ pseudoparticle holes respectively, to the
rotated-electron site distribution configurations of a given
energy eigenstate confirms that there are $N_c$ rotated-electron
singly occupied sites and $N^h_c=[N_a-N_c]$ rotated-electron
doubly-occupied and empty sites, as mentioned in Sec. II. From the
use of Eqs. (\ref{Mas}) and (\ref{Mcs}) we find that these numbers
can be expressed as follows,

\begin{eqnarray}
N^h_c=N_a - N_c & = & L_c + 2\sum_{\nu =1}^{\infty} \nu \,
N_{c,\,\nu}
\, ; \nonumber \\
N_c & = & L_s + 2\sum_{\nu =1}^{\infty} \nu \, N_{s,\,\nu} \, .
\label{LcsLWS}
\end{eqnarray}

The local charge sequences (and spin sequences) introduced in
property 3-III involve only the rotated-electron doubly-occupied
and empty site distribution configurations (and spin-down and
spin-up rotated-electron singly occupied site distribution
configurations) of $N^h_c=[N_a-N_c]$ sites (and $N_c$ sites) out
of a total number $N_a$ of sites. As a result of the independent
conservation of the charge and spin sequences, the
rotated-electron site distribution configurations of a charge (and
spin) sequence is for any energy eigenstate obtained simply by
omitting the $N_c$ singly occupied sites (and $N^h_c=[N_a-N_c]$
doubly-occupied and empty sites). Thus a local charge (and spin)
sequence corresponds to the rotated-electron occupancy
configurations of a site domain of a number $N^h_c=[N_a-N_c]$ of
sites (and $N_c$ of sites). It follows that the number of sites of
such a domain depends on the specific state under consideration.
In the case of the ground state numbers provided in Sec. II, the
number of sites of the effective electronic lattice which belong
to the charge and spin sequences is $[N_a-N^0]$ and $N^0$
respectively.

Also the effective $\alpha,\nu$ pseudoparticle lattices mentioned
in property 8-III and introduced below, result from independent
conservation laws associated with each $\alpha,\nu$ pseudoparticle
branch such that $\alpha=c,\,s$ and $\nu=1,2,...$ branch. These
effective $c,\nu$ pseudoparticle (and $s,\nu$ pseudoparticle)
lattices are generated below by omission of a number
$\sum_{\nu'=1}^{\infty} \Bigl[\nu + \nu' - \vert\nu -
\nu'\vert\Bigl] N_{c,\,\nu'}-N_{c,\,\nu}$ (and
$\sum_{\nu'=1}^{\infty} \Bigl[\nu + \nu' - \vert\nu -
\nu'\vert\Bigl] N_{s,\,\nu'}-N_{s,\,\nu}$) of sites out of the
total number $N^h_c=[N_a-N_c]$ (and $N_c$) of sites of the local
charge sequence (and spin sequence). Again the number of sites of
the local charge (and spin) sequence which contribute to an
effective $c,\nu$ pseudoparticle (and $s,\nu$ pseudoparticle)
lattice depends on the specific state under consideration. For
instance, in the case of a ground state the number of sites of the
local charge sequence (and spin sequence) which contribute to an
effective $c,\nu$ pseudoparticle (and $s,\nu$ pseudoparticle)
lattice is $[N_a-N^0]$ (and $N^0_{\uparrow}$ for the $s,1$
pseudoparticle branch and $[N^0_{\uparrow}-N^0_{\downarrow}]$ for
the $s,\nu$ pseudoparticle branches such that $\nu>1$) out of the
$[N_a-N^0]$ sites (and $N^0=[N^0_{\uparrow}+N^0_{\downarrow}]$
sites) of such a sequence. On the other hand, the number of sites
of the effective $c$ pseudoparticle lattice equals the number
$N_a$ of sites of the effective electronic lattice and is the same
for all energy eigenstates.

The expressions of Eq. (\ref{LcsLWS}) show that the value of the
number $L_c$ of Yang holons (and $L_s$ of HL spinons) of the local
charge (and spin) sequence is uniquely determined by the values of
the numbers of $c$ pseudoparticles and $c,\nu$ pseudoparticles
($c$ pseudoparticles and $s,\nu$ pseudoparticles) of the same
sequence. Thus for given number $N^h_c=[N_a+N_c]$ of
rotated-electron doubly-occupied and empty sites (and $N_c$ of
rotated-electron singly occupied sites) we can uniquely define the
rotated-electron site distribution configurations of the local
charge (and spin) sequence by providing the sites occupied by
local $c,\nu$ pseudoparticles (and $s,\nu$ pseudoparticles)
belonging to the branches $\nu=1,2,...$ with finite occupancy. The
charge-sequence (and spin-sequence) sites of the effective
electronic lattice left over by these local pseudoparticles define
the positions of the Yang holons (and HL spinons). It is useful to
introduce the charge-sequence site index $h$ and the spin-sequence
site index $l$ such that,

\begin{equation}
h=1,2,...,[N_a-N_c] \, ; \hspace{1cm} l=1,2,...,N_c \, .
\label{hl}
\end{equation}
The ordering of the charge-sequence (and spin-sequence) index
corresponds to the location order from the left to the right-hand
side of the corresponding site in the effective electronic
lattice. The position in the effective electronic lattice of a
spin-sequence site of index $l$ and of a charge-sequence site of
index $h$ are given by,

\begin{equation}
x_{j_{l}}=j_{l}\,a \, , \hspace{0.5cm} l=1,2,...,N_c \, ;
\hspace{1cm} x_{j_{h}}= j_{h}\,a \, , \hspace{0.5cm}
h=1,2,...,[N_a-N_c] \, , \label{chcl}
\end{equation}
where $j_l$ are the indices of Eq. (\ref{lcp}) which define the
location of the rotated-electron singly occupied sites and $j_h$
are the indices which define the location of the rotated-electron
doubly occupied and empty sites. The location of the latter sites
corresponds to the sites left over by the rotated-electron singly
occupied sites.

From both the above analysis and property 3-III we conclude that
one can uniquely specify a given rotated-electron site
distribution configuration by providing the position of the $N_c$
sites occupied by local $c$ pseudoparticles,
$2\sum_{\nu=1}^{\infty}\nu\,N_{c,\,\nu}$ sites occupied by local
$c,\nu$ pseudoparticles, and
$2\sum_{\nu=1}^{\infty}\nu\,N_{s,\,\nu}$ sites occupied by local
$s,\nu$ pseudoparticles. The $N_c$ sites occupied by $c$
pseudoparticles define the location of the rotated-electron singly
occupied sites. The $[N_a-N_c]$ rotated-electron doubly-occupied
and empty sites are the sites left over by the rotated-electron
singly occupied sites. The $L_c=2S_c$ sites occupied by Yang
holons and the $L_s=2S_s$ sites occupied by HL spinons are the
sites left over in the charge and spin sequences respectively, by
the sites occupied by local $c,\nu$ pseudoparticles and local
$s,\nu$ pseudoparticles respectively. As mentioned above, to start
with we consider that all Yang holons (and HL spinons) have
$\eta$-spin projection (and spin projection) $+1/2$ and thus
correspond to rotated-electron empty sites (and rotated-electron
spin-up singly occupied sites). The generalization to
rotated-electron site distribution occupancies associated with
states containing both $\pm 1/2$ Yang holons (and $\pm 1/2$ HL
spinons) is straightforward and is introduced later in this
section.

The Yang holons (and HL spinons) have no internal structure and
are the simplest of the quantum objects which occupy the charge
(and spin) sequence of the effective electronic lattice. According
to property 1-III, the $+1/2$ Yang holons (and $+1/2$ HL spinons)
correspond to rotated-electron empty sites (and spin-up
rotated-electron singly occupied sites) of these sequences. On the
other hand, a local $\alpha,\nu$ pseudoparticle has internal
structure and thus is a more involved quantum object than a Yang
holon or a HL spinon. This internal structure corresponds to the
rotated-electron distribution configurations of the $\nu$ doubly
occupied sites and $\nu$ empty sites ($\alpha =c$) or $\nu$
spin-down singly occupied sites and $\nu$ spin-up singly occupied
sites ($\alpha =s$) which according to property 3-III describe
such a local $\alpha,\nu$ pseudoparticle. If in the
rotated-electron site distribution configurations which describe
the local $c,\nu$ pseudoparticles we replace doubly occupied sites
$\bullet$ and empty sites $\circ$ by spin-down singly occupied
sites $\downarrow$ and spin-up singly occupied sites $\uparrow$
respectively, and omit the phase factors generate by the operator
$\exp ({i\pi\sum_{j}j\,{\hat{D}}_j})$ mentioned in property 3-III,
we obtain the corresponding distribution configurations of the
$s,\nu$ pseudoparticles. Therefore, often we consider the
rotated-electron site distribution configurations which describe
the local $c,\nu$ pseudoparticles only.

It is useful to classify the $2\nu$ lattice sites of the local
charge sequence involved in the description of a local $c,\nu$
pseudoparticle into two sets of $\nu$ sites each. Let us denote
the index of these two sets of $\nu$ sites by $h_{j,\,g}$ and
$h_{j,\,\nu +g}$ respectively. Here $g=1,2,...,\nu$ and the index
$j=1,2,...,N^*_{c,\,\nu}$ refers to the position $[h_j\,a]$ of the
local $c,\nu$ pseudoparticle in the effective electronic lattice
where,

\begin{equation}
h_j = {j_{h_{j,\,\nu}}+j_{h_{j,\,\nu +1}}\over 2} \, ,
\label{stringC}
\end{equation}
and the numbers $j_{h_{j,\,\nu}}$ are the indices $j_h$ of Eq.
(\ref{lcp}) which define the location of the rotated-electron
doubly occupied/empty sites in such a lattice. Combination of the
values of the indices $j=1,2,...,N^*_{c,\,\nu}$ and
$g=1,2,...,\nu$ fully defines the position of the above $2\nu$
sites. The index $g$ is such that
$h_{j,\,1}<h_{j,\,2}<...<h_{j,\,\nu}$ and $h_{j,\,\nu +1}
<h_{j,\,\nu +2}<...<h_{j,\,2\nu}$ respectively, where $h_{j,\,\nu}
<h_{j,\,\nu +1}$. Equivalently, often we denote these $2\nu$
indices simply by $h_{j,\,x}$, where $x=1,2,...,2\nu$ and
$h_{j,\,1}<h_{j,\,2}<...<h_{j,\,2\nu}$. The set of $2\nu$ indices
$\{h_{j,\,1},\,h_{j,\,2},...,h_{j,\,2\nu}\}$ is in general a
sub-set of the $[N_a-N_c]$ charge-sequence indices $h$ given in
Eq. (\ref{hl}). The latter indices define the location of the
charge-sequence rotated-electron doubly occupied and empty sites.
We can also define the {\it charge-sequence position} of the local
$c,\nu$ pseudoparticle which is defined as,

\begin{equation}
{\bar{h}}_j = {h_{j,\,\nu}+h_{j,\,\nu +1}-1\over 2} \, .
\label{bstringC}
\end{equation}
The same definitions hold for the local $s,\nu$ pseudoparticles
with the indices $h_{j,\,g}$ and $h_{j,\,\nu +g}$ replaced by the
indices $l_{j,\,g}$ and $l_{j,\,\nu +g}$ respectively, and thus
the equivalent indices $h_{j,\,x}$ replaced by $l_{j,\,x}$. The
position of the local $s,\nu$ pseudoparticle in the effective
electronic lattice is defined as $[l_j\,a]$ where,

\begin{equation}
l_j = {j_{l_{j,\,\nu}}+j_{l_{j,\,\nu +1}}\over 2} \, ,
\label{stringS}
\end{equation}
$j=1,2,...,N^*_{s,\,\nu}$, and $j_l$ are the indices of Eq.
(\ref{lcp}) which define the location of the rotated-electron
singly occupied sites. On the other hand, the {\it spin-sequence
position} of the local $s,\nu$ pseudoparticle is defined as,

\begin{equation}
{\bar{l}}_{j} = {l_{j,\,\nu}+l_{j,\,\nu +1}-1\over 2} \, ,
\label{bstringS}
\end{equation}
where again $j=1,2,...,N^*_{s,\,\nu}$. We note that the indices
${\bar{h}}_j$ (and ${\bar{l}}_{j}$) given in Eq. (\ref{bstringC})
(and Eq. (\ref{bstringS})) which define the charge-sequence (and
spin-sequence) position of the local $c,\nu$ pseudoparticle (and
local $s,\nu$ pseudoparticle) are always positive integer numbers
$1,\,2,\,3,...$. According to Eq. (\ref{stringC}) (and Eq.
(\ref{stringS})) the position $[h_j\,a]$ (and $[l_j\,a]$) of the
local $c,\nu$ pseudoparticle (and local $s,\nu$ pseudoparticle) in
the $N_a$-site effective electronic lattice refers to a single
point inside the $2\nu$-site domain associated with such a quantum
object. On the other hand, the charge-sequence position (and
spin-sequence position) of the local $c,\nu$ pseudoparticles (and
local $s,\nu$ pseudoparticles) defined by the index of Eq.
(\ref{bstringC}) (and Eq. (\ref{bstringS})) refers to the location
of that quantum object relative to the $[N_a-N_c]$ sites (and
$N_c$ sites) of the charge (and spin) sequence only.

Below we clarify the following two issues: First, we find the
rotated-electron distribution occupancies of the $2\nu$ sites
which describe a local $\alpha,\nu$ pseudoparticle; Second, we
find the rotated-electron site distribution occupancies which
define the $N^h_{\alpha,\,\nu}$ charge-sequence ($\alpha=c$) or
spin-sequence ($\alpha=s$) {\it empty sites} corresponding to the
local $\alpha,\nu$ pseudoparticle branch. This study reveals that
the position in the effective electronic lattice of the sites
associated with the $N^h_{\alpha,\,\nu}$ charge-sequence or
spin-sequence {\it empty sites} of each local $\alpha,\nu$
pseudoparticle branch is uniquely determined by the position of
the local $c$ pseudoparticles, local $c,\nu$ pseudoparticles, and
local $s,\nu$ pseudoparticles. The same holds for the locations of
the Yang holons and HL spinons, as discussed above. Below we also
confirm that for fixed values of the numbers $N_{\alpha,\,\nu}$
and $N^*_{\alpha,\,\nu}$ the number of occupancy configurations of
the local $\alpha,\nu$ pseudoparticles is given by,

\begin{equation}
{N_{\alpha ,\,\nu}^*\choose N_{\alpha ,\,\nu}} = {N_{\alpha
,\,\nu}^*!\over N_{\alpha ,\,\nu}!\,N^h_{\alpha ,\,\nu}!}  \, .
\label{NanNS}
\end{equation}
We classify each of the local $\alpha,\nu$ pseudoparticle
occupancy configurations by providing the indices $h_j$ or $l_j$
given in Eqs. (\ref{stringC}) or (\ref{stringS}) respectively,
corresponding to the $N_{c,\,\nu}$ and $N_{s,\,\nu}$
pseudoparticle locations in the effective electronic lattice,

\begin{equation}
(h_{1},\,h_2,...,h_{N_{c,\,\nu}}) \, , \label{localcnp}
\end{equation}
and

\begin{equation}
(l_{1},\,l_2,...,l_{N_{s,\,\nu}}) \, , \label{localsnp}
\end{equation}
respectively. From the combination of Eqs. (\ref{NcNhc}) and
(\ref{NanNS}) it follows that the number of occupancy
configurations of the local $c$ and $\alpha,\nu$ pseudoparticles
of a CPHS ensemble subspace is given by,

\begin{equation}
{\cal{N}}_{CPHS-es} = {N_a\choose N_c}\prod_{\alpha
=c,s}\prod_{\nu =1}^{\infty}\, {N_{\alpha ,\,\nu}^*\choose
N_{\alpha ,\,\nu}} \, . \label{Ncphs}
\end{equation}

Let us denote each local state representing a specific local
charge, spin, and $c$ pseudoparticle sequence whose number for a
given CPHS ensemble subspace is given in Eq. (\ref{Ncphs}) by,

\begin{equation}
\vert(j_{1},\,j_2,...,j_{N_c});\,
\{(h_{1},\,h_2,...,h_{N_{c,\,\nu}})\};\,\{(l_{1},\,l_2,...,l_{N_{s,\,\nu}})\}\rangle
\, . \label{localst}
\end{equation}
Here

\begin{equation}
\{(h_{1},\,h_2,...,h_{N_{c,\,\nu}})\} =
(h_1,\,h_2,...,h_{N_{c,\,1}});\,
(h_{1},\,h_2,...,h_{N_{c,\,2}});\,(h_{1},\,h_2,...,h_{N_{c,\,3}});...\,
, \label{argc}
\end{equation}
gives the positions in the effective electronic lattice of the
local $c,\nu$ pseudoparticles belonging to branches with finite
occupancy in the state under consideration and

\begin{equation}
\{(l_{1},\,l_2,...,l_{N_{s,\,\nu}})\} =
(l_{1},\,l_2,...,l_{N_{s,\,1}});\,
(l_{1},\,l_2,...,l_{N_{s,\,2}});\,(l_{1},\,l_2,...,l_{N_{s,\,3}});...\,
, \label{args}
\end{equation}
provides the positions in the same lattice of the local $s,\nu$
pseudoparticles belonging to branches with finite occupancy in the
same state.

We emphasize that the number (\ref{Ncphs}) of such local states
which equals the number of different local pseudoparticle
occupancy configurations of a CPHS ensemble subspace indeed equals
the dimension of such a subspace. It is straightforward to confirm
from the results of Ref. \cite{I} that the number of band-momentum
energy eigenstates that span such a CPHS ensemble subspace also
has the same value. This suggests that the set of local states of
form (\ref{localst}) is complete in that subspace, as is confirmed
below. We find later on in this section that the generalization of
the present analysis to rotated-electron site distribution
configurations describing energy eigenstates with finite
occupancies of $\pm 1/2$ Yang holons and $\pm 1/2$ HL spinons
requires the introduction of the numbers $L_{c,\,-1/2}$ of $-1/2$
Yang holons and $L_{s,\,-1/2}$ of $-1/2$ HL spinons in the
labelling of the local states (\ref{localst}).

In order to find the rotated-electron site distribution
configurations which describe the internal structure of a local
$\alpha,\nu$ pseudoparticle we use mainly the basic properties
3-III, 5-III, 6-III, and 7-III. Often we consider the specific
case of a local $c,\nu$ pseudoparticle. The internal structure of
such a quantum object is studied by considering first a charge
sequence constituted by $\nu$ rotated-electron doubly occupied
sites and $\nu$ rotated-electron empty sites. The use of
properties 5-III and 6-III leads to the finding of the specific
rotated-electron site distribution configurations which describe
the local $c,\nu$ pseudoparticle. Property 7-III then states that
the obtained rotated-electron site distribution configurations
describe a local $c,\nu$ pseudoparticle in any charge sequence.

According to property 3-III, the internal structure of a local
$c,\nu$ pseudoparticle involves a number $\nu$ of rotated-electron
doubly occupied sites and an equal number of rotated-electron
empty sites. There is a number ${2\nu\choose \nu}$ of different
distribution configurations of these $\nu$ rotated-electron doubly
occupied sites and $\nu$ rotated-electron empty sites. For
simplicity let us restrict the present preliminary analysis to
charge sequences with $\nu$ rotated-electron doubly occupied sites
and an equal number of rotated-electron empty sites and without
Yang holons. Below we find that a local $c,\nu$ pseudoparticle is
a superposition of a number $2^{\nu}$ of the above
rotated-electron site distribution configurations. For $\nu=1$ the
local $c,1$ pseudoparticle is a properly symmetrized superposition
of the two available rotated-electron site distribution
configurations. For $\nu=2$ the local $c,2$ pseudoparticle is a
properly symmetrized superposition of four out of the six
available rotated-electron site distribution configurations. Thus
states including all possible six different distribution
configurations are a superposition of two states including two
local $c,1$ pseudoparticles and a local $c,2$ pseudoparticle
respectively. For $\nu=3$ the local $c,3$ pseudoparticle is a
properly symmetrized superposition of eight out of the twenty
available rotated-electron site distribution configurations.
States including all possible twenty different distribution
configurations are a superposition of three states including (i)
three local $c,1$ pseudoparticles, (ii) a local $c,1$
pseudoparticle and a local $c,2$ pseudoparticle, and (iii) a local
$c,3$ pseudoparticle. In the general case one has that the local
$c,\nu$ pseudoparticle is a properly symmetrized superposition of
$2^{\nu}$ out of the ${2\nu\choose \nu}$ available
rotated-electron site distribution configurations. States
including all possible ${2\nu\choose \nu}$ different distribution
configurations are a superposition of several states including
different numbers $N_{c,\,\nu'}$ of $c,\nu'$ pseudoparticles such
that $\nu'=1,2,...,\nu$ and
$\sum_{\nu'=1}^{\nu}\nu'\,N_{c,\,\nu'}=\nu$.

Among the $2^{\nu}$ rotated-electron site distribution
configurations of a $c,\nu$ pseudoparticle there is always a
distribution configuration where the first $\nu$ sites of index
$h_{j,\,g}$ and $g=1,2,...,\nu$ are doubly occupied by rotated
electrons and the last $\nu$ sites of index $h_{j,\,\nu +g}$ and
$g=1,2,...,\nu$ are free of rotated electrons. We represent such a
local $c,\nu$ pseudoparticle rotated-electron site distribution
configuration of a charge sequence by,

\begin{equation}
(\bullet_{h_{j,\,1}},...,\bullet_{h_{j,\,\nu}},
\,\circ_{h_{j,\,1+\nu}},..., \circ_{h_{j,\,2\nu}}) \, .
\label{lcnp}
\end{equation}
The same applies to a local $s,\nu$ pseudoparticle, the
rotated-electron site distribution configuration of a spin
sequence corresponding to the one given in Eq. (\ref{lcnp}) being,

\begin{equation}
(\downarrow_{l_{j,\,1}},...,
\downarrow_{l_{j,\,\nu}},\,\uparrow_{l_{j,\,1+\nu}},...,
\uparrow_{l_{j,\,2\nu}}) \, . \label{lsnp}
\end{equation}
We confirm below that the $c,\nu$ pseudoparticle located in the
effective electronic lattice at position $[h_j\,a]$ and the
$s,\nu$ pseudoparticle located in the same lattice at position
$[l_j\,a]$ are described by the following properly symmetrized
superposition of $2^{\nu}$ rotated-electron site distribution
configurations,

\begin{equation}
\Bigl[\prod_{x=1}^{2\nu}e^{i\pi
h_{j,\,x}{\hat{D}}_{j,\,x}}\Bigr]\,
\Bigl[\prod_{g=1}^{\nu}(1-{\hat{\cal{T}}}_{c,\,\nu,\,j,\,g})\Bigr]
\,(\bullet_{h_{j,\,1}},...,\bullet_{h_{j,\,\nu}},
\,\circ_{h_{j,\,1+\nu}},..., \circ_{h_{j,\,2\nu}}) \, ;
\hspace{0.5cm} j=1,2,...,N_{c,\,\nu} \, , \label{cnp}
\end{equation}
and

\begin{equation}
\Bigl[\prod_{g=1}^{\nu}(1-{\hat{\cal{T}}}_{s,\,\nu,\,j,\,g})\Bigr]
\, (\downarrow_{l_{j,\,1}},...,
\downarrow_{l_{j,\,\nu}},\,\uparrow_{l_{j,\,1+\nu}},...,
\uparrow_{l_{j,\,2\nu}}) \, ; \hspace{0.5cm} j=1,2,...,N_{s,\,\nu}
\, , \label{snp}
\end{equation}
respectively. Here and according to property 3-III, the
site-$h_{j,\,x}$ rotated-electron double occupation operator
${\hat{D}}_{j,\,x}$ has eigenvalue $1$ and $0$ when that site is
doubly occupied by rotated electrons and free of rotated electrons
respectively, and the operator ${\hat{\cal{T}}}_{c,\,\nu,\,j,\,g}$
(and ${\hat{\cal{T}}}_{s,\,\nu,\,j,\,g}$) acts on the pair of
sites of indices $h_{j,\,g}$ and $h_{j,\,g+\nu}$ (and $l_{j,\,g}$
and $l_{j,\,g+\nu}$) only. This operator always acts onto
rotated-electron site distribution configurations of the
particular form illustrated in Eq. (\ref{lcnp}) (and in Eq.
(\ref{lsnp})). From the application of this operator onto such a
rotated-electron site distribution configuration a new
distribution configuration is generated where the site of index
$h_{j,\,g}$ is free of rotated electrons, the site of index
$h_{j,\,g+\nu}$ is doubly occupied by rotated electrons, and the
occupancy of the other $2(\nu-1)$ sites remains unchanged, {\it
i.e.}

\begin{equation}
{\hat{\cal{T}}}_{c,\,\nu,\,j,\,g}\,(...,\bullet_{h_{j,\,g}},...,\circ_{h_{j,\,g+\nu}},...)
= (...,\circ_{h_{j,\,g}},...,\bullet_{h_{j,\,g+\nu}},...) \, .
\label{Tc}
\end{equation}
The same transformation law,

\begin{equation}
{\hat{\cal{T}}}_{s,\,\nu,\,j,\,g}\,(...,\downarrow_{l_{j,\,g}},...,\uparrow_{l_{j,\,g+\nu}},...)
= (...,\uparrow_{l_{j,\,g}},...,\downarrow_{l_{j,\,g+\nu}},...) \,
, \label{Ts}
\end{equation}
is associated with the application of the operator
${\hat{\cal{T}}}_{s,\,\nu,\,j,\,g}$ onto an rotated-electron site
distribution configuration of the type illustrated in Eq.
(\ref{lsnp}).

The $2^{\nu}$ internal rotated-electron site distribution
configurations on the right-hand side of Eq. (\ref{cnp}) (and Eq.
(\ref{snp})) are generated by considering that in each of the
$g=1,2,...,\nu$ pairs of sites of indices $h_{j\,g}$ and
$h_{j,\,g+\nu}$, the site of index $h_{j\,g}$ is doubly occupied
by rotated electrons and the site of index $h_{j,\,g+\nu}$ is free
of rotated electrons and vice versa. In general we call $h_{j,\,x}
\leftrightarrow h_{j,\,x'}$ {\it site pair} a superposition of two
rotated-electron distribution configurations of a rotated-electron
doubly occupied site and a rotated-electron empty site where
$h_{j,\,x}$ and $h_{j,\,x'}$, such that $x,x'=1,...,2\nu$ and $x'>
x$, are the indices of the two lattice sites involved. In the
first rotated-electron site distribution configuration the sites
of indices $h_{j,\,x}$ and $h_{j,\,x'}$ are doubly occupied by
rotated electrons and free of rotated electrons respectively. In
the second rotated-electron site distribution configuration the
sites of indices $h_{j,\,x}$ and $h_{j,\,x'}$ are free of rotated
electrons and doubly occupied by rotated electrons respectively.
The rotated-electron distribution configurations of all the
remaining sites of the charge sequence except these two are
identical. The

\begin{equation}
h_{j,\,x} \leftrightarrow h_{j,\,x'} \, ; \hspace{1cm}
x,x'=1,...,2\nu \, ; \hspace{0.5cm} x'>x \, , \label{sitep}
\end{equation}
site pair is defined as a superposition of these two
rotated-electron site distributions configurations where the first
and the second distributions are multiplied by a phase factor of
$1$ and $-1$ respectively.

In figure 1 we represent a $h_{j,\,x} \leftrightarrow h_{j,\,x'}$
site pair by two vertical lines connected by a horizontal line.
The two vertical lines have the same height and connect the
corresponding site of the charge sequence to the horizontal line.
Thus the two sites of a charge sequence connected by three such
lines are assumed to be involved in a superposition of two
rotated-electron site distribution configurations multiplied by
the phase factors of $1$ and $-1$ respectively, where according to
the above recipe one of these sites is doubly occupied by rotated
electrons and the other one is free of rotated electrons and vice
versa. Similarly, we call

\begin{equation}
l_{j,\,x} \leftrightarrow l_{j,\,x'} \, ; \hspace{1cm}
x,x'=1,...,2\nu \, ; \hspace{0.5cm} x'>x \, , \label{lsitep}
\end{equation}
site pair an equivalent superposition of two reotated-electron
site distribution configurations of a spin sequence. The
definitions are the same provided that we replace rotated-electron
doubly occupied sites and rotated-electron empty sites by
spin-down rotated-electron singly occupied sites and spin-up
rotated-electron singly occupied sites respectively. Such a
rotated-electron singly occupied site pair is also graphically
represented by the two vertical lines connected by a horizontal
line plotted in Fig. 1. The concept of a site pair plays an
important role in the description of the rotated-electron site
distribution configurations of a local $\alpha,\nu$
pseudoparticle.

It is confirmed below that the rotated-electron site distribution
configurations which describe a local $\alpha,\nu$ pseudoparticle
always involve a number $\nu$ of site pairs such that $x=g$ and
$x'=g+\nu$ where $g=1,...,\nu$. Since there are two possible
rotated-electron site occupancies for each pair and the number of
pairs of each local $\alpha,\nu$ pseudoparticle is $\nu$, the
total number of different internal rotated-electron site
distribution configurations is indeed $2^{\nu}$. For example, in
the case of a $c,\nu$ pseudoparticle located in the effective
electronic lattice at the position $[h_j\,a]$, the $2^{\nu}$
rotated-electron site distribution configurations superposed in
expression (\ref{cnp}) are the following,

\begin{equation}
\Bigl[\prod_{x=1}^{2}e^{i\pi
h_{j,\,x}{\hat{D}}_{j,\,x}}\Bigr]\,(1-{\hat{\cal{T}}}_{c,\,1,\,j,\,g})
\,(\bullet_{h_{j,\,1}},\, \circ_{h_{j,\,2}}) = - e^{i\pi
h_{j,\,1}}\,(\bullet_{h_{j,\,1}},\, \circ_{h_{j,\,2}}) + e^{i\pi
h_{j,\,2}}\,(\circ_{h_{j,\,1}},\, \bullet_{h_{j,\,2}}) \label{c1p}
\end{equation}
in the case of the $c,1$ pseudoparticle,

\begin{eqnarray}
& & \Bigl[\prod_{x=1}^{4}e^{i\pi h_{j,\,x}{\hat{D}}_{j,\,x}}\Bigr]
\,\Bigl[\prod_{g=1}^{2}(1-{\hat{\cal{T}}}_{c,\,2,\,j,\,g})\Bigr]
\,(\bullet_{h_{j,\,1}},\,\bullet_{h_{j,\,2}},\,
\circ_{h_{j,\,3}},\,\circ_{h_{j,\,4}}) = \nonumber \\
& + & e^{i\pi [h_{j,\,1}+h_{j,\,2}]}\,
(\bullet_{h_{j,\,1}},\,\bullet_{h_{j,\,2}},\,
\circ_{h_{j,\,3}},\,\circ_{h_{j,\,4}})
 - e^{i\pi[h_{j,\,2}+h_{j,\,3}]}
\,(\circ_{h_{j,\,1}},\,\bullet_{h_{j,\,2}},\,
\bullet_{h_{j,\,3}},\,\circ_{h_{j,\,4}}) \nonumber \\
& + & e^{i\pi
[h_{j,\,3}+h_{j,\,4}]}\,(\circ_{h_{j,\,1}},\,\circ_{h_{j,\,2}},\,
\bullet_{h_{j,\,3}},\,\bullet_{h_{j,\,4}}) - e^{i\pi
[h_{j,\,1}+h_{j,\,4}]}\,
(\bullet_{h_{j,\,1}},\,\circ_{h_{j,\,2}},\,
\circ_{h_{j,\,3}},\,\bullet_{h_{j,\,4}}) \, , \label{c2p}
\end{eqnarray}
for the $c,2$ pseudoparticle, and

\begin{eqnarray}
& & \Bigl[\prod_{x=1}^{6}e^{i\pi h_{j,\,x}{\hat{D}}_{j,\,x}}\Bigr]
\,\Bigl[\prod_{g=1}^{3}(1-{\hat{\cal{T}}}_{c,\,3,\,j,\,g})\Bigr]\,
(\bullet_{h_{j,\,1}},\,\bullet_{h_{j,\,2}},\,
\bullet_{h_{j,\,3}},\,\circ_{h_{j,\,4}},\,
\circ_{h_{j,\,5}},\,\circ_{h_{j,\,6}}) = \nonumber \\
& - &
e^{i\pi[h_{j,\,1}+h_{j,\,2}+h_{j,\,3}]}\,(\bullet_{h_{j,\,1}},\,\bullet_{h_{j,\,2}},\,
\bullet_{h_{j,\,3}},\,\circ_{h_{j,\,4}},\,
\circ_{h_{j,\,5}},\,\circ_{h_{j,\,6}}) \nonumber \\
& + &
e^{i\pi[h_{j,\,2}+h_{j,\,3}+h_{j,\,4}]}\,(\circ_{h_{j,\,1}},\,\bullet_{h_{j,\,2}},\,
\bullet_{h_{j,\,3}},\,\bullet_{h_{j,\,4}},\,
\circ_{h_{j,\,5}},\,\circ_{h_{j,\,6}}) \nonumber \\
& - &
e^{i\pi[h_{j,\,3}+h_{j,\,4}+h_{j,\,5}]}\,(\circ_{h_{j,\,1}},\,\circ_{h_{j,\,2}},\,
\bullet_{h_{j,\,3}},\,\bullet_{h_{j,\,4}},\,
\bullet_{h_{j,\,5}},\,\circ_{h_{j,\,6}}) \nonumber \\
& + &
e^{i\pi[h_{j,\,1}+h_{j,\,3}+h_{j,\,5}]}\,(\bullet_{h_{j,\,1}},\,\circ_{h_{j,\,2}},\,
\bullet_{h_{j,\,3}},\,\circ_{h_{j,\,4}},\,
\bullet_{h_{j,\,5}},\,\circ_{h_{j,\,6}}) \nonumber \\
& - &
e^{i\pi[h_{j,\,1}+h_{j,\,5}+h_{j,\,6}]}\,(\bullet_{h_{j,\,1}},\,\circ_{h_{j,\,2}},\,
\circ_{h_{j,\,3}},\,\circ_{h_{j,\,4}},\,
\bullet_{h_{j,\,5}},\,\bullet_{h_{j,\,6}}) \nonumber \\
& + &
e^{i\pi[h_{j,\,1}+h_{j,\,2}+h_{j,\,6}]}\,(\bullet_{h_{j,\,1}},\,\bullet_{h_{j,\,2}},\,
\circ_{h_{j,\,3}},\,\circ_{h_{j,\,4}},\,
\circ_{h_{j,\,5}},\,\bullet_{h_{j,\,6}}) \nonumber \\
& - &
e^{i\pi[h_{j,\,2}+h_{j,\,4}+h_{j,\,6}]}\,(\circ_{h_{j,\,1}},\,\bullet_{h_{j,\,2}},\,
\circ_{h_{j,\,3}},\,\bullet_{h_{j,\,4}},\,
\circ_{h_{j,\,5}},\,\bullet_{h_{j,\,6}}) \nonumber \\
& + &
e^{i\pi[h_{j,\,4}+h_{j,\,5}+h_{j,\,6}]}\,(\circ_{h_{j,\,1}},\,\circ_{h_{j,\,2}},\,
\circ_{h_{j,\,3}},\,\bullet_{h_{j,\,4}},\,
\bullet_{h_{j,\,5}},\,\bullet_{h_{j,\,6}}) \, , \label{c3p}
\end{eqnarray}
in the case of a $c,3$ pseudoparticle. The $2^{4}=16$
rotated-electron site distribution configurations of a $c,4$
pseudoparticle located in the effective electronic lattice at
position $[h_j\,a]$ are shown and discussed in Appendix B. Similar
expressions for local $s,1$, $s,2$, $s,3$, or $s,4$
pseudoparticles are obtained by replacing in Eqs.
(\ref{c1p})-(\ref{c3p}) and in Eq. (\ref{c4p}) of Appendix B
respectively, the $\bullet$ and $\circ$ sites by $\downarrow$ and
$\uparrow$ sites respectively, the $h_{j,g}$ indices by $l_{j,g}$
indices, and the local rotated-electron double occupation operator
phase factors by one.

Before using properties 3-III, 6-III, and 7-III to confirm the
validity of the local pseudoparticle expressions
(\ref{cnp})-(\ref{c3p}), let us check whether these expressions
conform to the requirement of property 5-III. According to that
requirement, application of the off-diagonal generators of the
$SU(2)$ $\eta$-spin algebra given in Eq. (\ref{Sc}) onto the
superposition of the $2^{\nu}$ internal rotated-electron site
distribution configurations of a $c,\nu$ pseudoparticle located in
the effective electronic lattice at position $[h_{j}\,a]$ must
give zero. It is straightforward to confirm from the analysis of
the changes in the rotated-electron site distribution
configurations generated by these operators that this requirement
is fulfilled and then,

\begin{eqnarray}
& & {\hat{S}}^c_{\pm}\,\Bigl[\prod_{x=1}^{2\nu}e^{i\pi
h_{j,\,x}{\hat{D}}_{j,\,x}}\Bigr]\,
\Bigl[\prod_{g=1}^{\nu}(1-{\hat{\cal{T}}}_{c,\,\nu,\,j,\,g})\Bigr]
\nonumber \\
& \times & (\bullet_{h_{j,\,1}},...,\bullet_{h_{j,\,\nu}},
\,\circ_{h_{j,\,1+\nu}},..., \circ_{h_{j,\,2\nu}}) = 0 \, ;
\hspace{0.5cm} j=1,2,...,N_{c,\,\nu} \, . \label{cnpSU2}
\end{eqnarray}
The same is valid for application of the off-diagonal generators
of the $SU(2)$ spin algebra given in Eq. (\ref{Ss}) onto the
superposition of the $2^{\nu}$ internal rotated-electron site
distribution configurations of a $s,\nu$ pseudoparticle located in
the effective electronic lattice at position $[l_{j}\,a]$,

\begin{equation}
{\hat{S}}^s_{\pm}\,\Bigl[\prod_{g=1}^{\nu}(1-{\hat{\cal{T}}}_{s,\,\nu,\,j,\,g})\Bigr]
\, (\downarrow_{l_{j,\,1}},...,
\downarrow_{l_{j,\,\nu}},\,\uparrow_{l_{j,\,1+\nu}},...,
\uparrow_{l_{j,\,2\nu}}) = 0 \, ; \hspace{0.5cm}
j=1,2,...,N_{s,\,\nu} \, . \label{snpSU2}
\end{equation}

In order to confirm the validity of the pseudoparticle expressions
(\ref{cnp})-(\ref{c3p}), let us follow property 5-III and consider
again a local charge sequence with no Yang holons and constituted
by a number $\nu$ of rotated-electron doubly occupied sites and a
number $\nu$ of rotated-electron empty sites. If we request that
the rotated-electron distribution configurations of these $2\nu$
sites describe a single $c,\nu$ pseudoparticle, according to the
restrictions imposed by the Bethe-ansatz solution \cite{I}, the
associated local charge sequence must be characterized by the
following numbers,

\begin{equation}
M_c =2\nu \, ; \hspace{0.5cm} N_{c,\,\nu}=1 \, ; \hspace{0.5cm}
N^h_{c,\,\nu}=0 \, ; \hspace{0.5cm} N_{c,\,\nu'}=0 \, ,
\hspace{0.3cm} \nu'\neq\nu \, . \label{Ncn}
\end{equation}
Such a local charge sequence has a single $c,\nu$ pseudoparticle
and no $c,\nu$ pseudoparticle {\it empty sites}. This corresponds
to an {\it effective $c,\nu$ pseudoparticle lattice} constituted
by a single {\it site} located in the effective electronic lattice
at $[h_{1}\,a]$. That {\it site} is occupied by a $c,\nu$
pseudoparticle. In this limiting case in order to ensure periodic
boundary conditions for the original electronic problem it is
required that the charge sequence should be properly symmetrized
by the method used in Ref. \cite{Geb}. Such a symmetrization
involves the operator ${\hat{\cal{T}}}_C$ of Eq. (\ref{Tcseq}).

A result of key importance is that the fulfillment of the
requirement of property 5-III imposes that the superposition of
rotated-electron distribution configurations of the $\nu$ doubly
occupied sites and $\nu$ empty sites (and $\nu$ spin-down singly
occupied sites and $\nu$ spin-up singly occupied sites) of such a
$c,\nu$ pseudoparticle (and $s,\nu$ pseudoparticle) must be
expressed in the form of a number $\nu$ of $h_{1,\,x}
\leftrightarrow h_{1,\,x'}$ site pairs (and $l_{1,\,x}
\leftrightarrow l_{1,\,x'}$ site pairs) where $x$ and $x'$ are
such that $x'> x$ and have the values $x,x'=1,2,...,2\nu$. There
is a number $(2\nu -1)!!$ of possible different choices for these
$\nu$ $h_{1,\,x} \leftrightarrow h_{1,\,x'}$ (and $l_{1,\,x}
\leftrightarrow l_{1,\,x'}$) site pairs. Fortunately, the use of
requirement 6-III reveals that only one of these $(2\nu -1)!!$
choices corresponds to the $c,\nu$ pseudoparticle (and $s,\nu$
pseudoparticle). The remaining $(2\nu -1)!!-1$ choices can be
expressed as a superposition of several states described by
different numbers $N_{c,\,\nu'}$ of $c,\nu'$ pseudoparticles such
that $\sum_{\nu'=1}^{\nu}\nu'\,N_{c,\,\nu'}=\nu$ (and
$N_{s,\,\nu'}$ of $s,\nu'$ pseudoparticles such that
$\sum_{\nu'=1}^{\nu}\nu'\,N_{s,\,\nu'}=\nu$). Moreover, the fact
that the symmetrized rotated-electron site distribution
configuration which describes the $c,\nu$ pseudoparticle (and
$s,\nu$ pseudoparticle) involves a number $\nu$ of site pairs
implies that it corresponds to a state with charge (and spin)
momentum $k_C =\pi /a$ (and $k_S =\pi /a$). The possible values of
such a charge momentum are given in Eq. (\ref{kCkS}). For other
values of charge momentum that state cannot be described in terms
of a number $\nu$ of site pairs and thus the requirement
introduced in property 5-III is not fulfilled.

For instance, when $\nu=1$ application of the basic properties
3-III, 5-III, and 6-III leads uniquely to the following
description of the $c,1$ pseudoparticle,

\begin{equation}
\Bigl[\prod_{x=1}^{2}e^{i\pi
h_{1,\,x}{\hat{D}}_{1,\,x}}\Bigr]\,\sum_{\nu_C=0}^1\,e^{i\pi\nu_C}\,
\Bigl({\hat{\cal{T}}}_C\Bigr)^{\nu_C}
\,(\bullet_{h_{1,\,1}},\,\circ_{h_{1,\,2}}) \, . \label{c1SCS}
\end{equation}
This is the product of a rotated-electron double-occupation
operator phase factor with a symmetrized charge sequence. The
charge sequence defined by Eq. (\ref{c1SCS}) can also be written
as,

\begin{equation}
\Bigl[\prod_{x=1}^{2}e^{i\pi
h_{1,\,x}{\hat{D}}_{1,\,x}}\Bigr]\,(1-{\hat{\cal{T}}}_{c,\,1,\,1,\,1})\,
(\bullet_{h_{1,\,1}},\,\circ_{h_{1,\,2}}) \, , \label{c1alone}
\end{equation}
in agreement with the general expression (\ref{cnp}) and the local
$c,1$ pseudoparticle expression (\ref{c1p}). For $j=1$, $x=1$, and
$x'=2$ Fig. 1 describes the $h_{1,\,1} \leftrightarrow h_{1,\,2}$
site pair associated with this $c,1$ pseudoparticle.

Next we consider the case $\nu=2$. The fulfillment of the
requirement introduced in property 5-III alone imposes in this
case that the superposition of rotated-electron distribution
configurations of the two doubly occupied sites and two empty
sites representative of a $c,2$ pseudoparticle must be expressed
in the form of two $h_{1,\,x} \leftrightarrow h_{1,\,x'}$ site
pairs where $x$ and $x'$ are such that $x'>x$ and have the values
$x,x'=1,2,3,4$. There are three possible different choices for the
superposition of rotated-electron distribution configurations
formed by two site pairs. These choices correspond to the
following sets: (a) $\{h_{1,\,1} \leftrightarrow h_{1,\,4}\,
;h_{1,\,2} \leftrightarrow h_{1,\,3}\}$, (b) $\{h_{1,\,1}
\leftrightarrow h_{1,\,3}\, ;h_{1,\,2} \leftrightarrow
h_{1,\,4}\}$, and (c) $\{h_{1,\,1} \leftrightarrow h_{1,\,2}\,
;h_{1,\,3} \leftrightarrow h_{1,\,4}\}$. According to property
7-III, the charge sequence (c) is excluded because it describes
two $c,1$ pseudoparticles and thus there remain two possible
choices for the $c,2$ pseudoparticle. Figure 2 represents the
charge sequences corresponding to the choices (a) and (b). Each of
the $h_{1,\,x} \leftrightarrow h_{1,\,x'}$ site pairs are
represented as in Fig. 1 with $j=1$. The figure shows how the
rotated-electron site distribution configurations of these two
charge sequences change as a result of a cyclic permutation which
transforms the first site of the charge sequence to the last site
and multiplies the final state by a phase factor of $-1$. This
phase factor arises from the required $k_C =\pi /a$ charge
momentum. We note that the single $h_{1,\,1} \leftrightarrow
h_{1,\,2}$ site pair descriptive of the $c,1$ pseudoparticle given
in expressions (\ref{c1SCS}) and (\ref{c1alone}) transforms into
itself under such a transformation. On the other hand, in the
present case the rotated-electron site distribution configuration
(a) transforms onto two $c,1$ pseudoparticles whereas the
rotated-electron site distribution configuration (b) transforms
onto itself. Thus according to property 6-III only the
rotated-electron site distribution configuration (b) is properly
symmetrized and describes the $c,2$ pseudoparticle. Moreover, only
this sequence can be written as a symmetrized charge sequence of
charge momentum $k_C =\pi /a$. Again, such a symmetrized charge
sequence can also be written in the form given in the general Eq.
(\ref{cnp}) and in the local $c,2$ pseudoparticle expression
(\ref{c2p}). The two equivalent expressions read,

\begin{eqnarray}
& & \Bigl[\prod_{x=1}^{4}e^{i\pi
h_{1,\,x}{\hat{D}}_{j,\,x}}\Bigr]\,\sum_{\nu_C=0}^3\,e^{i\pi\nu_C}\,
\Bigl({\hat{\cal{T}}}_C\Bigr)^{\nu_C}
(\bullet_{h_{1,\,1}},\,\bullet_{h_{1,\,2}},\,
\circ_{h_{1,\,3}},\,\circ_{h_{1,\,4}})\nonumber \\
& = & \Bigl[\prod_{x=1}^{4}e^{i\pi
h_{1,\,x}{\hat{D}}_{j,\,x}}\Bigr]
\,\Bigl[\prod_{g=1}^{2}(1-{\hat{\cal{T}}}_{c,\,2,\,1,\,g})\Bigr]\,
(\bullet_{h_{1,\,1}},\,\bullet_{h_{1,\,2}},\,
\circ_{h_{1,\,3}},\,\circ_{h_{1,\,4}}) \, . \label{c2alone}
\end{eqnarray}

Let us now consider the case $\nu=3$. Again the fulfillment of the
requirement introduced in property 5-III imposes that the
superposition of rotated-electron distribution configurations of
the three doubly occupied sites and three empty sites
representative of a $c,3$ pseudoparticle should be expressed in
the form of three $h_{1,\,x} \leftrightarrow h_{1,\,x'}$ site
pairs where $x$ and $x'$ are such that $x'>x$ and have the values
$x,x'=1,2,3,4,5,6$. There are fifteen different possible choices
for superposition of rotated-electron site distribution
configurations formed by three site pairs. However, only one of
these charge sequences can be written in the form of symmetrized
charge sequence of charge momentum $k_C =\pi /a$. For example, in
Fig. 3 we represent the charge sequences (a) $\{h_{1,\,1}
\leftrightarrow h_{1,\,3}\, ;h_{1,\,2} \leftrightarrow h_{1,\,5}\,
; h_{1,\,4} \leftrightarrow h_{1,\,6}\}$ and (b) $\{h_{1,\,1}
\leftrightarrow h_{1,\,4}\, ;h_{1,\,2} \leftrightarrow h_{1,\,5}\,
; h_{1,\,3} \leftrightarrow h_{1,\,6}\}$. The figure shows how the
rotated-electron site distribution configurations of these two
charge sequences change as a result of a cyclic permutation which
transforms the first site of the charge sequence to the last site
and multiplies the final state by a phase factor of $-1$. Both the
rotated-electron site distribution configuration (a) and all
remaining possible rotated-electron site distribution
configurations constituted by three $h_{1,\,x} \leftrightarrow
h_{1,\,x'}$ site pairs except the rotated-electron site
distribution configuration (b) do not transform onto themselves.
The latter rotated-electron site distribution configuration
transforms onto itself, as confirmed by analysis of the figure.
Thus according to property 6-III only the rotated-electron site
distribution configuration (b) is properly symmetrized and
describes the local $c,3$ pseudoparticle. Only this sequence can
be written in terms of a symmetrized charge sequence of charge
momentum $k_C =\pi /a$. Again, such symmetrized charge sequence
can also be written in the form given in the general Eq.
(\ref{cnp}) and in the local $c,3$ expression (\ref{c3p}). The two
equivalent expressions are the following,

\begin{eqnarray}
\nonumber \\
& & \Bigl[\prod_{x=1}^{6}e^{i\pi
h_{1,\,x}{\hat{D}}_{j,\,x}}\Bigr]\,\{\sum_{\nu_C=0}^5\,e^{i\pi\nu_C}\,
\Bigl({\hat{\cal{T}}}_C\Bigr)^{\nu_c}\,
(\bullet_{h_{1,\,1}},\,\bullet_{h_{1,\,2}},\,
\bullet_{h_{1,\,3}},\,\circ_{h_{1,\,4}},\,\circ_{h_{1,\,5}},\,
\circ_{h_{1,\,6}})
\nonumber \\
& + &
\sum_{\nu_C=0}^1\,e^{i\pi\nu_C}\,\Bigl({\hat{\cal{T}}}_C\Bigr)^{\nu_c}\,
(\circ_{h_{1,\,1}},\,\bullet_{h_{1,\,2}},\,
\circ_{h_{1,\,3}},\,\bullet_{h_{1,\,4}},\,\circ_{h_{1,\,5}},\,
\bullet_{h_{1,\,6}})\} \nonumber \\
& = & \Bigl[\prod_{x=1}^{6}e^{i\pi
h_{1,\,x}{\hat{D}}_{j,\,x}}\Bigr]\,
\Bigl[\prod_{g=1}^{3}(1-{\hat{\cal{T}}}_{c,\,3,\,1,\,g})\Bigr]\,
(\bullet_{h_{1,\,1}},\,\bullet_{h_{1,\,2}},\,
\bullet_{h_{1,\,3}},\,\circ_{h_{1,\,4}},\,\circ_{h_{1,\,5}},\,
\circ_{h_{1,\,6}}) \, . \label{c3alone}
\end{eqnarray}
Note that in this case one needs two symmetrized charge sequences
to reach the $c,3$ expression (\ref{c3p}) upon application of the
operator ${\hat{\cal{T}}}_C$ onto these charge sequences two and
six times respectively. However, in agreement with properties
5-III and 6-III, the state obtained from the superposition of
these two charge sequences is an eigenstate of the charge momentum
operator with eigenvalue $k_C =\pi /a$. Also the $c,\nu$
pseudoparticles belonging to branches such that $\nu>3$ are a
superposition of several such symmetrized charge sequences.
However, always the state obtained from the superposition of
several charge sequences is an eigenstate of the charge momentum
operator with eigenvalue $k_C =\pi /a$. That state can also be
expressed in the general form given in Eq. (\ref{cnp}).

As a last example we provide the expression in terms of
symmetrized charge sequences of the rotated-electron site
distribution configurations representative of a $c,4$
pseudoparticle. This can again be written both (i) in the form
given in Eqs. (\ref{cnp}) and in Eq. (\ref{c4p}) of Appendix B and
(ii) as a superposition of symmetrized charge sequences. The two
equivalent expressions are in this case given by,

\begin{eqnarray}
& & \Bigl[\prod_{x=1}^{8}e^{i\pi
h_{1,\,x}{\hat{D}}_{j,\,x}}\Bigr]\,\{\sum_{\nu_C=0}^7\,e^{i\pi\nu_C}\,
\Bigl({\hat{\cal{T}}}_C\Bigr)^{\nu_C}\,
(\bullet_{h_{1,\,1}},\,\bullet_{h_{1,\,2}},\,
\bullet_{h_{1,\,3}},\,\bullet_{h_{1,\,4}},\,\circ_{h_{1,\,5}},\,
\circ_{h_{1,\,6}},\,\circ_{h_{1,\,7}},\, \circ_{h_{1,\,8}})
\nonumber \\
& + &
\sum_{\nu_C=0}^7\,e^{i\pi\nu_C}\,\Bigl({\hat{\cal{T}}}_C\Bigr)^{\nu_C}\,
(\bullet_{h_{1,\,1}},\,\circ_{h_{1,\,2}},\,
\bullet_{h_{1,\,3}},\,\circ_{h_{1,\,4}},\,\circ_{h_{1,\,5}},\,
\bullet_{h_{1,\,6}},\,\circ_{h_{1,\,7}},\, \bullet_{h_{1,\,8}})\}
\nonumber \\
& = & \Bigl[\prod_{x=1}^{8}e^{i\pi
h_{1,\,x}{\hat{D}}_{j,\,x}}\Bigr]\,
\Bigl[\prod_{g=1}^{4}(1-{\hat{\cal{T}}}_{c,\,4,\,1,\,g})\Bigr]\,
(\bullet_{h_{1,\,1}},\,\bullet_{h_{1,\,2}},\,
\bullet_{h_{1,\,3}},\,\bullet_{h_{1,\,4}},\,\circ_{h_{1,\,5}},\,
\circ_{h_{1,\,6}},\,\circ_{h_{1,\,7}},\, \circ_{h_{1,\,8}})
 \, . \label{c4alone}
\end{eqnarray}
Also in this case one needs more than one symmetrized charge
sequence to reach the $c,4$ expression (\ref{c4p}) given in
Appendix A. Again, the state obtained from the superposition of
the involved charge sequences is an eigenstate of the charge
momentum operator with eigenvalue $k_C =\pi /a$.

For the case of charge (and spin) sequences constituted by several
local $c,\nu$ pseudoparticles (and local $s,\nu$ pseudoparticles)
belonging to one or several $\nu=1,2,...$ branches and by Yang
holons (and HL spinons) the internal structure of the local
$c,\nu$ pseudoparticles (and local $s,\nu$ pseudoparticles) is
given by Eq. (\ref{cnp}) (and Eq. (\ref{snp})). This result
follows from the basic property 7-III. According to that property
the rotated-electron distribution configurations of the $\nu$
doubly occupied sites and $\nu$ empty sites (and $\nu$ spin-down
singly occupied sites and $\nu$ spin-up singly occupied sites)
which describe the internal structure, a $c,\nu$ pseudoparticle
(and $s,\nu$ pseudoparticle) located in the effective electronic
lattice at position $[h_{j}\,a]$ (and $[l_j\,a]$) are the same as
for the above limiting charge (and spin) sequence constituted by a
single pseudoparticle. We will in the ensuing section confirm that
in agreement with property 8-III, this ensures the periodic
boundary conditions of the original electronic problem provided
that the band-momentum energy eigenstates are suitable
Fourier-transform superpositions of the local charge sequences
with band-momentum values obeying Eqs. (\ref{pbccp}) and
(\ref{pbcanp}).

However, whether inside the domain of $2\nu$ sites of a local
$c,\nu$ pseudoparticle there can be sites with index $j_h$ such
that $j_{h_{j,\,1}}<j_h<j_{h_{j,\,2\nu}}$ and which sites are
occupied by Yang holons or other local $c,\nu'$ pseudoparticles
remains an open question. This problem is absent in the limiting
case we considered here where the charge sequence is constituted
by a number $2\nu$ of sites of the effective electronic lattice.
This issue is closely related to the definition of the local
$c,\nu$ pseudoparticle charge-sequence {\it empty sites} which is
discussed in the following. A similar problem occurs for the spin
sequences.

\subsection{THE LOCAL $\alpha,\nu$ PSEUDOPARTICLE {\it EMPTY SITES} AND THE
MOVING PSEUDOPARTICLE ELEMENTARY STEPS}

In the definition of the $N^h_{c,\,\nu}$ charge-sequence {\it
empty sites} of a local $c,\nu$ pseudoparticle it is useful to
consider different charge sequences where all remaining local
$c,\nu'$ pseudoparticles except that specific local $c,\nu$
pseudoparticle keep their charge-sequence positions ${\bar{h}}_j$
defined in Eq. (\ref{bstringC}). Below we call {\it steady} local
$c,\nu'$ pseudoparticles the former local pseudoparticles whose
charge-sequence positions ${\bar{h}}_j$ remain unchanged.
Obviously, any local pseudoparticle can be chosen to be the {\it
moving} pseudoparticle. This just refers to a set of different
charge sequences where the charge-sequence position of that
quantum object changes whereas the charge-sequence positions of
the remaining local pseudoparticles remain unchanged. We note that
according to Eq. (\ref{bstringC}) the charge-sequence position of
a local $c,\nu$ pseudoparticle can remain unchanged in spite of
the {\it movements} of some of its $2\nu$ sites, as we confirm
below. Concerning the definition of the $N^h_{c,\,\nu}$
charge-sequence {\it empty sites} of a local $c,\nu$
pseudoparticle, from the use of properties 1-III to 8-III and of
other properties of the 1D Hubbard model we find the following
distribution configuration properties:\vspace{0.5cm}

1-IV Let us consider a set of rotated-electron doubly occupied
and/or rotated-electron empty sites located at positions
$[j_h\,a]$ where according to Eqs. (\ref{hl}) and (\ref{chcl}) the
index $h$ has the values $h=1,2,...,[N_a-N_c]$. Moreover, we
consider that all of these sites are located inside the domain of
sites of a local $c,\nu$ pseudoparticle and thus their indices
$j_h$ are such that $j_{h_{j,\,1}}<j_h<j_{h_{j,\,2\nu}}$ and
$j_h\neq j_{h_{j,\,x}}$ where $x=1,2,...,2\nu$. This property
states that the set of rotated-electron doubly occupied/empty
sites of index $j_h$ obeying these inequalities cannot correspond
to sites occupied by Yang holons or to any of the $2\nu'$ sites of
the charge sequence associated with local $c,\nu'$ pseudoparticles
belonging to branches such that $\nu'>\nu$. Thus such a set of
rotated-electron doubly occupied or empty sites can only
correspond to sub-sets of $2\nu'$ sites of the charge sequence
associated with local $c,\nu'$ pseudoparticles belonging to
branches such that $\nu'\leq\nu$.\vspace{0.5cm}

2-IV Let us consider a local $c,\nu$ pseudoparticle and a local
$c,\nu'$ pseudoparticle belonging to the same charge sequence and
to branches such that $\nu\leq\nu'$. Moreover, we consider that
there are no charge-sequence local pseudoparticles inside the
$2\nu'$-site domain of the $c,\nu'$ pseudoparticle other than the
$c,\nu$ pseudoparticle. The description of cases involving more
than two local pseudoparticles can be easily reached by
generalization of the two-pseudoparticle situation considered
here. Then the $2\nu$ sites of the charge sequence associated with
the local $c,\nu$ pseudoparticle are located either outside the
domain of $2\nu'$ sites of the local $c,\nu'$ pseudoparticle
limited in the left and right ends by the sites of index
$h_{j,\,1}$ and $ h_{j,\,2\nu'}$ respectively, or are all located
inside that domain. If the $2\nu$ sites are located inside the
domain of these $2\nu'$ sites there is a number $2\nu'-1$ of
permitted positions for the $c,\nu$ pseudoparticle. In each of
these permitted positions all of the $2\nu$ sites of the local
$c,\nu$ pseudoparticle are located between two first neighboring
sites of indices $h_{j,x}$ and $h_{j',x+1}$ of the local $c,\nu'$
pseudoparticle. These permitted positions correspond to the
neighboring sites of indices $h_{j,x}$ and $h_{j',x+1}$ such that
$x=1,...,2\nu'-1$.\vspace{0.5cm}

3-IV Let us consider again the local $c,\nu$ pseudoparticle and
the local $c,\nu'$ pseudoparticle referred to in property 2-IV.
The charge-sequence position ${\bar{h}}_j$ given in Eq.
(\ref{bstringC}) of such a steady local $c,\nu'$ pseudoparticle
remains unchanged in all possible $2\nu'-1$ positions of the local
$c,\nu$ pseudoparticle while passing its $2\nu'$-site domain. The
inverse statement is even stronger. It is required that when a
local $c,\nu'$ pseudoparticle belonging to a branch such that
$\nu'\geq\nu$ passes a steady local $c,\nu$ pseudoparticle, the
position in the effective electronic lattice of all the $2\nu$
sites of the latter pseudoparticle must remain unchanged.
Moreover, the $2\nu'-1$ possible different positions of the $2\nu$
sites of the local $c,\nu$ pseudoparticle inside the site domain
of the local $c,\nu'$ pseudoparticle must be the same when the
local $c,\nu$ pseudoparticle passes the steady local $c,\nu'$
pseudoparticle from its left to its right-hand side and vice
versa. Also the relative positions of the two quantum objects when
the local $c,\nu$ pseudoparticle passes the steady local $c,\nu'$
pseudoparticle and when the local $c,\nu'$ pseudoparticle passes
the steady local $c,\nu$ pseudoparticle must be the same. However,
note that although these relative positions are the same, when a
local $c,\nu'$ pseudoparticle belonging to a branch such that
$\nu'\geq\nu$ passes a steady local $c,\nu$ pseudoparticle it does
not occupy any of the $2\nu$ sites of the effective electronic
lattice occupied by the latter local pseudoparticle. Thus the
positions in the lattice of these $2\nu$ sites remain indeed
unchanged. A last requirement is that the positions of all $2\nu'$
sites of a steady local $c,\nu'$ pseudoparticle belonging to a
branch such that $\nu'\geq\nu$ must be the same in the
rotated-electron site distribution configurations before and after
it is passed by the moving local $c,\nu$ pseudoparticle. If the
local $c,\nu$ pseudoparticle is moving from the left to the
right-hand side of the steady local $c,\nu'$ pseudoparticle, {\it
before} and {\it after} is meant here as the rotated-electron site
distribution configurations where the $2\nu$ sites of the former
quantum object are located outside the domain of $2\nu'$ sites of
the steady local pseudoparticle and on its right and left hand
sides respectively.\vspace{0.5cm}

4-IV Corresponding rules are also valid for the case of the spin
sequences and local $s,\nu$ pseudoparticles.\vspace{0.5cm}

A simple example of a domain constituted by fourteen sites of a
permitted charge sequence is represented in Fig. 4. In such a
figure the $h_{j,\,g} \leftrightarrow h_{j,\,g+\nu}$ site pairs of
the local $c,\nu$ pseudoparticles are represented as in Figs. 1 to
3. The six sites with no vertical lines correspond to empty sites
associated with $+1/2$ Yang holons. There are two local $c,1$
pseudoparticles and a local $c,2$ pseudoparticle in the
charge-sequence domain represented in the figure.

In order to confirm the validity of the counting of
rotated-electron site distribution occupancies leading to
expression (\ref{NanNS}), let us use the properties 1-IV to 4-IV
in the classification of the different $N^*_{c,\,\nu}$ possible
positions of a local $c,\nu$ pseudoparticle in the effective
electronic lattice. Our analysis can be easily generalized to the
case of a $s,\nu$ pseudoparticle. Our study includes the
introduction of the concept of {\it $2\nu$-leg caterpillar step},
where the $2\nu$-leg caterpillar is nothing but the $2\nu$-site
pseudoparticle block associated with the moving $c,\nu$
pseudoparticle. For simplicity we consider a charge sequence such
that $N_{c,\,\nu}=1$ for the $\nu$ branch. However, the
generalization of our results to branches with finite occupancy of
several local pseudoparticles is straightforward. As mentioned
above, the possible positions of such a pseudoparticle correspond
to a number $N^*_{c,\,\nu}$ of different charge sequences where
the charge-sequence positions of all remaining quantum objects
remain unchanged. Thus these $N^*_{c,\,\nu}$ charge sequences
differ in the position of the $c,\nu$ pseudoparticle relative to
the other quantum objects. The above property 3-IV states that the
charge-sequence position ${\bar{h}}_j$ given in Eq.
(\ref{bstringC}) of all quantum objects except that of the {\it
moving} $c,\nu$ pseudoparticle must remain unchanged in all these
$N^*_{c,\,\nu}$ charge sequences. Let us confirm that the
charge-sequence position of a steady local $c,\nu'$ pseudoparticle
belonging to a branch such that $\nu'>\nu$ indeed remains
unchanged when it is passed by a local $c,\nu$ pseudoparticle, in
spite of changes of the positions in the effective electronic
lattice of some of its $2\nu'$ sites.

We start by considering the case of a local $c,{\bar{\nu}}$
pseudoparticle where $\nu={\bar{\nu}}$ denotes the largest $\nu$
value of the $c,\nu$ pseudoparticle branches with finite occupancy
in the charge sequence. According to Eq. (\ref{N*sum}), in this
case the number of local $c,{\bar{\nu}}$ pseudoparticle {\it empty
sites} is given by $N^h_{c,\,{\bar{\nu}}} = L_{c}$ and equals the
number of Yang holons. The different possible positions of a
$c,{\bar{\nu}}$ pseudoparticle can be achieved by elementary steps
in the charge sequence where each of its $2{\bar{\nu}}$ sites
moves forward by a single site of the charge sequence. If the
movement is from the left to the right-hand side, the
$2{\bar{\nu}}\,th$ pseudoparticle site of index
$h_{j,\,2{\bar{\nu}}}$ moves into the site previously occupied by
a Yang holon. The remaining $2{\bar{\nu}}-1$ pseudoparticle sites
of index $h_{j,\,x}$ move into the site previously associated with
the pseudoparticle site of index $h_{j,\,x+1}$ where
$x=1,...,2{\bar{\nu}}-1$. Finally, the Yang holon removed from the
site newly associated with the pseudoparticle site of index
$h_{j,\,2{\bar{\nu}}}$ moves into the site previously associated
with the first pseudoparticle site of index $h_{j,\,1}$. We call
such an elementary collective step of the $2{\bar{\nu}}$
pseudoparticle sites {\it $2{\bar{\nu}}$-leg caterpillar step}.
According to such an analogy the $2{\bar{\nu}}$ {\it legs} refer
to these sites and the {\it caterpillar} refers to the local
pseudoparticle. While each of these legs moves forward by a
charge-sequence site, the whole caterpillar itself also moves
forward by a single charge-sequence site. The net result of such a
$2{\bar{\nu}}$-leg caterpillar step is that the compact domain of
$2{\bar{\nu}}$ pseudoparticle sites interchanges position with a
Yang holon. When such a moving $c,{\bar{\nu}}$ pseudoparticle
passes a steady local $c,\nu$ pseudoparticle, each of its
$2{\bar{\nu}}$ sites jump the $2\nu$ sites of the latter local
pseudoparticle. Thus the $2{\bar{\nu}}$-leg caterpillar moves
forward like the $2\nu$ sites of such a steady local $c,\nu$
pseudoparticle do not exist in the charge sequence. As we discuss
below, the construction of the corresponding effective
$c,{\bar{\nu}}$ pseudoparticle lattice mentioned in property 8-III
involves omission of the sites of the effective electronic lattice
belonging to the $2\nu$-site domains of such a local $c,\nu$
pseudoparticles. Moreover, for such an effective $c,{\bar{\nu}}$
pseudoparticle lattice the $2{\bar{\nu}}$-site domain of the
$c,{\bar{\nu}}$ pseudoparticles is {\it seen} as a point-like
occupied site. Thus in the particular case of the ${\bar{\nu}}$
branch such an effective lattice has
$N^*_{c,\,{\bar{\nu}}}=L_c+N_{c,\,{\bar{\nu}}}$ sites and a number
$N^h_{c,\,{\bar{\nu}}}=L_c$ free of $c,{\bar{\nu}}$
pseudoparticles, as we confirm below.

Let us consider now the case of a general local $c,\nu$
pseudoparticle belonging to a branch such that
$1\leq\nu\leq{\bar{\nu}}$. For simplicity we keep the assumption
that $N_{c,\,\nu}=1$ for the $\nu$ branch. We consider a given
charge sequence where the $c,\nu$ pseudoparticle is located in the
effective electronic lattice at position $[h_j\,a]$. It follows
from property 4-III that the $N^h_{c,\,\nu}$ remaining possible
positions of the local $c,\nu$ pseudoparticle in the charge
sequence define the positions of the local $c,\nu$ pseudoparticle
charge-sequence {\it empty sites}. Out of these $N^h_{c,\,\nu} =
L_{c} + 2\sum_{\nu'=\nu +1}^{\infty} (\nu' -\nu) N_{c,\,\nu'}$
possible positions, $L_c$ is the number of positions associated
with Yang holons. Again, the different positions of the moving
local $c,\nu$ pseudoparticle can be achieved by $2\nu$-leg
caterpillar steps. The net result of such a $2\nu$-leg caterpillar
step is that the compact domain constituted by the $2\nu$ $c,\nu$
pseudoparticle sites interchanges position with a Yang holon or
with a site of a $h_{j',\,g'}\leftrightarrow h_{j',\,g'+\nu'}$
pair of a steady local $c,\nu'$ pseudoparticle belonging to a
branch such that $\nu'>\nu$. A $2\nu$-leg caterpillar step is
defined here as in the case of the above moving $c,{\bar{\nu}}$
pseudoparticle. If the movement is from the left to the right-hand
side, the local $c,\nu$ pseudoparticle site of index
$h_{j,\,2\nu}$ moves into the site previously occupied by a Yang
holon or into one of the two sites previously associated with a
local $c,\nu'$ pseudoparticle $h_{j',\,g'} \leftrightarrow
h_{j',\,g'+\nu'}$ site pair. The remaining $2\nu-1$ sites of the
local $c,\nu$ pseudoparticle of index $h_{j,\,x}$ such that
$x=1,...,2\nu-1$, move into the site previously associated with
the local $c,\nu$ pseudoparticle site of index $h_{j,\,x+1}$ where
again $x=1,...,2\nu-1$. Finally, the Yang holon or $c,\nu'$
pseudoparticle site associated with the local $h_{j',\,g'}
\leftrightarrow h_{j',\,g'+\nu'}$ site pair which was removed from
the site newly associated with the local $c,\nu$ pseudoparticle
site of index $h_{1,\,2\nu}$, moves into the site previously
associated with the local $c,\nu$ pseudoparticle site of index
$h_{j,\,1}$. Again, the net result of such a $2\nu$-leg
caterpillar step is that the local $c,\nu$ pseudoparticle moves
forward by a single charge-sequence site.

Let us consider that in the initial rotated-electron site
distribution configuration the moving $c,\nu$ pseudoparticle is
located on the left-hand side of the steady local $c,\nu'$
pseudoparticle. Thus in the final state just after passing the
latter pseudoparticle, the $c,\nu$ pseudoparticle will occupy in
the charge sequence a domain of $2\nu$ sites located on the
right-hand side of the $c,\nu'$ pseudoparticle site domain.
According to property 3-IV, in both the initial and final
rotated-electron site distribution configurations all $2\nu'$
sites of the latter quantum object remain the same. For simplicity
we consider that in the initial rotated-electron site distribution
configuration, the $2\nu$ charge-sequence sites located just on
the right-hand side of the steady local $c,\nu'$ pseudoparticle
are occupied by Yang holons. However, our results can be
generalized to rotated-electron site distribution configurations
where these sites are occupied by a third pseudoparticle. Another
case which we could consider is when the moving local
pseudoparticle passes two steady local $c,\nu'$ and $c,\nu''$
pseudoparticles such that one of them is located within the site
domain of the other one. From the systematic use of the properties
1-IV to 4-IV one can describe the movements of a local $c,\nu$
pseudoparticle when it passes any rotated-electron site
distribution configuration involving several steady local
pseudoparticles.

According to property 2-IV the $2\nu$ sites of a local $c,\nu$
pseudoparticle can have $2\nu'-1$ different positions inside the
$2\nu'$-site domain of a local $c,\nu'$ pseudoparticle belonging
to a branch such that $\nu'>\nu$. We consider that in the initial
rotated-electron site distribution configuration, the local
$c,\nu$ and $c,\nu'$ pseudoparticles are located side by side in
the charge sequence. Thus these two quantum objects occupy
together a compact $2(\nu+\nu')$-site domain. By simple counting
arguments it is straightforward to realize that in order to reach
each of these permitted positions each of the $2\nu$ sites of the
local $c,\nu$ pseudoparticle must perform a jump of $2\nu$ sites
in the charge sequence somewhere inside the $2\nu'$-site domain of
the steady local $c,\nu'$ pseudoparticle. This result is
consistent with the expression of the number $2\sum_{\nu'=\nu
+1}^{\infty} (\nu' -\nu) N_{\alpha,\,\nu'}$ of the available
positions for the moving local $c,\nu$ pseudoparticle involving
sites of steady local $c,\nu'$ pseudoparticles belonging to
branches such that $\nu'>\nu$. This expression reveals that when
located inside the site domain of such a pseudoparticle each of
the $2\nu$ sites of the local $c,\nu$ pseudoparticle can move into
a possible $2(\nu' -\nu)$ sites out of the $2\nu'$ total number of
sites of the steady local $c,\nu'$ pseudoparticle. This agrees
with the above result obtained from counting arguments alone. We
note that the number of charge-sequence sites jumped by each local
$c,\nu$ caterpillar leg equals precisely $2\nu$, the number of
sites of such a $c,\nu$ pseudoparticle. Thus these jumps ensure
that the position in the effective electronic lattice of all the
$2\nu'$ sites of the steady local $c,\nu'$ pseudoparticle remains
unchanged after the $c,\nu$ passes it, as requested by the
property 3-IV. This result also applies to the general situation
when a moving local $c,\nu$ pseudoparticle passes a domain
containing several local steady $c,\nu'$ pseudoparticles belonging
to branches such that $\nu'>\nu$. Also in the general situation
that each of the legs of the moving caterpillar uses $2(\nu'
-\nu)$ possible sites of each steady local $c,\nu'$
pseudoparticle, out of a total of $2\nu'$ sites.

The question which arises from the occurrence of the above jump
mechanism is which are the $2\nu$ sites of the steady local
$c,\nu'$ pseudoparticle that are jumped by each site of the moving
local $c,\nu$ pseudoparticle. Fortunately, the requirements
mentioned in the property 3-IV provide the needed information for
the precise definition of the $2\nu'-1$ permitted positions of the
local $c,\nu$ pseudoparticle inside the steady local $c,\nu'$
pseudoparticle. We find that in order to reach these permitted
positions each site of the local $c,\nu$ pseudoparticle must
perform two independent jumps of $\nu$ sites in the charge
sequence. Let us suppose that the local $c,\nu$ pseudoparticle is
moving from the left to the right-hand side. These jumps bring the
local $c,\nu$ pseudoparticle to and from respectively, a position
where its charge-sequence location ${\bar{h}}_j$, as given by Eq.
(\ref{bstringC}), coincides with that of the steady local $c,\nu'$
pseudoparticle. In such an event each of the $2\nu$ legs of the
caterpillar jumps a number $\nu$ of sites and reaches a position
where its $2\nu$ legs occupy $2\nu$ sites of the charge sequence
located around the center of the steady local $c,\nu'$
pseudoparticle. In order to move forward to the right-hand side
from that position, each of the $2\nu$ legs of the caterpillar
jump again a number of $\nu$ sites of the charge sequence. Thus
the possible positions of a local $c,\nu$ pseudoparticle when it
passes from the left to the right-hand side of a steady local
$c,\nu'$ pseudoparticle are the following: First the local $c,\nu$
pseudoparticle performs a $2\nu$-leg caterpillar step. The net
result of such an elementary collective step is that the compact
domain of $2\nu$ sites of the local $c,\nu$ pseudoparticle
interchanges position with the site of index $h_{j',\,1}$ of the
local $c,\nu'$ pseudoparticle. Thus this first elementary
collective step brings the local $c,\nu$ pseudoparticle inside the
domain of $2\nu'$ sites of the steady local $c,\nu'$
pseudoparticle. In the reached rotated-electron site distribution
configuration all the $2\nu$ sites of the local $c,\nu$
pseudoparticle are located between the sites of index $h_{j',\,1}$
and $h_{j',\,2}$ of the steady local $c,\nu'$ pseudoparticle. This
procedure is repeated $\nu'-1$ times until the $2\nu$ sites of the
$c,\nu$ pseudoparticle are located between the sites of index
$h_{j',\,\nu'-1}$ and $h_{j',\,\nu'}$ of the steady local $c,\nu'$
pseudoparticle. In each of the corresponding $2\nu$-leg
caterpillar steps the local $c,\nu$ pseudoparticle moves forward
by a single lattice site of the charge sequence. However, in the
$\nu'\,th$ step each site of the $c,\nu$ pseudoparticle jumps
$\nu$ sites of the charge sequence. This jump brings it to a
rotated-electron site distribution configuration where the
charge-sequence position ${\bar{h}}_j$ of the local $c,\nu$
pseudoparticle given in Eq. (\ref{bstringC}) coincides with that
of the steady local $c,\nu'$ pseudoparticle. Let $h_{j,\,x}$ where
$x=1,...,2\nu$ denote the indices of the $2\nu$ sites of the
$c,\nu$ pseudoparticle while it was located between the sites of
index $h_{j',\,\nu'-1}$ and $h_{j',\,\nu'}$ of the local $c,\nu'$
pseudoparticle. In the next collective movement the local $c,\nu$
pseudoparticle site of index $h_{j,\,2\nu}$ jumps into the site
which was associated with the local $c,\nu'$ pseudoparticle site
of index $h_{j',\,\nu'+\nu}$. Out of the remaining $2\nu-1$ sites
of the local $c,\nu$ pseudoparticle, $\nu$ of these sites of
indices $h_{j,\,x}$ such that $x=\nu,...,2\nu-1$ jump into the
local $c,\nu'$ pseudoparticle sites of indices
$h_{j',\,\nu'-\nu+x}$ where again $x=\nu,...,2\nu-1$. The
remaining $\nu-1$ sites of the local $c,\nu$ pseudoparticle of
indices $h_{j,\,x}$ such that $x=1,...,\nu-1$ jump into the sites
which were occupied by the local $c,\nu$ pseudoparticle sites of
indices $h_{j,\,x}$ where $x=\nu +2 ,...,2\nu$. As a result of
these events the local $c,\nu'$ pseudoparticle site of index
$h_{j',\,\nu'}$ jumps $\nu-1$ charge-sequence sites from the right
to the left-hand side and thus moves by a block of $\nu$ sites in
such a direction. In contrast, all the remaining $c,\nu'$
pseudoparticle sites of indices $h_{j',\,x}$ such that
$x=1,...,\nu'-1,\,\nu'+1,...,2\nu'$ jump $\nu-1$ charge-sequence
sites from the left to the right-hand side and move by a block of
$\nu$ sites in such a direction. These collective jumps reach a
configuration where the $2\nu$ sites of the $c,\nu$ pseudoparticle
are located between the sites of index $h_{j',\,\nu'}$ and
$h_{j',\,1+\nu'}$ of the $c,\nu'$ pseudoparticle. In order to
reach the next permitted position each of the $2\nu$ local $c,\nu$
pseudoparticle sites jump again a number $\nu$ of lattice sites of
the charge sequence. This brings the local $c,\nu$ pseudoparticle
to a distribution configuration where its $2\nu$ sites are located
between the sites of index $h_{j',\,1+\nu'}$ and $h_{j',\,2+\nu'}$
of the steady local $c,\nu'$ pseudoparticle. These two jumps are
followed by $\nu'-1$ $2\nu$-leg caterpillar steps where the
$c,\nu$ pseudoparticle moves again by a single lattice site of the
charge sequence. The net result of each of these elementary
collective steps is that the compact domain of $2\nu$ $c,\nu$
pseudoparticle sites interchanges position with one of the sites
of the $c,\nu'$ pseudoparticle. Step number $\nu'-2$ of these
$\nu'-1$ $2\nu$-leg caterpillar steps leads to a distribution
configuration where the $2\nu$ sites of the $c,\nu$ pseudoparticle
are located between the sites of index $h_{j',\,2\nu'-1}$ and
$h_{j',\,2\nu'}$ of the steady local $c,\nu'$ pseudoparticle.
Finally, the last step brings the local $c,\nu$ pseudoparticle to
a distribution configuration where it is located outside and on
the right-hand side of the domain of $2\nu'$ sites of the steady
local $c,\nu'$ pseudoparticle. The same intermediate positions are
reached when the $c,\nu$ pseudoparticle passes the domain of
$2\nu'$ sites of the $c,\nu'$ pseudoparticle from the right to the
left-hand side.

We have just studied a situation where a local $c,\nu$
pseudoparticle passes a steady local $c,\nu'$ pseudoparticle
belonging to a branch such that $\nu'>\nu$. For simplicity we
assumed that in the initial rotated-electron site distribution
configuration the domain of $2\nu$ charge-sequence sites located
just on the right-hand side of the steady local $c,\nu'$
pseudoparticle are occupied by Yang holons. However, our results
can be generalized to more complicated situations. An example is a
situation where a local $c,\nu$ pseudoparticle passes a steady
local $c,\nu'$ pseudoparticle which has as first neighbor in the
charge sequence a second steady local $c,\nu''$ pseudoparticle.
Another example is when a local $c,\nu$ pseudoparticle passes a
steady local $c,\nu'$ pseudoparticle containing inside its
$2\nu'$-site domain a third steady local $c,\nu''$ pseudoparticle
belonging to a branch such that $\nu''\leq\nu'$. These more
general situations can be described by the use of the properties
1-IV to 4-IV. One always finds that in order to move forward, the
$2\nu$ sites of the local $c,\nu$ pseudoparticle use $2(\nu'
-\nu)$ sites out of the total number of $2\nu'$ sites of each
steady local $c,\nu'$ pseudoparticle belonging to branches such
that $\nu'>\nu$. Moreover, when passing the central region of the
$2\nu'$-site domain of such a local $c,\nu'$ pseudoparticle, the
$2\nu$ legs of the $c,\nu$ caterpillar always perform the two
$\nu$-site jumps described above.

In summary, the number of charge-sequence {\it empty sites} of a
$c,\nu$ pseudoparticle equals the number of sites occupied by Yang
holons plus a number of sites given by $2(\nu'-\nu)$ for each
$c,\nu'$ pseudoparticle of the charge sequence belonging to
branches such that $\nu'>\nu$. The number of permitted positions
of the local $c,\nu$ pseudoparticle inside each of the site
domains of such a $c,\nu'$ pseudoparticle is $2\nu'-1$. In these
permitted positions the $c,\nu$ pseudoparticle is located between
the sites $h_{j',x}$ and $h_{j',x+1}$ such that $x=1,...,2\nu'-1$
of the local $c,\nu'$ pseudoparticle. In order to reach or to
leave a position between the sites of index $h_{j',\,\nu'}$ and
$h_{j',\,1+\nu'}$ of a steady local $c,\nu'$ pseudoparticle
belonging to a branch such that $\nu'>\nu$, each of the $2\nu$
legs of the moving caterpillar always performs a jump of $\nu$
sites in the charge sequence. Otherwise it moves throughout the
charge sequence by $2\nu$-leg caterpillar steps. In such steps the
$2\nu$-site block of the local $c,\nu$ pseudoparticle moves
forward by a charge-sequence site. The net result of such an event
is that the $2\nu$-site domain of the moving quantum object
interchanges position with either a Yang holon or with a site of a
$c,\nu'$ pseudoparticle belonging to a branch such that
$\nu'>\nu$. The present definition of the permitted positions of a
local $c,\nu$ pseudoparticle when passing the remaining quantum
objects which occupy a charge sequence provides a consistent,
complete, and unique description for the rotated-electron
distribution configurations of doubly occupied sites and empty
sites which describe the charge sequence. The results obtained for
the charge sequences and corresponding $c,\nu$ pseudoparticles are
also valid for the spin sequences and their $s,\nu$
pseudoparticles.

The events studied above are illustrated in Figs. 5 and 6 for the
case when a local $c,1$ pseudoparticle passes a steady local $c,1$
pseudoparticle and a steady local $c,2$ pseudoparticle
respectively, from the left to the right-hand side. If instead we
consider that the local $c,1$ pseudoparticle is steady, Fig. 6
illustrates the possible positions of a local $c,2$ pseudoparticle
when it passes from the right to the left-hand side a steady local
$c,1$ pseudoparticle. In this case the position in the effective
electronic lattice of the two sites of the $c,1$ pseudoparticle
remains unchanged and the different relative positions of the two
quantum objects are reached by movements of the $c,2$
pseudoparticle. Importantly, we emphasize that in this case the
$c,2$ pseudoparticle does not occupy the sites occupied by the
$c,1$ pseudoparticle. Again this is consistent with property 3-IV.

The charge sequence (and the spin sequence) is obtained by
omission of the rotated-electron singly occupied sites (and
rotated-electron doubly occupied/empty sites). Such a procedure
follows from the independent conservation of the charge and spin
sequence. We note that each of the $c,\nu$ pseudoparticle (and
$s,\nu$ pseudoparticle) branch occupancy configurations are also
independently conserved. Thus it follows our above analysis that
the construction of the effective $c,\nu$ pseudoparticle lattice
mentioned in property 8-III also involves omission of the sites of
the effective electronic lattice which are jumped by the movements
of a $c,\nu$ pseudoparticle, since these sites are not {\it seen}
by such a moving quantum object. The omitted sites correspond to
the $2\nu'$-site domains of local $c,\nu'$ pseudoparticles of
branches such that $\nu'<\nu$ and to $2\nu$-site domains belonging
to the $2\nu'$ sites of each $c,\nu'$ pseudoparticle such that
$\nu'>\nu$. Moreover, for such an effective $c,\nu$ pseudoparticle
lattice the $2\nu$-site domain of the $c,\nu$ pseudoparticles is
{\it seen} as a point-like occupied site. Thus such an effective
lattice has
$N^*_{c,\,\nu}=L_c+\sum_{\nu'>\nu}2(\nu'-\nu)\,N_{c,\,\nu'}+N_{c,\,\nu}$
sites, a number
$N^h_{c,\,\nu}=L_c+2\sum_{\nu'>\nu}(\nu'-\nu)\,N_{c,\,\nu'}$ of
which are empty of $c,\nu$ pseudoparticles, as we confirm below. A
similar analysis holds for the $s,\nu$ pseudoparticles.

\subsection{COMPLETE SET OF CHARGE, SPIN, AND $c$ PSEUDOPARTICLE
SEQUENCE LOCAL STATES}

We have just found the rotated-electron distribution
configurations of the $2\nu$ sites which describe a local
$\alpha,\nu$ pseudoparticle. Moreover, we also defined the
$N^h_{\alpha,\,\nu}$ charge-sequence {\it empty sites} of such a
local $\alpha,\nu$ pseudoparticle. Although our results can be
generalized to the case of the local spin sequences and
corresponding local $s,\nu$ pseudoparticles, often our analysis
focused explicitly on the case of the local charge sequences and
associated local $c,\nu$ pseudoparticles. An important result
which follows from our findings is that indeed the positions of
the $N^h_{c,\,\nu}$ charge-sequence {\it empty sites} of each
local $c,\nu$ pseudoparticle branch are fully determined by the
location of the $c$ pseudoparticles and $c,\nu$ pseudoparticles
alone. Our results thus confirm that for fixed values of the
numbers $N_{c,\,\nu}$ and $N^*_{c,\,\nu}$ the number of occupancy
configurations of the local $c,\nu$ pseudoparticles is given by
Eq. (\ref{NanNS}) for $\alpha =c$. It follows that we can classify
the local $c,\nu$ pseudoparticle occupancy configurations for
$\nu=1,2,...$ branches with finite occupancy in a given state by
providing the indices $h_j$ given in Eqs. (\ref{stringC}). These
indices correspond to the $N_{c,\,\nu}$ pseudoparticle locations
provided in Eq. (\ref{localcnp}). Our results also confirm that
the number of local pseudoparticle occupancy configurations of a
CPHS ensemble subspace is indeed given by Eq. (\ref{Ncphs}).
Finally, since our results can be generalized to spin sequences as
well, they also validate the use of the representation associated
with the local state defined in Eqs. (\ref{localst})-(\ref{args}).
Such a local state represents each of the different local
pseudoparticle occupancy configurations whose number for a
specific CPHS ensemble subspace is given in Eq. (\ref{Ncphs}). Our
results also confirm that the set of these local states whose
number is given in Eq. (\ref{Ncphs}) constitutes a complete set of
states in the corresponding CPHS ensemble subspace.

Thus in the case of CPHS ensemble subspaces with no $-1/2$ Yang
holons and no $-1/2$ HL spinons the local states defined in Eqs.
(\ref{localst})-(\ref{args}) constitute a complete set of states.
Our main goal here is the generalization of these local states to
CPHS ensemble subspaces with finite occupancies of both $\pm 1/2$
Yang holons and $\pm 1/2$ HL spinons. We start by expressing the
complete set of local states defined in Eqs.
(\ref{localst})-(\ref{args}) in terms of the following more basic
states,

\begin{equation}
\vert(j_{1},\,j_2,...,j_{N_c});\,
\{(\bullet_{h_{j,\,1}},...,\bullet_{h_{j,\,\nu}},
\,\circ_{h_{j,\,1+\nu}},...,
\circ_{h_{j,\,2\nu}})\};\,\{(\downarrow_{l_{j,\,1}},...,
\downarrow_{l_{j,\,\nu}},\,\uparrow_{l_{j,\,1+\nu}},...,
\uparrow_{l_{j,\,2\nu}})\}\rangle \, , \label{initial}
\end{equation}
where the charge sequence refers to the following sets of
rotated-electron distribution configurations of doubly-occupied
and empty sites,

\begin{eqnarray}
& & \{(\bullet_{h_{j,\,1}},...,\bullet_{h_{j,\,\nu}},
\,\circ_{h_{j,\,1+\nu}},..., \circ_{h_{j,\,2\nu}})\} =
(\bullet_{h_{1,\,1}},\,\circ_{h_{1,\,2}}),\,(\bullet_{h_{2,\,1}},\,\circ_{h_{2,\,2}}),...,
(\bullet_{h_{N_{c,\,1}},\,1},\,\circ_{h_{N_{c,\,1},\,2}});
\nonumber \\
& & (\bullet_{h_{1,\,1}},\,\bullet_{h_{1,\,2}},\,
\circ_{h_{1,\,3}},\,\circ_{h_{1,\,4}}),\,(\bullet_{h_{2,\,1}},\,\bullet_{h_{2,\,2}},\,
\circ_{h_{2,\,3}},\,\circ_{h_{2,\,4}})
,\,...,(\bullet_{h_{N_{c,\,2},\,1}},\,\bullet_{h_{N_{c,\,2},\,2}},\,
\circ_{h_{N_{c,\,2},\,3}},\,\circ_{h_{N_{c,\,2},\,4}});
...;\nonumber \\
& & (\bullet_{h_{1,\,1}},...,\bullet_{h_{1,\,\nu}},
\,\circ_{h_{1,\,1+\nu}},..., \circ_{h_{1,\,2\nu}}),\,
(\bullet_{h_{2,\,1}},...,\bullet_{h_{2,\,\nu}},
\,\circ_{2_{j,\,1+\nu}},..., \circ_{h_{2,\,2\nu}}),...,\nonumber \\
& &
(\bullet_{h_{N_{c,\,\nu},\,1}},...,\bullet_{h_{N_{c,\,\nu},\,\nu}},\,
\circ_{h_{N_{c,\,\nu},\,\nu+1}},...,\circ_{h_{N_{c,\,\nu},\,2\nu}})
;\,... \label{argcn}
\end{eqnarray}
and the same follows for the spin sequence with the index $h$
replaced by the index $l$, rotated-electron doubly occupied sites
$\bullet$ replaced by spin-down rotated-electron singly occupied
sites $\downarrow$, and sites free of rotated electrons $\circ$
replaced by spin-up rotated-electron singly occupied sites
$\uparrow$. The states defined in Eqs.
(\ref{initial})-(\ref{argcn}) provide the location of the $2\nu$
sites associated with each of the $\alpha,\nu$ pseudoparticles
whose branches have finite occupancy in the corresponding charge
and spin sequence. However, we note that these states include only
rotated-electron site distribution configurations of the simple
type given in Eqs. (\ref{lcnp}) and (\ref{lsnp}). The general
charge and spin sequences can be obtained from these states by
application of suitable operators, as we discussed above. These
are thus the most basic and simple states. Our first task is
expressing the states defined in Eqs. (\ref{localst})-(\ref{args})
in terms of these simple states.

Each local $c,\nu$ and $s,\nu$ pseudoparticle is described by the
superposition of $2^{\nu}$ rotated-electron site distribution
configurations given in Eqs. (\ref{cnp}) and (\ref{snp})
respectively. The type of configuration superposition given in
these equations is common to the description of all local
$\alpha,\nu$ pseudoparticles of a charge or spin sequence. Thus
the proper description of the the charge and spin sequence of each
of the states defined in Eqs. (\ref{localst})-(\ref{args})
involves the superposition of $2^{\textstyle\sum_{\alpha
=c,s}\sum_{\nu=1}^{\infty}\nu\,N_{\alpha,\,\nu}}$ such a
rotated-electron site distribution configurations. It follows that
these normalized local states can be expressed as follows,

\begin{eqnarray}
& & 2^{\textstyle[\sum_{\alpha
=c,s}\sum_{\nu=1}^{\infty}\nu\,N_{\alpha,\,\nu}]/2}\,\vert(j_{1},\,j_2,...,j_{N_c});\,
\{(h_{1},\,h_2,...,h_{N_{c,\,\nu}})\};\,\{(l_{1},\,l_2,...,l_{N_{s,\,\nu}})\}\rangle
\nonumber \\
& = &
\Bigl[\prod_{\nu'=1}^{\infty}\prod_{j'=1}^{N_{c,\nu'}}\prod_{x=1}^{2\nu'}e^{i\pi
h_{j',\,x}{\hat{D}}_{j',\,x}}\Bigr]\,\Bigl[\prod_{\alpha
=c,s}\prod_{\nu''=1}^{\infty}\prod_{j''=1}^{N_{\alpha,\nu''}}
\prod_{g''=1}^{\nu''}(1-{\hat{\cal{T}}}_{\alpha,\,\nu'',\,j'',\,g''})\Bigr]\nonumber
\\
& \times & \vert(j_{1},\,j_2,...,j_{N_c});\,
\{(\bullet_{h_{j,\,1}},...,\bullet_{h_{j,\,\nu}},
\,\circ_{h_{j,\,1+\nu}},...,
\circ_{h_{j,\,2\nu}})\};\,\{(\downarrow_{l_{j,\,1}},...,
\downarrow_{l_{j,\,\nu}},\,\uparrow_{l_{j,\,1+\nu}},...,
\uparrow_{l_{j,\,2\nu}})\}\rangle \, , \label{prima}
\end{eqnarray}
where the operator ${\hat{D}}_{j',\,x}$ measures the number of
rotated-electron doubly occupied sites in the charge-sequence site
of index $h_{j',\,x}$ (whose value is given by $1$ or $0$) and the
transformation laws for application of the operator
${\hat{\cal{T}}}_{\alpha,\,\nu'',\,j'',\,g''}$ were given above
and are illustrated in Eqs. (\ref{Tc}) and (\ref{Ts}).

The local states given in both Eqs. (\ref{localst})-(\ref{args})
and in Eq. (\ref{prima}) provide a complete basis of states for
CPHS ensemble subspaces spanned by lowest-weight states of both
the $SU(2)$ $\eta$-spin and spin algebras. Let us now consider the
general case of states associated with CPHS ensemble subspaces
spanned by states with both finite occupancies of $\pm 1/2$ Yang
holons and $\pm 1/2$ HL spinons. The description of the local
charge and spin sequences with finite occupancies of $\pm 1/2$
Yang holons and $\pm 1/2$ HL spinons respectively, is quite
similar to the ones we studied above. For simplicity let us
consider first the case of the charge sequences. According to
property 1-III the $-1/2$ Yang holons and $+1/2$ Yang holons
correspond to single rotated-electron doubly occupied lattice
sites and single rotated-electron empty lattice sites of such a
charge sequence respectively. In the case of a lowest weight state
(and highest weight state) of the $\eta$-spin algebra all Yang
holons have the same $\eta$-spin projection $\sigma_c =+1/2$ (and
$\sigma_c =-1/2$) and correspond to rotated-electron empty sites
(and rotated-electron doubly occupied sites). The application of
the off-diagonal generators of the $SU(2)$ $\eta$-spin algebra
given in Eq. (\ref{Sc}) leads to flips of the Yang holons
$\eta$-spins \cite{I}. Following property 5-III, in order to
achieve the Yang holon occupancy configurations required by the
$SU(2)$ $\eta$-spin algebra, we find that each local charge
sequence is described by a superposition

\begin{equation}
{L_c\choose L_{c,\,-1/2}} = {L_c!\over
L_{c,\,-1/2}!\,L_{c,\,+1/2}!} \, , \label{NLc}
\end{equation}
of Yang holon occupancy configurations. Here $L_{c}=2S_c$ stands
for the total number of Yang holons in the charge sequence. All
the ${L_c\choose L_{c,\,-1/2}}$ states associated with these
configurations have the same set of $L_c$ lattice sites occupied
by Yang holons and the same numbers of $-1/2$ Yang holons and
$+1/2$ Yang holons. However, these states differ in the
distributions of the $-1/2$ Yang holons and $+1/2$ Yang holons
over the $L_c$ lattice sites. The same analysis holds for spin
sequences with finite occupancies of $\pm 1/2$ HL spinons. Thus
there are $\prod_{\alpha =c,s}(L_{\alpha}+1)$ states with
precisely the same occupancy configurations of local $c$
pseudoparticles and local $\alpha,\nu$ pseudoparticles. Each of
these occupancy configurations is described by a state of form
given in Eq. (\ref{prima}). On the other hand, we denote each of
the $\prod_{\alpha =c,s}(L_{\alpha}+1)$ general states by,

\begin{equation}
\vert(L_{c,-1/2},\,L_{s,-1/2});\,(j_{1},\,j_2,...,j_{N_c});\,
\{(h_{1},\,h_2,...,h_{N_{c,\,\nu}})\};\,\{(l_{1},\,l_2,...,l_{N_{s,\,\nu}})\}\rangle
\, , \label{nlwss}
\end{equation}
where $L_{c,-1/2}$ and $L_{s,-1/2}$ are the corresponding numbers
of $-1/2$ Yang holons and $-1/2$ HL spinons respectively. We note
that Eq. (\ref{LcsLWS}) remains valid for the general local states
(\ref{nlwss}). Thus the values of the numbers $L_c$ and $L_s$ are
determined by the values of the numbers of $c$ pseudoparticles and
of $\alpha,\nu$ pseudoparticles. The values of the numbers
$L_{c,-1/2}$ and $L_{s,-1/2}$ were added to the local states
(\ref{nlwss}) because they are needed for the specification of the
corresponding CPHS ensemble subspace. In contrast, the values of
the numbers $L_{c,+1/2}$ and $L_{s,+1/2}$ are dependent and given
by $L_{c,+1/2}=L_c-L_{c,-1/2}$ and $L_{s,+1/2}=L_s-L_{c,-1/2}$ and
are not explicitly provided in the expression of the local states
(\ref{nlwss}). Each of these normalized states is given by,

\begin{eqnarray}
& & \sqrt{\Bigl[\prod_{\alpha =c,s}{L_{\alpha}\choose
L_{\alpha,\,-1/2}}\Bigr]\,\Bigl[\prod_{\alpha'
=c,s}\prod_{\nu'=1}^{\infty}
2^{\nu'\,N_{\alpha',\,\nu'}}\Bigr]}\nonumber \\
& \times &
\vert(L_{c,-1/2},\,L_{s,-1/2});\,(j_{1},\,j_2,...,j_{N_c});\,
\{(h_{1},\,h_2,...,h_{N_{c,\,\nu}})\};\,\{(l_{1},\,l_2,...,l_{N_{s,\,\nu}})\}\rangle
\nonumber \\
& = & \Bigl[\prod_{\alpha
=c,s}\Bigl({\hat{S}}^{\alpha}_{+}\Bigr)^{L_{\alpha,\,-1/2}}\Bigr]\,
\Bigl[\prod_{\nu'=1}^{\infty}\prod_{j'=1}^{N_{c,\nu'}}\prod_{x=1}^{2\nu'}e^{i\pi
h_{j',\,x}{\hat{D}}_{j',\,x}}\Bigr]\,\Bigl[\prod_{\alpha'
=c,s}\prod_{\nu''=1}^{\infty}\prod_{j''=1}^{N_{\alpha',\nu''}}
\prod_{g''=1}^{\nu''}(1-{\hat{\cal{T}}}_{\alpha',\,\nu'',\,j'',\,g''})\Bigr]\nonumber
\\
& \times & \vert(j_{1},\,j_2,...,j_{N_c});\,
\{(\bullet_{h_{j,\,1}},...,\bullet_{h_{j,\,\nu}},
\,\circ_{h_{j,\,1+\nu}},...,
\circ_{h_{j,\,2\nu}})\};\,\{(\downarrow_{l_{j,\,1}},...,
\downarrow_{l_{j,\,\nu}},\,\uparrow_{l_{j,\,1+\nu}},...,
\uparrow_{l_{j,\,2\nu}})\}\rangle \, , \label{completes}
\end{eqnarray}
where the operators ${\hat{S}}^{\alpha}_{+}$ with $\alpha =c,s$
are the off-diagonal generators defined in Eqs. (\ref{Sc}) and
(\ref{Ss}). Importantly, application of these operators onto the
local states given in Eq. (\ref{prima}) leaves the
rotated-electron site distribution configurations of the
$\alpha,\nu$ pseudoparticles invariant and generates $\eta$-spin
and spin flips in the $+1/2$ Yang holons and $+1/2$ HL spinons
respectively, as requested by property 5-III.

The set of all local states of general form given in Eq.
(\ref{completes}) provides a complete basis of states in any CPHS
ensemble subspace of the Hilbert space. However, these states do
not ensure the periodic boundary conditions of the original
electronic problem. Such conditions are ensured by the
construction of the energy eigenstates as Fourier-transform
superpositions of local states of general form given in Eq.
(\ref{completes}), as discussed in the ensuing section.

\section{SEPARATION OF THE $\alpha,\nu$ PSEUDOPARTICLE TRANSLATIONAL -
INTERNAL DEGREES OF FREEDOM: EFFECTIVE $\alpha,\nu$ PSEUDOPARTICLE
LATTICES AND THE ENERGY EIGENSTATES}

In this section we finally reach a precise definition for the
concept of effective $\alpha,\nu$ pseudoparticle lattices. The
precise definition of such a concept is a necessary condition for
the construction of the energy eigenstates as Fourier-transform
superpositions of the local states introduced in the previous
section. The concept of effective $\alpha,\nu$ pseudoparticle
lattice is related to the separation of the translational and
internal degrees of freedom of these quantum objects, as we
explain below. As mentioned in property 8-III, in addition to the
effective $c$ pseudoparticle lattice, there is an effective
$\alpha,\nu$ pseudoparticle lattice for each $\alpha =c,s$ and
$\nu =1,2,...$ pseudoparticle branch. The spatial coordinate
associated with these lattices is the conjugate of the
pseudoparticle band-momentum $q_j$, which is a good quantum number
of the many-electron problem. Thus the Fourier transforms which
relate the local pseudoparticles introduced in this paper to the
band-momentum pseudoparticles obtained from analysis of the
Bethe-ansatz solution in Ref. \cite{I} involve the spatial
coordinate of the effective pseudoparticle lattices.

Besides the introduction of the effective pseudoparticle lattices,
in this section we also provide explicit expressions for the
energy eigenstates in terms of Slater determinants which involve
Fourier-transform superpositions of the local states introduced in
the previous section. This clarifies the relation of these energy
eigenstates to the rotated-electron site distribution
configurations. Indeed, analysis of the new found expressions of
the energy eigenstates provides useful information on the
processes involved in the diagonalization of the non-perturbative
quantum problem after the electron - rotated-electron
Hilbert-space unitary rotation is performed. Such a
non-perturbative diagonalization leads to nothing but the energy
eigenstates also reached by the Bethe-ansatz solution.

\subsection{TRANSLATIONAL - INTERNAL DEGREES OF FREEDOM SEPARATION
AND THE EFFECTIVE $\alpha,\nu$ PSEUDOPARTICLE LATTICES}

Since the present problem is treated in a similar manner for the
$c,\nu$ and $s,\nu$ pseudoparticles, for simplicity let us again
concentrate namely on the case of the $c,\nu$ pseudoparticles. The
separation of the $\alpha,\nu$ pseudoparticle translational and
internal degrees of freedom can be defined in terms of the
following general properties which are obtained from the above
properties 1-III to 8-III and 1-IV to 3-IV, and other features of
the 1D Hubbard model:\vspace{0.5cm}

1-V One can separate the $2\nu$-site internal structure of a local
$c,\nu$ pseudoparticle from the problem of the description of the
movements of these quantum objects in the effective electronic
lattice. Such separation leads to the concept of effective
pseudoparticle lattice, whose spatial coordinates are associated
with the pseudoparticle translational degrees of freedom only.
According to the properties 1-IV to 4-IV, in addition to the
$2\nu$-leg caterpillar steps associated with the motion of a
$c,\nu$ pseudoparticle, such a quantum object jumps each of the
$2\nu$-site domains of the remaining local pseudoparticles of the
same branch. Thus in the description of the motion of a local
$c,\nu$ pseudoparticle, the $2\nu$-site domains of these objects
play the role of the occupied {\it sites} of an effective $c,\nu$
pseudoparticle lattice. From the point of view of the motion of
these objects in such an effective $c,\nu$ pseudoparticle lattice
these $2\nu$-site domains are point-like sites without internal
structure. Roughly speaking, these point-like pseudoparticle
occupied sites correspond to the {\it center of mass} of the
$2\nu$-site block associated with each local $c,\nu$
pseudoparticle. This description corresponds to a separation of
the pseudoparticle translational and internal degrees of freedom.
The latter degrees of freedom are described by the $2\nu$-site
rotated-electron distribution configurations studied in the
previous section. Such internal structure plays a key role both in
the fulfilment of the $c,\nu$ pseudoparticle transformations under
application of the off-diagonal generators of the $SU(2)$
$\eta$-spin algebra. In addition, such an internal structure is a
necessary condition for the fulfilment of the periodic boundary
conditions of the original electronic problem. On the other hand,
the $c,\nu$ pseudoparticle translational degrees of freedom are
closely related to the spatial coordinates of the effective
pseudoparticle lattice. These spatial coordinates play the role of
conjugate variable relative to the band momentum $q_j$. Such a
band momentum obeys Eq. (\ref{pbcanp}) to ensure periodic boundary
conditions for the original electronic problem. An important point
is that the pseudoparticle spatial coordinate occupancy
configuration of each $c,\nu$ pseudoparticle branch with finite
occupancy in a given energy eigenstate is independently conserved.
As the charge (and spin) sequence corresponds to the
rotated-electron doubly occupied and empty sites (and
rotated-electron singly occupied sites) only, one can also
introduce an independent effective $c,\nu$ pseudoparticle lattice
for each of these occupied pseudoparticle branches, whose
coordinates correspond to some of the sites of the local charge
sequence only. \vspace{0.5cm}

2-V Within the above separation of the pseudoparticle
translational and internal degrees of freedom, the moving $c,\nu$
pseudoparticle {\it sees} the $2\nu$-site domains of each of the
local pseudoparticles belonging to the same branch as point-like
occupied pseudoparticle sites. On the other hand, according to the
property 3-IV a moving $c,\nu$ pseudoparticle jumps the
$2\nu'$-site domains representative of each local $c,\nu'$
pseudoparticles belonging to branches such that $\nu'\leq\nu$.
This property can be understood as follows: Since in its movements
the $c,\nu$ pseudoparticle {\it sees} the $2\nu$-site domains of
the remaining $c,\nu$ pseudoparticles of the same branch as
point-like unreachable occupied pseudoparticle sites, such a
$2\nu$-site domain width is the smallest width {\it seen} by the
$c,\nu$ pseudoparticle. Thus such a width plays the role of a site
domain {\it width uncertainty}. As a result, pseudoparticle site
domains of width smaller than the $2\nu$-site domain are not {\it
seen} by the $c,\nu$ pseudoparticle. This is consistent with the
exact property that the moving $c,\nu$ pseudoparticle does not
{\it see} ({\it i.e.} jumps) smaller $2\nu'$-site $c,\nu'$
pseudoparticle domains corresponding to branches such that
$\nu'<\nu$. On the other hand, again according to property 3-IV,
the $2\nu$-leg caterpillar step movements of the $c,\nu$
pseudoparticle uses $2(\nu'-\nu)$ sites only, out of the
$2\nu'$-site domain of each local $c,\nu'$ pseudoparticle
belonging to branches such that $\nu'>\nu$. Thus, again
consistently with the above site domain {\it width uncertainty},
in their motion throughout the effective electronic lattice, a
local $c,\nu$ pseudoparticle does not {\it see} a number $2\nu$ of
sites out of the $2\nu'$ sites of such a $\nu'>\nu$ local $c,\nu'$
pseudoparticle.\vspace{0.5cm}

3-V Let us introduce the concepts of empty sites and lattice
constant of the effective $c,\nu$ pseudoparticle lattice which
follow from the above analysis. For the moving $c,\nu$
pseudoparticle all the $N^h_{c,\,\nu}$ charge-sequence {\it empty
sites} corresponding to a number $L_c$ of $\pm 1/2$ Yang holons
and a number $2\sum_{\nu'=\nu+1}^{\infty}(\nu'-\nu)\,N_{c,\,\nu'}$
of sites belonging to the $2\nu'$-site domains of local $c,\nu'$
pseudoparticles of the charge sequence such that $\nu'>\nu$ are
seen as equivalent and indiscernible point-like empty sites of a
{\it effective $c,\nu$ pseudoparticle lattice}. Moreover,
according to the property 1-V such a quantum object also {\it
sees} the $N_{c,\,\nu}$ $2\nu$-site domains of pseudoparticles
belonging to the $c,\nu$ pseudoparticle branch as equivalent and
indiscernible point-like occupied sites. Thus while running
through all its possible positions in the effective electronic
lattice the moving $c,\nu$ pseudoparticle {\it sees} all the
$N^*_{c,\,\nu}=N_{c,\,\nu}+N^h_{c,\,\nu}$ sites of the {\it
effective $c,\nu$ pseudoparticle lattice} as equivalent and
indiscernible point-like sites. Therefore, these sites are equally
spaced for the effective $c,\nu$ pseudoparticle lattice. The value
of the corresponding lattice constant is uniquely defined as
follows: In its movements a $c,\nu$ pseudoparticle {\it does not
see} (i) the $2\sum_{\nu'=1}^{\nu}\nu'\,N_{c,\,\nu'}$ sites
occupied by $c,\nu'$ pseudoparticles belonging to branches such
that $\nu'\leq\nu$; (ii) a number
$2\nu\,\sum_{\nu'=\nu+1}^{\infty}N_{c,\,\nu'}$ of sites belonging
to the $2\nu'$-site domains of $c,\nu'$ pseudoparticles belonging
to branches such that $\nu'>\nu$; and (iii) the $N_c$ sites singly
occupied by rotated electrons. Thus the the $c,\nu$ pseudoparticle
{\it jumps} all the above sites of the effective electronic
lattice and {\it sees} the $2\nu$-site pseudoparticle domains of
the $c,\nu$ pseudoparticle branch as point-like occupied sites. On
the other hand, in order to pass all the
$N^*_{c,\,\nu}=N_{c,\,\nu}+N^h_{c,\,\nu}$ sites of the effective
$c,\nu$ pseudoparticle lattice and return to its original
position, the $c,\nu$ pseudoparticle should run through a distance
which equals the length $L$ of the real-space and effective
electronic lattice. Therefore, a necessary condition to ensure the
periodic boundary conditions of the original electronic problem is
that the length of the effective $c,\nu$ pseudoparticle lattice
must equal the length $L$ of the effective electronic lattice.
This determines uniquely the value of the lattice constant of the
effective $c,\nu$ pseudoparticle lattice which reads
$a_{c,\,\nu}=L/N^*_{c,\,\nu}$. That the length of the effective
$c,\nu$ pseudoparticle lattice is $L$ is consistent with the
spacing $q_{j+1}-q_j=2\pi/L$ given in Eq. (\ref{qj1j}) for the
corresponding pseudoparticle discrete band-momentum values
provided by the exact Bethe-ansatz solution \cite{I}. Such a
discrete band momentum $q_j$ is the conjugate of the coordinate
$x_j=a_{c,\,\nu}\,j$ of the effective $c,\nu$ pseudoparticle
lattice where $j=1,2,...,N^*_{c,\,\nu}$. \vspace{0.5cm}

4-V A similar analysis is valid for the local $s,\nu$
pseudoparticles, provided that the above mentioned $N_c$ sites
singly occupied by rotated electrons are replaced by the
$[N_a-N_c]$ sites doubly occupied by rotated electrons and free of
rotated electrons. The lattice constant of the effective $s,\nu$
pseudoparticle lattice is given by $a_{s,\,\nu}=L/N^*_{s,\,\nu}$.
On the other hand, the effective $c$ pseudoparticle lattice has
$N_a$ lattice sites and its lattice constant $a$ is the same as
for the effective electronic lattice. The positions of the $N_c$
$c$ pseudoparticles and $N^h_c=[N_a-N_c]$ $c$ pseudoparticle holes
in this effective lattice equal the corresponding positions of the
rotated-electron singly occupied sites and rotated-electron doubly
occupied/empty sites, as already discussed in previous
sections.\vspace{0.5cm}

5-V The set of pseudoparticle occupancy configurations of the
effective $c$ pseudoparticle, $c,\nu$ pseudoparticle, and $s,\nu$
pseudoparticle lattices belonging to branches $\nu=1,2,...$ with
finite occupancy in a given local state of form (\ref{completes})
together with the numbers $L_{c,\,-1/2}$ and $L_{s,\,-1/2}$ of
such a state contain the same information as the corresponding
description of the same local state in terms of the
rotated-electron site distribution configurations.\vspace{0.5cm}

Since the value of the number $N^*_{\alpha ,\,\nu}$ given in Eqs.
(\ref{N*sum}) and (\ref{Nhag}) is distinct for different CPHS
ensemble subspaces, it follows from the expression of the
effective $\alpha,\nu$ pseudoparticle lattice constants,

\begin{equation}
a_{\alpha ,\,\nu} = a\,{N_a\over N^*_{\alpha ,\,\nu}} =  {L\over
N^*_{\alpha ,\,\nu}} \, ; \hspace{1cm} \alpha = c,\,s \, ;
\hspace{0.5cm} \nu = 1,2,3,... \, , \label{aan}
\end{equation}
that the value of such constants changes accordingly.

Let us introduce the following notation for the spatial
coordinates of the effective $c$ pseudoparticle lattice,

\begin{equation}
x_j =a\,j \, , \hspace{0.5cm} j=1,2,3,...,N_a \, , \label{xc}
\end{equation}
where the index $j$ was called $j_l$ and $j_h$ in Eq. (\ref{chcl})
for the $c$ pseudoparticle occupied sites ($l=1,2,3,...,N_c$) and
empty sites ($h=1,2,3,...,[N_a-N_c]$) respectively. We recall that
the number of these occupied and empty sites of the effective $c$
pseudoparticle lattice equals the number of rotated-electron
singly occupied sites and rotated-electron doubly occupied/empty
sites respectively. Moreover, let us introduce the following
notation for the spatial coordinates of the effective $c ,\nu$
pseudoparticle and $s,\nu$ pseudoparticle lattices,

\begin{equation}
x_j =a_{c ,\,\nu}\, j \, , \hspace{0.5cm} j=1,2,3,...,N^*_{c
,\,\nu} \, , \label{xe}
\end{equation}
and

\begin{equation}
x_j =a_{s ,\,\nu}\, j \, , \hspace{0.5cm} j=1,2,3,...,N^*_{s
,\,\nu} \, , \label{xg}
\end{equation}
respectively.

The band-momentum limiting values given in Eq. (\ref{qag}) for the
$\alpha,\nu$ pseudoparticle bands can be expressed in terms of the
corresponding lattice constants $a_{\alpha ,\,\nu}$ as follows,

\begin{equation}
q_{\alpha ,\,\nu} = {\pi\over a_{\alpha ,\,\nu}}[1-1/N_a] \approx
{\pi\over a_{\alpha ,\,\nu}} \, . \label{qagaan}
\end{equation}
Also the $c$ pseudoparticle limiting band-momentum value
(\ref{qc}) can be written as,

\begin{equation}
q_c = {\pi\over a}[1-{1/N_a}] \approx {\pi\over a} \, .
\label{qceac}
\end{equation}
Thus the domain of available pseudoparticle band-momentum values
corresponds to an effective {\it first-Brillouin zone} associated
with an underlying effective pseudoparticle lattice. The fact that
the spatial coordinate introduced in Eqs. (\ref{xc}), (\ref{xe}),
and (\ref{xg}) is the conjugate of the pseudoparticle band
momentum given in Eq. (\ref{qj}) is used in Refs. \cite{IIIb,V} in
Fourier analysis involving pseudoparticle creation and
annihilation operators.

The ground-state plays an important role in the study of
finite-energy few-electron spectral functions \cite{IIIb,V}. Thus
let us use the ground-state Eqs. (\ref{N*csnu}) and (\ref{qcanGS})
in order to find the corresponding values for the effective
$\alpha,\nu$ pseudoparticle lattice constants in the case of
electronic densities $n$ and spin densities $m$ such that $0< n<
1/a$ and $0< m< n$ respectively. We find that in the case of a
ground state corresponding to an electronic density and a spin
density whose values are within these domains these constants
read,

\begin{equation}
a_{c,\,\nu}^0 = {a\over \delta} \, ; \hspace{1cm} a_{s,\,1}^0 =
{a\over n_{\uparrow}} \, ; \hspace{1cm} a_{s,\,\nu}^0 = {a\over m}
\, , \label{acanGS}
\end{equation}
where $\delta =[1/a-n]$ is the doping concentration away from half
filling.

Moreover, the ground-state number $N^*_{\alpha,\,\nu}$ given in
Eq. (\ref{N*csnu}) and the ground-state limiting band-momentum
values (\ref{qcanGS}) can be written in terms of the effective
pseudoparticle lattice constants as follows,

\begin{equation}
N^*_{\alpha,\,\nu}={L\over a_{\alpha,\,\nu}^0} \, ,
\label{N*anuGS}
\end{equation}
and

\begin{equation}
q^0_{\alpha,\,\nu} = {\pi\over a_{\alpha,\,\nu}^0} \, .
\label{qcanGSefa}
\end{equation}
respectively.

We note that when the effective pseudoparticle lattice constants
given in Eq. (\ref{acanGS}) diverge, as is the case for
$a_{c,\,\nu}^0$ as $\delta =[1/a-n]\rightarrow 0$ and for
$a_{s,\,\nu}^0$ as $m=[n_{\uparrow}-n_{\downarrow}]\rightarrow 0$
when $\nu >1$, the corresponding number $N^*_{\alpha,\,\nu}$ is
zero. This just means that in these limits these bands and
corresponding effective lattices do not exist for states belonging
to the ground-state CPHS ensemble subspace.

Finally, let us relate the new introduced effective pseudoparticle
lattices to previous results on the 1D Hubbard model in the limit
$U/t\rightarrow\infty$. Indeed the effective $c$ and $s,1$
pseudoparticle lattices introduced here are related to known
properties of the model in that limit. For instance, it is well
known that in such a limit and at zero spin density the charge and
spin degrees of freedom of the 1D Hubbard model can be described
by two independent systems of $N$ spin-less fermions and $N/2$
spin-down spins \cite{Ogata,Penc95,Penc96,Penc97,I,II}. The
spin-less fermions can be associated with an effective lattice
with $j=1,2,3,...,N_a$ sites, whereas the $N/2$ spin-down spins
correspond to a squeezed effective lattice with $j=1,2,3,...,N/2$
sites. The spin-less fermion and spin occupancy configurations of
these effective lattices describe electron occupancy
configurations of the associated real-space lattice. In the limit
$U/t\rightarrow\infty$ such spin-less fermion and spin effective
lattices are directly related to the effective $c$ pseudoparticle
and $s,1$ pseudoparticle lattices respectively, introduced above.
On the other hand, within our pseudoparticle description of the
quantum problem these effective pseudoparticle lattices are well
defined for all finite values of $U/t$. In that case the
pseudoparticle occupancy configurations of these effective
lattices describe rotated-electron site distribution
configurations. In the limit $U/t\rightarrow\infty$ the energy
eigenstates with finite occupancies in the effective $c,\nu$
pseudoparticle lattices have finite values of electron double
occupation. Thus according to the energy spectrum given in Eq.
(\ref{EHUinf}) these states have in such a limit an infinite
excitation energy relative to the ground state and do not
contribute to the finite-energy physics. This explains why these
states are unimportant for the $U/t\rightarrow\infty$ physics and
are in general not considered in that limit
\cite{Ogata,Penc95,Penc96,Penc97}. At zero spin density the
effective $s,\nu$ pseudoparticle lattices such that $\nu>1$ do not
exist for the ground state because $N^*_{s,\,\nu}=0$ for such a
state. However, at zero spin density excitations with finite
occupancy of $s,\nu$ pseudoparticles belonging to branches such
that $\nu>1$ have a gapless energy spectrum. In the limit
$U/t\rightarrow\infty$ the spin excitations involving both
occupancy configurations of $s,1$ pseudoparticles and $s,\nu$
pseudoparticles belonging to branches such that $\nu>1$ are often
described by the isotropic Heisenberg chain which describes these
excitations \cite{Ogata,Penc95,Penc96,Penc97}.

\subsection{CONSTRUCTION OF THE ENERGY EIGENSTATES}

Our last goal is the construction of the energy eigenstates in
terms of the local charge, spin, and $c$ pseudoparticle sequences
studied in previous sections. The energy eigenstates are described
in terms of band-momentum pseudoparticle, Yang holon, and HL
spinon occupancy configurations in Ref. \cite{I} by direct use of
the Bethe-ansatz solution and $SO(4)$ symmetry of the model. In
the previous section we expressed the local states given in both
Eqs. (\ref{localst})-(\ref{args}) and Eq. (\ref{prima}) in terms
of rotated-electron site distribution configurations and showed
that they provide a complete basis of states. In order to relate
these local states to the energy eigenstates we should first
express the former states in terms of local $c$ and $\alpha,\nu$
pseudoparticle occupancy configurations in the corresponding
effective pseudoparticle lattices. We then Fourier transform the
obtained states into band-momentum space $q_j$ with respect to the
spatial coordinates of the effective $c$ and $\alpha,\nu$
pseudoparticle lattices.

As in the case of Eq. (\ref{chcl}) for the case of the effective
$c$ pseudoparticle lattice, we denote by $j_l$ and $j_h$ where
$l=1,2,...,N_{\alpha,\,\nu}$ and $h=1,2,...,N^h_{\alpha,\,\nu}$
respectively, the occupied and empty sites respectively, of the
effective $\alpha,\nu$ pseudoparticle lattices. The spatial
coordinate of these occupied and empty pseudoparticle sites is
given by,

\begin{eqnarray}
x_{j_{l}} & = & a_{\alpha,\,\nu}\,j_l \, , \hspace{0.5cm}
l=1,2,...,N_{\alpha,\,\nu} \, ; \hspace{1cm} \alpha = c,\,s \, ; \nonumber \\
x_{j_{h}} & = & a_{\alpha,\,\nu}\,j_h \, , \hspace{0.5cm}
h=1,2,...,N^h_{\alpha,\,\nu} \, ; \hspace{1cm} \alpha = c,\,s \, ,
\label{aeloe}
\end{eqnarray}
where $j_l$ is the index which defines the position of the
occupied pseudoparticle sites and $j_h$ the location of the empty
pseudoparticle sites. We note that the empty sites of the
effective pseudoparticle lattices correspond to the sites left
over by the occupied sites and vice versa. Thus we can uniquely
specify a given effective pseudoparticle lattice site distribution
configuration by providing the location of the occupied sites (or
empty sites) only. Here we choose the representation in terms of
the locations in the effective $c$ and $\alpha,\nu$ pseudoparticle
lattices of the occupied sites $x_{j_{l}}$ of Eqs. (\ref{chcl})
and (\ref{aeloe}). In such a representation the states given in
Eq. (\ref{completes}) are denoted as,

\begin{equation}
\vert(L_{c,-1/2},\,L_{s,-1/2});\,(x_{j_{1}},\,x_{j_2},...,x_{j_{N_c}});\,
\{(x_{j_{1}},\,x_{j_2},...,x_{j_{N_{c,\,\nu}}})\};\,\{(x_{j_{1}},\,x_{j_2},...,x_{j_{N_{s,\,\nu}}})\}\rangle
\, . \label{ellss}
\end{equation}
Here $(x_{j_{1}},\,x_{j_2},...,x_{j_{N_c}})$ is the set of spatial
coordinates associated with the set of indices
$(j_{1},\,j_2,...,j_{N_c})$ given in Eq. (\ref{lcp}). These
spatial coordinates correspond to the set of sites occupied by $c$
pseudoparticles in the corresponding effective lattice. Moreover,

\begin{eqnarray}
\{(x_{j_{1}},\,x_{j_2},...,x_{j_{N_{\alpha,\,\nu}}})\} & = &
(x_{j_1},\,x_{j_2},...,x_{j_{N_{\alpha,\,1}}});\,
(x_{j_{1}},\,x_{j_2},...,x_{j_{N_{\alpha,\,2}}});\,(x_{j_{1}},\,x_{j_2},...,x_{j_{N_{\alpha,\,3}}});...
\, , \nonumber \\
\alpha & = & c,\,s \, , \label{eanel}
\end{eqnarray}
gives the local pseudoparticle coordinates of each of the
effective $\alpha,\nu$ pseudoparticle lattices with finite
occupancy in the associated effective electronic lattice.

We emphasize that from analysis of the specific rotated-electron
site distribution configurations relative to a given state of form
(\ref{prima}) one can construct by use of the above properties 1-V
to 3-V the $c$ pseudoparticle, $c,\nu$ pseudoparticle, and $s,\nu$
pseudoparticle occupancy configurations in the effective $c$,
$c,\nu$, and $s,\nu$ pseudoparticle lattices respectively, of the
corresponding state of form (\ref{ellss}). This relation was
already mentioned in the above property 5-V and the inverse
statement is obviously also true. Thus Eqs. (\ref{prima}) and
(\ref{ellss}) refer to two different representations of the same
local states. These local states constitute a complete set of
states. However, the form of these local states does not ensure
the periodic boundary conditions of the original electronic
problem.

The representation of the local states in terms of the effective
pseudoparticle lattice occupancy configurations given in Eq.
(\ref{ellss}) is the most suitable starting point for construction
of the energy and momentum eigenstates. Such a construction is
fulfilled by Fourier transforming the local states given in Eq.
(\ref{ellss}) into band-momentum space with respect to the spatial
coordinates of the effective pseudoparticle lattices given in Eqs.
(\ref{xc}), (\ref{xe}), and (\ref{xg}). Provided that the discrete
band-momentum values obey the boundary conditions associated with
Eqs. (\ref{pbccp}) and (\ref{pbcanp}), the form of the obtained
states ensures the periodic boundary conditions of the original
electronic problem. Such procedures lead to the following
expression of the energy eigenstates in terms of the local states
of Eq. (\ref{ellss}),

\begin{eqnarray}
& & N_a^{N_c/2}\Bigl(\prod_{\alpha =c,s}\prod_{\nu'=1}^{\infty}
[N^*_{\alpha,\,\nu'}]^{N_{\alpha,\,\nu'}/2}\Bigr)\nonumber \\
& \times & \vert
(L_{c,-1/2},\,L_{s,-1/2});\,(q_{j_1},\,q_{j_2},...,q_{j_{N_c}});\,
\{(q_{j_1},\,q_{j_2},...,q_{j_{N_{c,\,\nu}}})\};\,\{(q_{j_1},\,q_{j_2},...,q_{j_{N_{s,\,\nu}}})\}\rangle
\nonumber \\
& = & \sum_{j_{1}<j_2<...<j_{N_c}}\sum_{{\cal{P}}}\,(-1)^{\cal{P}}
\,e^{(ia\sum_{l=1}^{N_c}j_{{\cal{P}}(l)}\,q_{j_{l}})}\nonumber
\\
& \times &
\Bigl[\prod_{\nu'=1}^{\infty}\,\Bigl(\sum_{j_{1}<j_2<...<j_{N_{c,\,\nu'}}}\sum_{{\cal{P}}}\,(-1)^{\cal{P}}
\,e^{(ia_{c,\,\nu}\sum_{l=1}^{N_{c,\,\nu'}}j_{{\cal{P}}(l)}\,[{\pi\over
a} - q_{j_{l}}])}\Bigr)\Bigr]\nonumber
\\
& \times &
\Bigl[\prod_{\nu''=1}^{\infty}\,\Bigl(\sum_{j_{1}<j_2<...<j_{N_{s,\,\nu''}}}\sum_{{\cal{P}}}\,(-1)^{\cal{P}}
\,e^{(ia_{s,\,\nu}\sum_{j'=1}^{N_{s,\,\nu''}}j_{{\cal{P}}(l)}\,q_{j_{l}})}\Bigl)\Bigr]\nonumber
\\
& \times &
\vert(L_{c,-1/2},\,L_{s,-1/2});\,(x_{j_{1}},\,x_{j_2},...,x_{j_{N_c}});\,\{(x_{j_{1}},\,x_{j_2},...,
x_{j_{N_{c,\,\nu}}})\}
;\,\{(x_{j_{1}},\,x_{j_2},...,x_{j_{N_{s,\,\nu}}})\}\rangle \, .
\label{wffin}
\end{eqnarray}
The permutations ${\cal{P}}$ on the right-hand side of Eq.
(\ref{wffin}) generate a Slater determinant of the band momenta
and the spatial coordinates of the $N_c$ $c$ pseudoparticles and
$N_{\alpha,\,\nu}$ $\alpha,\nu$ pseudoparticles belonging to
branches $\alpha =c,\,s$ and $\nu =1,2,...$ with finite occupancy
in the corresponding local charge and spin sequences. On the
left-hand side of Eq. (\ref{wffin}) the set of pseudoparticle
occupied band-momentum values specify the energy eigenstate. There
$(q_{j_1},\,q_{j_2},...,q_{j_{N_c}})$ are the set of $N_c$
occupied band momentum values out of the $N_a$ available discrete
$q_j$ values of the $c$ pseudoparticle band such that
$j=1,...,N_a$. Moreover,

\begin{eqnarray}
\{(q_{j_1},\,q_{j_2},...,q_{j_{N_{\alpha,\,\nu}}})\} & = &
(q_{j_1},\,q_{j_2},...,q_{j_{N_{\alpha,\,1}}});\,
(q_{j_1},\,q_{j_2},...,q_{j_{N_{\alpha,\,2}}});\,(q_{j_1},\,q_{j_2},...,
q_{j_{N_{\alpha,\,3}}});...\, , \nonumber \\
\alpha & = & c,\,s \, , \label{eanqj}
\end{eqnarray}
are the set of $N_{\alpha,\,\nu}$ occupied band momentum values
out of the $N^*_{\alpha,\,\nu}$ available discrete $q_j$ values of
the $\alpha,\nu$ pseudoparticle bands of branches $\alpha =c,\,s$
and $\nu=1,2,...$ such that $j=1,...,N^*_{\alpha,\,\nu}$. Such
band-momentum pseudoparticle occupancy configurations are
discussed in Ref. \cite{I}.

Since equations (\ref{prima}) and (\ref{ellss}) refer to two
different representations of the same states, the energy
eigenstates (\ref{wffin}) are Fourier-transform superpositions
involving permutations of the local charge-sequence and
spin-sequence states (\ref{ellss}). The latter local states
correspond to rotated-electron site distribution configurations.
Thus combination of the general expressions (\ref{completes}) and
(\ref{wffin}) provides important information about the
relationship between the band-momentum pseudoparticles and the
rotated-electron occupancy configurations. We emphasize that the
permutations on the right-hand side of Eq. (\ref{wffin}) are
associated with different locations for the local $\alpha,\nu$
pseudoparticles but that the internal structure of these quantum
objects remains invariant under these permutations.  Both the
Fourier transforms and permutations of the general expression
(\ref{wffin}) do not touch the internal structure of the local
$\alpha,\nu$ pseudoparticles and refer only to their translational
degrees of freedom. This results from the point-like character of
the occupied sites of the effective pseudoparticle lattices. Thus
the local and band-momentum $\alpha,\nu$ pseudoparticles have the
same internal structure.

While the complete set of symmetrized states used in Ref.
\cite{Geb} are energy eigenstates of the 1D Hubbard model in the
limit $U/t\rightarrow\infty$ only, the states given in Eqs.
(\ref{wffin}) and (\ref{eanqj}) are energy eigenstates of that
model for all values of $U/t$. The local charge, spin, and $c$
pseudoparticle sequences defined in expressions (\ref{prima}) and
(\ref{completes}) and used in the construction of the latter
states, describe rotated-electron site distribution configurations
which in the limit $U/t\rightarrow\infty$ refer to similar
electron configurations. Thus in contrast to the complete set of
symmetrized energy eigenstates used in Ref. \cite{Geb}, the
complete set of energy eigenstates of form (\ref{wffin}) are in
the limit $U/t\rightarrow\infty$ the band-momentum states
associated with the unitary operator $\hat{V}(U/t)$.

In the case of the energy eigenstates given in Eq. (\ref{wffin}),
a first condition for the fulfilment of the periodic boundary
conditions of the original electronic problem is that in all the
local charge and spin sequences described by the states
(\ref{completes}), the local $c,\nu$ pseudoparticles and $s,\nu$
pseudoparticles are described by the same rotated-electron
$2\nu$-site distribution configurations given in Eqs. (\ref{cnp})
and (\ref{snp}) respectively. These are obtained from the proper
symmetrization of the corresponding single pseudoparticle
$2\nu$-site sequences. A second condition to ensure these periodic
boundary conditions is that energy eigenstates given in Eq.
(\ref{wffin}) are Fourier transform superpositions of these charge
and spin sequences of the form given on the right-hand side of
that equation. Provided that the discrete values of the band
momentum $q_j$ of these Slater determinant superpositions obey
Eqs. (\ref{pbccp}) and (\ref{pbcanp}), the periodic boundary
conditions of the original electronic problem are ensured. Note
that the general expression provided in Eq. (\ref{wffin}) includes
suitable permutations of the local sequences. In the same equation
the rotated-electron site distribution configurations of the local
charge and spin sequences associated with the states (\ref{prima})
or (\ref{completes}) are reexpressed in terms of pseudoparticle
occupancy configurations of the corresponding effective
pseudoparticle lattices.

Let us consider the specific case of a ground state with
electronic densities and spin densities belonging to the domains
$0\leq n\leq 1/a$ and $0\leq m\leq n$ respectively. Such a state
is a superposition of local normalized states (\ref{completes})
which have no finite occupancies of $-1/2$ Yang holons, $-1/2$ HL
spinons, $c,\nu$ pseudoparticles, and $s,\nu$ pseudoparticles
belonging to branches such that $\nu>1$ \cite{II,III}. Thus in the
case of such a ground state these local states are of the
following simplified form,

\begin{eqnarray}
& & 2^{N_{s,\,1}/2}\,
\vert(0,\,0);\,(x_{j_{1}},\,x_{j_2},...,x_{j_{N_c}});\,(x_{j_{1}},\,x_{j_2},...,
x_{j_{N_{s,\,1}}})\rangle
\nonumber \\
& = & \Bigl[\prod_{j'=1}^{N_{s,1}}
(1-{\hat{\cal{T}}}_{s,\,1,\,j',\,1})\Bigr]\,\vert(0,\,0);\,(j_{1},\,j_2,...,j_{N_c});
\,\{(\downarrow_{l_{j',\,1}},\,\uparrow_{l_{j',\,2}})\}\rangle \,
, \label{localGS}
\end{eqnarray}
where

\begin{equation}
\{(\downarrow_{l_{j,\,1}},\,\uparrow_{l_{j,\,2}})\} =
(\downarrow_{l_{1,\,1}},\,\uparrow_{l_{1,\,2}}),\,(\downarrow_{l_{2,\,1}},\,\uparrow_{l_{2,\,2}}),...,
(\downarrow_{l_{N_{s,\,1}},\,1},\,\uparrow_{l_{N_{s,\,1},\,2}}) \,
, \label{argGSl}
\end{equation}
and the operators ${\hat{\cal{T}}}_{s,\,1,\,j'',\,1}$ are the same
as in the general Eq. (\ref{completes}). The states defined in
Eqs. (\ref{localGS}) and (\ref{argGSl}) provide the position of
the two rotated-electron singly occupied sites associated with the
internal structure of each local $s,1$ pseudoparticle.

These local states have an alternative representation in terms of
the $c$ and $s,1$ pseudoparticle occupancy configurations of the
effective $c$ and $s,1$ pseudoparticle lattices respectively.
Within such an effective pseudoparticle lattice representation the
local states (\ref{localGS}) are denoted by
$\vert(0,\,0);\,(x_{j_{1}},\,x_{j_2},...,x_{j_{N_c}})
;\,(x_{j_{1}},\,x_{j_2},...,x_{j_{N_{s,\,1}}})\rangle$. Here
$(x_{j_{1}},\,x_{j_2},...,x_{j_{N_{s,\,1}}})$ are the spatial
coordinates which define the positions of the local $s,1$
pseudoparticles in the effective $s,1$ pseudoparticle lattice.
These spatial coordinates correspond to the indices
$(l_{1},\,l_2,...,l_{N_{s,\,1}})$ on the left-hand side of Eq.
(\ref{localGS}). The sites of such an effective $s,1$
pseudoparticle lattice are equally spaced, the corresponding
ground-state lattice constant $a_{s,\,1}^0$ being given in Eq.
(\ref{acanGS}). The ground state expression is a particular case
of the Fourier-transform superposition of local states given in
Eq. (\ref{wffin}). In the case of the ground-state such an
expression simplifies to,

\begin{eqnarray}
\vert GS\rangle & = &
N_a^{N^0/2}\,(N^0_{\uparrow})^{N^0_{\downarrow}/2}\vert(0,\,0);\,(q_{j_1},\,q_{j_2},...,q_{j_{N^0}})
;\,(q_{j_1},\,q_{j_2},...,q_{j_{N^0_{\downarrow}}})\rangle
\nonumber \\
& = & \sum_{c_{1}<c_2<...<c_{N^0}}\sum_{{\cal{P}}}\,(-1)^{\cal{P}}
\,e^{(ia\sum_{l=1}^{N^0}c_{{\cal{P}}(l)}\,q_{j_{l}})}\,
\sum_{g_{1}<g_2<...<g_{N^0_{\downarrow}}}\sum_{{\cal{P}}}\,(-1)^{\cal{P}}
\,e^{(ia_{s,\,1}\sum_{j'=1}^{N^0_{\downarrow}}g_{{\cal{P}}(l)}\,q_{j_{l}})}\nonumber
\\
& \times & \vert(0,\,0);\,(x_{j_{1}},\,x_{j_2},...,x_{j_{N^0}})
;\,(x_{j_{1}},\,x_{j_2},...,x_{j_{N^0_{\downarrow}}})\rangle \, ,
\label{GSfin}
\end{eqnarray}
where $N_c =N^0$, $N_{s,\,1}=N^0_{\downarrow}$, and
$N^*_{s,\,1}=N^0_{\uparrow}$.

\section{PRACTICAL APPLICATION OF THE DERIVED LOCAL PSEUDOPARTICLE REPRESENTATION}

The concepts of {\it local pseudoparticle} and of {\it effective
pseudoparticle lattice} introduced in this paper are applied in
the studies of Refs. \cite{IIIb,V} to the evaluation of
finite-energy few-electron spectral function expressions. In this
paper we identified the conjugate variable of the pseudoparticle
band momentum obtained naturally from the Bethe-ansatz solution in
Ref. \cite{I}. Such a conjugate variable is nothing but the
spatial coordinate of the effective pseudoparticle lattice. We
found that there is one of such lattices for each pseudoparticle
branch. Moreover, we studied the pseudoparticle site distribution
configurations which describe the energy eigenstates. We have also
related these configurations to the rotated-electron site
distribution configurations which describe the same states. Since
electrons and rotated electrons are related by a mere unitary
transformation, our results provide useful information about the
relation between electronic excitations and energy eigenstates.

The results obtained in this paper are used in Refs. \cite{IIIb,V}
in the evaluation of the overlap between few-electron excitations
and energy eigenstates. As discussed in Ref. \cite{IIIb}, the
pseudoparticles studied here have residual interactions. In that
reference it is found that a canonical unitary transformation maps
the pseudoparticles onto non-interacting {\it pseudofermions}.
(This transformation is other than the electron - rotated-electron
canonical unitary transformation associated with the operator
(\ref{SV}).) Interestingly, the latter quantum objects are more
suitable for the evaluation of matrix elements between the ground
state and excited states. The pseudoparticle - pseudofermion
unitary transformation maps the $c$ pseudoparticle (and
$\alpha,\nu$ pseudoparticle) discrete band momentum value $q_j$
onto the $c$ pseudofermion (and $\alpha,\nu$ pseudofermion)
discrete {\it shifted momentum} value ${\bar{q}}_j = q_j
+Q_c(q_j)/L$ (and $\bar{q} = q_j +Q_{\alpha,\,\nu}(q_j)/L$). Here
$Q_c(q)/L$ and $Q_{\alpha,\,\nu}(q)/L$ are momentum functionals
which depend on the excited state occupancy configurations and
involve the two-pseudofermion phase shifts. Although the momenta
$Q_c(q_j)/L$ and $Q_{\alpha,\,\nu}(q_j)/L$ are of order $1/L$ and
play a crucial role in the spectral properties of the quantum
problem, the above transformation is such that the differences
$[Q_c(q_{j+1})-Q_c(q_j)]/L$ and
$[Q_{\alpha,\,\nu}(q_{j+1})-Q_{\alpha,\,\nu}(q_j)]/L$ are of order
$[1/L]^2$ and thus vanish within our large-$L$ description.
Therefore, to first order in $1/L$ the shifted-momentum values are
such that ${\bar{q}}_{j+1}-{\bar{q}}_j=2\pi/L$, like the
corresponding band-momentum values. It follows that there is no
level crossing between the sequence of $\{q_j\}$ discrete values
and the corresponding sequence of $\{{\bar{q}}_j\}$ discrete
values such that $j=1,2,...,N_a$ and $j=1,2,...,N_{\alpha,\,\nu}$
for the $c$ and $\alpha,\nu$ branches respectively. Moreover, the
presence of the extra momentum term $Q_c(q_j)/L$ or
$Q_{\alpha,\,\nu}(q_j)/L$ in the shifted-momentum expression does
not affect the underlying effective $\alpha ,\nu$ pseudoparticle
lattice. Indeed, that momentum just imposes a twisted boundary
condition such that each $c$ pseudofermion (and $\alpha ,\nu$
pseudofermion) hopping from site $N_a$ to site $0$ (and site
$N^*_{\alpha ,\,\nu}$ to site $0$) of such an effective lattice
will acquire a phase $e^{iQ_c(q)}$ (and
$e^{iQ_{\alpha,\,\nu}(q)}$). Thus the effective pseudoparticle
lattice introduced here is invariant under the pseudoparticle -
pseudofermion transformation studied in Ref. \cite{IIIb}.

In the case of the Hilbert subspace where the few-electron
excitations are contained there is an one-to-one correspondence
between the pseudofermion shifted-momentum occupancy
configurations (and pseudofermion site distribution
configurations) and the corresponding pseudoparticle band-momentum
occupancy configurations (and pseudoparticle site distribution
configurations) that describe the energy eigenstates. Thus the
results found in the present paper are useful for the description
of these states in terms of pseudofermion occupancy configurations
as well. Since the pseudofermions are non-interacting, it is found
in Ref. \cite{IIIb} that the wave function of the excited states
of the ground-state normal-ordered 1D Hubbard model factorizes for
all values of $U/t$. Importantly, such a factorization is used in
Ref. \cite{V}, where the few-electron spectral functions are
expressed as a convolution of single pseudofermion spectral
functions. This factorization involves the pseudofermion branches
with finite occupancy in the excited states. Thus due to such a
factorization the problem of the evaluation of finite-energy
few-electron spectral functions is reduced to the problem of the
evaluation of pseudofermion spectral functions. The latter problem
involves the derivation of matrix elements between the ground
state and excited states for pseudofermion operators.

The effective pseudoparticle lattices introduced in this paper
play an important role in the evaluation of the pseudofermion
matrix elements. It turns out that the derivation of the above
matrix elements involves the anticommutator $\{f^{\dag
}_{{\bar{q}},\,\alpha,\,\nu},\,f_{{\bar{q}}',\,\alpha' ,\,\nu'}\}$
where the operators $f^{\dag }_{{\bar{q}},\,\alpha,\,\nu}$ and
$f_{{\bar{q}},\,\alpha,\,\nu}$ create and annihilate respectively,
a $\alpha,\,\nu$ pseudofermion of sifted momentum ${\bar{q}}$, and
similar operators can be introduced for the $c$ pseudofermions. We
note that because of the functional character of the momenta
$Q_c(q)/L$ and $Q_{\alpha,\,\nu}(q)/L$ these anticommutation
relations have not the usual fermionic form. This justifies the
designation {\it pseudofermion}. Furthermore, without the concept
of local pseudoparticle and effective pseudoparticle lattice
introduced in this paper the evaluation of these anticommutators
would be a complex open problem. However, our results imply the
existence of {\it local pseudofermions} which correspond to the
local pseudoparticles introduced in this paper. Fortunately both
the invariance of the pseudoparticle effective lattice under the
pseudoparticle - pseudofermion transformation and the associated
concept of local pseudofermion allows the expression of the above
anticommutators in terms of the local pseudofermion operator
anticommutators $\{f^{\dag
}_{x_j,\,\alpha,\,\nu},\,f_{x_{j'},\,\alpha' ,\,\nu'}\}$
associated with spatial coordinates $x_j$ and $x_{j'}$ of the
effective $\alpha ,\nu$ and $\alpha' ,\nu'$ lattices respectively.
Such an expression is very useful for the evaluation of the
pseudofermion matrix elements between the ground state and excited
states \cite{IIIb,V}. As in the case of the pseudoparticles, also
the shifted-momentum and local pseudofermion descriptions are
related by a simple Fourier transform \cite{IIIb}. Thus one can
express the above shifted-momentum pseudofermion operator
anticommutators in terms of the local pseudofermion operator
anticommutators as follows,

\begin{equation}
\{f^{\dag }_{{\bar{q}},\,\alpha,\,\nu},\,f_{{\bar{q}}',\,\alpha'
,\,\nu'}\} = {1\over
L}\,\sum_{j=1}^{N^*_{\alpha,\,\nu}}\,\sum_{j'=1}^{N^*_{\alpha',\,\nu'}}\,
e^{i{\bar{q}}\,x_j-i{\bar{q}}'\,x_{j'}}\, \{f^{\dag
}_{x_j,\,\alpha,\,\nu},\,f_{x_{j'},\,\alpha' ,\,\nu'}\} \, .
\label{pfacrjj}
\end{equation}
As a result of the introduction of the concept of effective
pseudoparticle (and pseudofermion) lattice the derivation of the
local operator anticommutators on the right-hand side of this
equation is trivial and expressions for the band-momentum
anticommutators operators (\ref{pfacrjj}) can be easily obtained
by performing the $j$ and $j'$ summations \cite{IIIb}.
Corresponding expressions are found for the $c$ pseudofermions
such that,

\begin{equation}
\{f^{\dag }_{{\bar{q}},\,c},\,f_{{\bar{q}}',\,c}\} = {1\over
L}\,\sum_{j=1}^{N_a}\,\sum_{j'=1}^{N_a}\,
e^{i{\bar{q}}\,x_j-i{\bar{q}}'\,x_{j'}}\, \{f^{\dag
}_{x_j,\,c},\,f_{x_{j'},\,c}\} \, . \label{pfacrjjc}
\end{equation}
and the anticommutators involving $\alpha,\nu$ pseudofermions and
$c$ pseudofermions vanish. (Also the anticommutators given in Eq.
(\ref{pfacrjj}) vanish for $\alpha,\nu\neq \alpha',\nu'$
\cite{IIIb}.)

We note that the pseudofermion description introduced in Ref.
\cite{IIIb} is a generalization for all values of $U/t$ of the
well known $U/t\rightarrow\infty$ quantum-object description used
in Refs. \cite{Ogata,Penc95,Penc96,Penc97}. For instance, in the
limit $U/t\rightarrow\infty$ the $c$ pseudofermions become the
spin-less fermions of Refs.
\cite{Ogata,Ricardo,Penc95,Penc96,Penc97} and the spinons become
the spins of these references. The $s,1$ pseudofermions describe a
spin pair of opposite projections. Also the remaining $s,\nu$
pseudofermions describe $2\nu$ spins, $\nu$ of each direction. In
the limit $U/t\rightarrow\infty$ the general $c$ pseufofermion
anticommutation relations (\ref{pfacrjjc}) become nothing but the
spin-less fermion anticommutation relations provided in the first
(unnumbered) equation of Sec. IV of Ref. \cite{Penc97}. As
discussed in Ref. \cite{IIIb} and confirmed in Ref. \cite{V},
except for the $U/t$ dependence of the two-pseudofermion phase
shifts involved in the general expressions of the momentum
functionals $Q_c(q)/L$ and $Q_{\alpha,\,\nu}(q)/L$, the evaluation
of the pseudofermion matrix elements for finite values of $U/t$ is
formally similar to the derivation of the spin-less fermion matrix
elements presented in Ref. \cite{Penc97} for
$U/t\rightarrow\infty$. This similarity results from the facts
that both the energy-eigenstate rotated-electron occupancy
configurations and corresponding pseudoparticle, holon, and spinon
occupancy configurations introduced in this paper are independent
of the value of $U/t$ and that in the limit $U/t\rightarrow\infty$
considered in Ref. \cite{Penc97} electrons and rotated electrons
are the same quantum objects. This example reveals how the local
pseudoparticle (and pseudofermion) representation introduced in
this paper is used for practical calculations at finite $U/t$. The
only formal difference between the finite $U/t$ studies of Refs.
\cite{IIIb,V} and those of Ref. \cite{Penc97} for
$U/t\rightarrow\infty$ is that in the former references the
evaluation of the spectral function for each pseudofermion branch
proceeds as in latter reference for the spin-less fermions. In
contrast, the studies of Ref. \cite{Penc97} use the well known
$U/t\rightarrow\infty$ connection between the spin degrees of
freedom of the 1D Hubbard model and the spin $1/2$ isotropic
Heisenberg chain  \cite{Emery} to evaluate the spin part of the
one-electron spectral function. Such direct connection does not
exist for finite values of $U/t$. However, the method used in
Refs. \cite{IIIb,V} for finite values of $U/t$ is also valid for
$U/t\rightarrow\infty$ and leads to the same final results as the
analysis of Ref. \cite{Penc97}.

\section{CONCLUDING REMARKS}

In this paper we studied the relation between the description of
the energy eigenstates of the 1D Hubbard model in terms of
rotated-electron occupancy configurations and of Yang holon, HL
spinons, and pseudoparticle band momentum occupancy
configurations. Such a relation involves the concepts of {\it
local pseudoparticle} and {\it effective pseudoparticle lattices}.
We introduced the pseudoparticle spatial coordinates $x_j$
associated with and conjugate of the pseudoparticle band-momentum
$q_j$. The band-momentum pseudoparticle description is provided by
the Bethe-ansatz Takahashi's thermodynamic equations
\cite{Takahashi,I}. The pseudoparticle spatial coordinates
introduced in this paper correspond to effective pseudoparticle
lattices whose length and lattice constant are independent of the
value of $U/t$. The introduction of the concept of pseudoparticle
effective lattice involved the study rotated-electron distribution
configurations of doubly occupied and empty sites (and of
spin-down and spin-up singly occupied sites) which describe the
internal structure of the local $c,\nu$ pseudoparticles (and local
$s,\nu$ pseudoparticles). In what the translational degrees of
freedom are concerned, the local $\alpha,\nu$ pseudoparticle is a
point-like quantum object, its internal structure being the same
as that of the corresponding band-momentum pseudoparticle. We also
found that the spatial coordinates of the occupied and empty sites
of the effective $c$ pseudoparticle lattice are the same as the
coordinates of the sites singly occupied by rotated electrons and
rotated-electron doubly occupied/empty sites respectively. The
energy eigenstates can be described in terms of pseudoparticle
site distribution configurations in the corresponding effective
pseudoparticle lattice. Our results reveal that there is an
one-to-one correspondence between the local pseudoparticle site
distribution configurations in the effective pseudoparticle
lattices which describe a given energy eigenstate and the
rotated-electron site distribution configurations which describe
the same state.

As mentioned in Sec. II, the electron site distribution
configurations which describe the energy eigenstates are very
complex and dependent on the value of $U/t$. This is confirmed by
the $U/t$ dependence of the double occupation expectation value
studied in Ref. \cite{II}. However, the electron -
rotated-electron canonical unitary transformation is such that it
maps these complex and $U/t$ dependent electron site distribution
configurations onto the corresponding $U/t$ independent
rotated-electron site distribution configurations studied in this
paper. This is an important property of the non-perturbative
diagonalization of the 1D Hubbard model. According to the results
obtained in this paper, this property implies that the local
pseudoparticle site distribution configurations of the effective
pseudoparticle lattices which describe the energy eigenstates are
also independent of the value of $U/t$. Moreover, since the
spatial coordinate of these effective lattices is the conjugate of
the pseudoparticle band-momentum obtained from analysis of the
Bethe-ansatz solution \cite{I}, it is again such a property which
justifies why the pseudoparticle band-momentum occupancies which
describe the energy eigenstates are independent of the values of
$U/t$. The invariance associated with the commutation of the
momentum operator with the electron - rotated-electron unitary
operator is related to the $U/t$ independence of the band-momentum
pseudoparticle, rotated-electron, and local pseudoparticle
occupancy configurations which describe the energy eigenstates. In
this paper we profitted from such a $U/t$ independence of the
pseudoparticle and rotated-electron occupancy configurations and
extracted some of our results from analysis of the
$U/t\rightarrow\infty$ problem. In that limit the quantum problem
simplifies because the $\eta$-spin and spin occupancy
configurations are degenerated and electrons and rotated electrons
are the same quantum object and thus electron double occupation is
a good quantum number.

We also provided preliminary information about the application of
the concepts of local pseudoparticle and effective pseudoparticle
lattice introduced in this paper to the evaluation of the spectral
properties of few-electron operators. The study of the
finite-energy few-electron spectral functions is required for the
further understanding of the unusual finite-energy/frequency
spectral properties observed in quasi-1D materials
\cite{Menzel,Hasan,Ralph,Gweon,Takenaka,Vescoli}, which are far
from being well understood. Indeed there are clear indications
that electronic correlations effects might play an important role
in the finite-energy physics of these low-dimensional materials
\cite{Menzel,Hasan,Ralph,Gweon,Takenaka,Vescoli}. For low values
of the excitation energy the microscopic electronic properties of
these materials are usually described by systems of coupled
chains. On the other hand, for finite values of the excitation
energy larger than the transfer integrals for electron hopping
between the chains, 1D lattice models like the 1D Hubbard model
are expected to provide a good description of the physics of these
materials. The studies of the present paper are a first necessary
step for the evaluation of few-electron spectral function
expressions for finite values of excitation energy, as confirmed
in Refs. \cite{IIIb,V}. As mentioned in Sec. I, application of a
preliminary version of the theory introduced here and in these
references to the study of the one-electron removal spectral
properties of quasi-1D materials leads to a quantitative and
successful description of the exotic independent charge and spin
spectral lines observed in photoemission experiments for all
values of excitation energy \cite{spectral0}.

\begin{acknowledgments}
We thank Nuno Peres for useful discussions related to the
occupancy configurations represented in the figures of this paper
and for valuable support in the production of these figures . We
also thank Jim W. Allen, Ant\^onio Castro Neto, Ralph Claessen,
Ricardo Dias, Francisco (Paco) Guinea, Katrina Hibberd, Peter
Horsch, Lu\'{\i}s Miguel Martelo, Karlo Penc, and Pedro Sacramento
for stimulating discussions.
\end{acknowledgments}
\appendix

\section{RELATION TO THE LARGE $U/t$ PHYSICS AND OTHER CHOICES OF ENERGY
EIGENSTATES FOR $U/t\rightarrow\infty$}

In this Appendix we review some aspects of the large $U/t$ physics
that are useful for the contextualization of the problems studied
in this paper. We start by considering the simple 1D Hubbard
Hamiltonian given in Eq. (\ref{HH}) which differs from the
Hamiltonians given in Eqs. (\ref{H}) and (\ref{HSO4}) in the
definition of the zero-energy level. The energy spectrum of the
stationary states of the model (\ref{HH}) can written as a
functional of the pseudoparticle band-momentum distribution
functions and $-1/2$ Yang holon numbers such that \cite{I,II},

\begin{eqnarray}
E_H & = & -2t \sum_{j=1}^{N_a}N_c (q_j)\, \cos k(q_j)
\nonumber \\
& + & 4t \sum_{\nu=1}^{\infty} \, \sum_{j=1}^{N^*_{c,\,\nu}}
N_{c,\,\nu} (q_j)\,{\rm Re}\,\Bigl\{\sqrt{1 - (\Lambda_{c,\,\nu}
(q_j) + i\,\nu\, U/4t)^2}\Bigr\} + U\,L_{c,\,-1/2} \, . \label{EH}
\end{eqnarray}
This energy functional depends on these distribution functions
through the momentum-rapidity functional $k(q_j)$ and rapidity
functional $\Lambda_{c,\,\nu}(q_j)$ \cite{I,II}. The momentum
spectrum $P$ of the energy eigenstates has a simpler expression,
being the linear superposition of the pseudoparticle band-momentum
distribution functions and $-1/2$ Yang holon number. It reads,

\begin{eqnarray}
P & = &  \sum_{j=1}^{N_a}N_c (q_j)\, q_j +
\sum_{j=1}^{N^*_{s,\,\nu}}
N_{s,\,\nu} (q_j)\,q_j  \nonumber \\
& + & \sum_{j=1}^{N^*_{c,\,\nu}} N_{c,\,\nu} (q_j)\, [{\pi\over a}
-q_j] + {\pi\over a}\,M_{c,\,-1/2} \, . \label{Pp}
\end{eqnarray}
As in the case of the momentum values of the quasi-particles of a
Fermi liquid, the momentum $P$ given in Eq. (\ref{Pp}) is additive
relative to the band-momentum values of the pseudoparticles.

The momentum rapidity functional $k(q_j)$ and the rapidity
functional $\Lambda_{c,\,\nu} (q_j)$ on the right-hand side of Eq.
(\ref{EH}) can be defined in terms of the Takahasi's thermodynamic
equations \cite{Takahashi} rewritten in functional form
\cite{I,II}. These quantities are functionals of the set of
band-momentum distribution functions $N_c(q_j)$ and
$\{N_{\alpha,\,\nu}(q_j)\}$, where $\alpha=c,s$ and $\nu=1,2,...$,
whose possible different occupancy configurations classify the
energy eigenstates. In the limit $U/t\rightarrow\infty$ these
functionals have the same value for all energy eigenstates and
read $k(q_j)=q_j$ and $\Lambda_{c,\,\nu} (q_j)=0$. Thus in such a
limit the energy and momentum expressions (\ref{EH}) and
(\ref{Pp}) respectively, simplify to expression (\ref{EHUinf}) and

\begin{eqnarray}
P & = &  \sum_{j=1}^{N_a}N_c (q_j)\, q_j +
\sum_{j=1}^{N^*_{s,\,\nu}}
N_{s,\,\nu} (q_j)\,q_j  \nonumber \\
& + & \sum_{j=1}^{N^*_{c,\,\nu}} N_{c,\,\nu} (q_j)\, [{\pi\over a}
-q_j] + {\pi\over a}\,D \, , \label{PpUinf}
\end{eqnarray}
respectively, where the electron double occupation $D$ is a good
quantum number such that $D=M_{c,\,-1/2}$ in that limit. For
finite values of $U/t$ the latter relation is replaced by
$D_r=M_{c,\,-1/2}$, where $D_r$ is the rotated-electron double
occupation. According to Eq. (\ref{Mas}) the number $M_{c,\,-1/2}$
of $-1/2$ holons can be written as

\begin{equation}
M_{c,\,-1/2} = D_r = L_{c,\,-1/2} + \sum_{\nu
=1}^{\infty}\nu\,N_{c,\,\nu} \, . \label{Mc-1/2}
\end{equation}

In the case of zero spin density, electronic densities $0\leq
n\leq 1/a$, and on-site repulsion $U>>t$ the ground state energy
$E_0$ of the 1D Hubbard model is given by \cite{Carmelo88},

\begin{equation}
E_0 = -{2N_a t\over\pi}\sin (\pi n)-{t^2\over U}\,4N\,n\,\ln
2\,\Bigl(1-{\sin (2\pi n)\over 2\pi n}\Bigr) \, , \label{E0}
\end{equation}
and the ground state electron double occupation $D_0$ can be
written as \cite{II,Carmelo88},

\begin{equation}
D_0 = {\partial E_0\over\partial U} =\Bigl({t\over
U}\Bigr)^2\,4N\, n\,\ln 2\,\Bigl(1-{\sin (2\pi n)\over 2\pi
n}\Bigr) \, . \label{D0}
\end{equation}
Here the term $-{2N_a t\over\pi}\sin (\pi n)$ is the kinetic
energy associated with the hopping processes which do not change
electron double occupation. On the other hand, the energy term
$-{t^2\over U}\,4N\,n\,\ln 2\,\Bigl(1-{\sin (2\pi n)\over 2\pi
n}\Bigr)$ includes both kinetic and potential energy
contributions. It arises from hopping processes which change
electron double occupation and lead to the ground-state electron
double occupation value given in Eq. (\ref{D0}). Thus the physics
associated with this second energy term corresponds to excitation
processes of higher order in $t/U$ relatively to the
$t/U\rightarrow 0$ limit where electron double occupation is a
good quantum number and the ground-state electron double
occupation of the 1D Hubbard model is exactly zero. If we include
energy contributions of the order $t(t/U)^1$ the ground state
contains a small but finite electron double occupation expectation
value given in Eq. (\ref{D0}) and the corresponding quantum
problem is not equivalent to the physical situation of interest
for the rotated-electron studies of this paper. Indeed, electrons
equal rotated electrons when the limit $t/U\rightarrow 0$ is
reached and double occupation is a good quantum number. Only in
that limit do the electron occupancy configurations which describe
the band-momentum energy eigenstates equal the corresponding
rotated-electron configurations which are valid for all values of
$U/t$.

Note that at half filling the electronic density reads $n=1/a$,
the energy term of order $t(t/U)^0$ on the right-hand side of Eq.
(\ref{E0}) vanishes, and the ground-state energy and electron
double occupation expressions given in Eqs. (\ref{E0}) and
(\ref{D0}) simplify to $E_0=-{t^2\over U}\,4N_a\, \ln 2$ and $D_0
= \Bigl({t\over U}\Bigr)^2\,4N_a\, \ln 2$ respectively. It is well
known that this energy can be associated with an isotropic
Heisenberg model \cite{Emery}. The corresponding ground state
leads to energy contributions of the order $t(t/U)^1$ and thus
contains a small but finite electron double occupation expectation
value, $D_0 = \Bigl({t\over U}\Bigr)^2\,4N_a\, \ln 2$. It follows
that the usual description of the large-$U/t$ half-filling Hubbard
model in terms of an isotropic Heisenberg model is not equivalent
to our limit. In that limit only hopping processes which do not
change the value of electron double occupation should be
considered.

To leading order in the parameter $t/U$, the energy spectrum of
the 1D Hubbard model in the limit $t/U\rightarrow 0$ is of the
form given in Eq. (\ref{EHUinf}). The permitted hopping processes
lead to contributions in the eigenstate energies of order
$t(t/U)^{-1}$ and $t(t/U)^0$. These contributions are associated
with the energy terms ${N_a\over 2\pi} \int_{q_c^{-}}^{q_c^{+}}
dq\, N_c (q) [-2t\cos q]$ and $U\,D$ respectively, on the
right-hand side of Eq. (\ref{EHUinf}). In the particular case of
the ground state there are no contributions of order $t(t/U)^{-1}$
because the electron double occupation eigenvalue $D_0$ is zero.
In the limit $t/U\rightarrow 0$ only hopping processes such that
the electron singly occupied sites can move relatively to the
electron doubly-occupied and empty site distribution
configurations without changing these configurations are
permitted. In such a limit there is a huge degeneracy of
$\eta$-spin and spin occupancy configurations. Thus there are
several choices for complete sets of energy eigenstates with the
same energy and momentum spectra given in Eqs. (\ref{EHUinf}) and
(\ref{PpUinf}) respectively. This is because in this limit there
are also many choices for complete sets of compatible observables.
The 1D Hubbard model in the limit of $U/t\rightarrow\infty$ has
been studied in the literature by many authors
\cite{Ogata,Penc95,Penc96,Penc97,Harris,Beni,Klein,Carmelo88,Parola,Ricardo,Eskes,Geb}.
In the case of evaluation of quantities describing the physics of
the model in the limit of $U/t\rightarrow\infty$, the alternative
use of different complete sets of states leads to the same final
expressions for correlation functions and other quantities of
physical interest.

In this paper we are interested in one of these choices of energy
eigenstates only. It corresponds to the complete set of $4^{N_a}$
energy eigenstates generated from the corresponding energy
eigenstates of the finite-$U/t$ 1D Hubbard model by turning off
adiabatically the parameter $t/U$. Only that set of states
correspond to the states obtained from the finite $t/U$ states by
the electron - rotated-electron unitary transformation. These
states are common eigenstates of both the 1D Hubbard model as
$U/t\rightarrow\infty$ and of the set of number operators
$\{{\hat{L}}_{\alpha ,\,-1/2}\}$, $\{{\hat{N}}_c (q_j)\}$, and
$\{{\hat{N}}_{\alpha,\,\nu}(q_j)\}$ with $\alpha =c,s$ and
$\nu=1,2,3,...$ in the same limit. Together with the Hamiltonian
these operators constitute a complete set of compatible and
commuting hermitian operators \cite{III}. These operators also
commute with the momentum operator $\hat{P}$. For these energy
eigenstates the band momentum $q_j$ of the $c$ pseudoparticles and
$c,\nu$ and $s,\nu$ pseudoparticles such that $\nu=1,2,...$ is a
good quantum number. This justifies the designation of
band-momentum energy eigenstates.

However, in the limit $t/U\rightarrow 0$ there are other choices
for complete sets of energy and momentum eigenstates which are due
to the periodic boundary conditions of the original electronic
problem. In such a limit the 1D Hubbard model can be mapped onto a
problem for which the number of doubly occupied sites is conserved
\cite{Harris,Eskes}. Since in the limit $t/U\rightarrow 0$ the
ground state has zero double occupation, one usually introduces
the concepts of lower and upper Hubbard bands which are associated
with the Hilbert subspaces spanned by states with zero and finite
values of the double occupation respectively \cite{Harris,Eskes}.
In the particular case of excitations involving creation of an
electron the upper Hubbard band corresponds to the Hilbert
subspace spanned by states with a single doubly occupied site.
Within the $t/U\rightarrow 0$ scheme used in Refs.
\cite{Harris,Eskes} one spectrally decomposes the elementary
electronic operators into those which solely act in the upper or
lower Hubbard bands, and eliminate perturbatively those parts
which couple the two bands. This is achieved by application of a
suitable transformation that eliminates these parts to a given
order in $t/U$. Such a transformation is nothing but the electron
- rotated-electron unitary transformation in the limit of small
values of $t/U$. This procedure leads to an effective Hamiltonian
which is equivalent to the 1D Hubbard model to lowest order in
$t/U$. In reference \cite{Geb} that effective Hamiltonian was
called {\it Harris-Lange model}. In reference \cite{III} the
concepts of lower and upper Hubbard bands were generalized to all
values of $t/U$ and associated with the rotated-electron double
occupation.

\section{DESCRIPTION OF A LOCAL $c,4$ PSEUDOPARTICLE IN TERMS OF
ROTATED-ELECTRON SITE DISTRIBUTION CONFIGURATIONS}

In this Appendix we illustrate for the specific case of a local
$c,4$ pseudoparticle how the rotated-electron site distribution
configurations which describe a local $\alpha,\nu$ pseudoparticle
always involve a number $\nu$ of site pairs such that $x=g$ and
$x'=g+\nu$ where $g=1,...,\nu$. In the present case we have that
$\alpha =c$ and thus the rotated-electron site distribution
configurations correspond to doubly occupied and empty sites.
Moreover, since $\nu=4$ a local $c,4$ pseudoparticle involves four
site pairs such that $x=g$ and $x'=g+4$ where $g=1,2,3,4$. Once
there are two possible rotated-electron site occupancies for each
pair and the number of pairs of each local $c,4$ pseudoparticle is
four, the total number of different internal rotated-electron site
distribution configurations is $2^4=16$. According to the general
expression (\ref{cnp}) these $16$ different internal
rotated-electron site distribution configurations of doubly
occupied and empty sites are such that,

\begin{eqnarray}
& & \Bigl[\prod_{x=1}^{8}e^{i\pi h_{j,\,x}{\hat{D}}_{j,\,x}}\Bigr]
\,\Bigl[\prod_{g=1}^{4}(1-{\hat{\cal{T}}}_{c,\,4,\,j,\,g})\Bigr]
\nonumber \\
& \times & (\bullet_{h_{j,\,1}},\,\bullet_{h_{j,\,2}},\,
\bullet_{h_{j,\,3}},\,\bullet_{h_{j,\,4}},\,
\circ_{h_{j,\,5}},\,\circ_{h_{j,\,6}},\,
\circ_{h_{j,\,7}},\,\circ_{h_{j,\,8}}) = \nonumber \\
& + & e^{i\pi[h_{j,\,1}+h_{j,\,2}+h_{j,\,3}+h_{j,\,4}]}\,
(\bullet_{h_{j,\,1}},\,\bullet_{h_{j,\,2}},\,
\bullet_{h_{j,\,3}},\,\bullet_{h_{j,\,4}},\,
\circ_{h_{j,\,5}},\,\circ_{h_{j,\,6}},\,
\circ_{h_{j,\,7}},\,\circ_{h_{j,\,8}})\nonumber \\
& - &
e^{i\pi[h_{j,\,2}+h_{j,\,3}+h_{j,\,4}+h_{j,\,5}]}\,(\circ_{h_{j,\,1}},\,\bullet_{h_{j,\,2}},\,
\bullet_{h_{j,\,3}},\,\bullet_{h_{j,\,4}},\,
\bullet_{h_{j,\,5}},\,\circ_{h_{j,\,6}},\,
\circ_{h_{j,\,7}},\,\circ_{h_{j,\,8}})\nonumber \\
& + &
e^{i\pi[h_{j,\,1}+h_{j,\,3}+h_{j,\,6}+h_{j,\,8}]}\,(\bullet_{h_{j,\,1}},\,\circ_{h_{j,\,2}},\,
\bullet_{h_{j,\,3}},\,\circ_{h_{j,\,4}},\,
\circ_{h_{j,\,5}},\,\bullet_{h_{j,\,6}},\,
\circ_{h_{j,\,7}},\,\bullet_{h_{j,\,8}})\nonumber \\
& - &
e^{i\pi[h_{j,\,1}+h_{j,\,3}+h_{j,\,4}+h_{j,\,6}]}\,(\bullet_{h_{j,\,1}},\,\circ_{h_{j,\,2}},\,
\bullet_{h_{j,\,3}},\,\bullet_{h_{j,\,4}},\,
\circ_{h_{j,\,5}},\,\bullet_{h_{j,\,6}},\,
\circ_{h_{j,\,7}},\,\circ_{h_{j,\,8}})\nonumber \\
& + &
e^{i\pi[h_{j,\,1}+h_{j,\,2}+h_{j,\,7}+h_{j,\,8}]}\,(\bullet_{h_{j,\,1}},\,\bullet_{h_{j,\,2}},\,
\circ_{h_{j,\,3}},\,\circ_{h_{j,\,4}},\,
\circ_{h_{j,\,5}},\,\circ_{h_{j,\,6}},\,
\bullet_{h_{j,\,7}},\,\bullet_{h_{j,\,8}}) \nonumber \\
& - &
e^{i\pi[h_{j,\,1}+h_{j,\,2}+h_{j,\,4}+h_{j,\,7}]}\,(\bullet_{h_{j,\,1}},\,\bullet_{h_{j,\,2}},\,
\circ_{h_{j,\,3}},\,\bullet_{h_{j,\,4}},\,
\circ_{h_{j,\,5}},\,\circ_{h_{j,\,6}},\,
\bullet_{h_{j,\,7}},\,\circ_{h_{j,\,8}}) \nonumber \\
& + &
e^{i\pi[h_{j,\,1}+h_{j,\,4}+h_{j,\,6}+h_{j,\,7}]}\,(\bullet_{h_{j,\,1}},\,\circ_{h_{j,\,2}},\,
\circ_{h_{j,\,3}},\,\bullet_{h_{j,\,4}},\,
\circ_{h_{j,\,5}},\,\bullet_{h_{j,\,6}},\,
\bullet_{h_{j,\,7}},\,\circ_{h_{j,\,8}}) \nonumber \\
& - &
e^{i\pi[h_{j,\,1}+h_{j,\,2}+h_{j,\,3}+h_{j,\,8}]}\,(\bullet_{h_{j,\,1}},\,\bullet_{h_{j,\,2}},\,
\bullet_{h_{j,\,3}},\,\circ_{h_{j,\,4}},\,
\circ_{h_{j,\,5}},\,\circ_{h_{j,\,6}},\,
\circ_{h_{j,\,7}},\,\bullet_{h_{j,\,8}})\nonumber \\
& + &
e^{i\pi[h_{j,\,2}+h_{j,\,3}+h_{j,\,5}+h_{j,\,8}]}\,(\circ_{h_{j,\,1}},\,\bullet_{h_{j,\,2}},\,
\bullet_{h_{j,\,3}},\,\circ_{h_{j,\,4}},\,
\bullet_{h_{j,\,5}},\,\circ_{h_{j,\,6}},\,
\circ_{h_{j,\,7}},\,\bullet_{h_{j,\,8}})\nonumber \\
& - &
e^{i\pi[h_{j,\,1}+h_{j,\,6}+h_{j,\,7}+h_{j,\,8}]}\,(\bullet_{h_{j,\,1}},\,\circ_{h_{j,\,2}},\,
\circ_{h_{j,\,3}},\,\circ_{h_{j,\,4}},\,
\circ_{h_{j,\,5}},\,\bullet_{h_{j,\,6}},\,
\bullet_{h_{j,\,7}},\,\bullet_{h_{j,\,8}})\nonumber \\
& + &
e^{i\pi[h_{j,\,3}+h_{j,\,4}+h_{j,\,5}+h_{j,\,6}]}\,(\circ_{h_{j,\,1}},\,\circ_{h_{j,\,2}},\,
\bullet_{h_{j,\,3}},\,\bullet_{h_{j,\,4}},\,
\bullet_{h_{j,\,5}},\,\bullet_{h_{j,\,6}},\,
\circ_{h_{j,\,7}},\,\circ_{h_{j,\,8}})\nonumber \\
& - &
e^{i\pi[h_{j,\,2}+h_{j,\,5}+h_{j,\,7}+h_{j,\,8}]}\,(\circ_{h_{j,\,1}},\,\bullet_{h_{j,\,2}},\,
\circ_{h_{j,\,3}},\,\circ_{h_{j,\,4}},\,
\bullet_{h_{j,\,5}},\,\circ_{h_{j,\,6}},\,
\bullet_{h_{j,\,7}},\,\bullet_{h_{j,\,8}})\nonumber \\
& + &
e^{i\pi[h_{j,\,2}+h_{j,\,4}+h_{j,\,5}+h_{j,\,7}]}\,(\circ_{h_{j,\,1}},\,\bullet_{h_{j,\,2}},\,
\circ_{h_{j,\,3}},\,\bullet_{h_{j,\,4}},\,
\bullet_{h_{j,\,5}},\,\circ_{h_{j,\,6}},\,
\bullet_{h_{j,\,7}},\,\circ_{h_{j,\,8}})\nonumber \\
& - &
e^{i\pi[h_{j,\,3}+h_{j,\,5}+h_{j,\,6}+h_{j,\,8}]}\,(\circ_{h_{j,\,1}},\,\circ_{h_{j,\,2}},\,
\bullet_{h_{j,\,3}},\,\circ_{h_{j,\,4}},\,
\bullet_{h_{j,\,5}},\,\bullet_{h_{j,\,6}},\,
\circ_{h_{j,\,7}},\,\bullet_{h_{j,\,8}})\nonumber \\
& + &
e^{i\pi[h_{j,\,5}+h_{j,\,6}+h_{j,\,7}+h_{j,\,8}]}\,(\circ_{h_{j,\,1}},\,\circ_{h_{j,\,2}},\,
\circ_{h_{j,\,3}},\,\circ_{h_{j,\,4}},\,
\bullet_{h_{j,\,5}},\,\bullet_{h_{j,\,6}},\,
\bullet_{h_{j,\,7}},\,\bullet_{h_{j,\,8}})\nonumber \\
& - &
e^{i\pi[h_{j,\,4}+h_{j,\,5}+h_{j,\,6}+h_{j,\,7}]}\,(\circ_{h_{j,\,1}},\,\circ_{h_{j,\,2}},\,
\circ_{h_{j,\,3}},\,\bullet_{h_{j,\,4}},\,
\bullet_{h_{j,\,5}},\,\bullet_{h_{j,\,6}},\,
\bullet_{h_{j,\,7}},\,\circ_{h_{j,\,8}}) \, . \label{c4p}
\end{eqnarray}
On the left-hand side of this equation the site-$h_{j,\,x}$
rotated-electron double occupation operator ${\hat{D}}_{j,\,x}$
has eigenvalue $1$ and $0$ when that site is doubly occupied by
rotated electrons and free of rotated electrons respectively, and
the operator ${\hat{\cal{T}}}_{c,\,4,\,j,\,g}$ acts on the pair of
sites of indices $h_{j,\,g}$ and $h_{j,\,g+4}$ only. This operator
is designed to act onto the rotated-electron eight-site
distribution configuration of the particular form illustrated on
the left-hand side of Eq. (\ref{c4p}). This specific configuration
is such that the first four sites are doubly occupied by rotated
electrons and the following four sites are free of rotated
electrons. From the application of this operator onto such a
rotated-electron eight-site distribution configuration a new
distribution configuration is generated where the site of index
$h_{j,\,g}$ is free of rotated electrons, the site of index
$h_{j,\,g+4}$ is doubly occupied by rotated electrons, and the
occupancy of the other six sites remains unchanged. This operation
is repeated according to the products on the left-hand side of Eq.
(\ref{c4p}) and leads to the $16$ internal rotated-electron site
distribution configurations on the right-hand side of the same
equation. These configurations are generated by considering that
in each of the $g=1,2,3,4$ pairs of sites of indices $h_{j\,g}$
and $h_{j,\,g+4}$, the site of index $h_{j\,g}$ is doubly occupied
by rotated electrons and the site of index $h_{j,\,g+4}$ is free
of rotated electrons and vice versa. The phase factors appearing
on both sides of the above equation result from the momentum
$\pi/a$ associated with each rotated-electron doubly occupied
site. This is the momentum of the $-1/2$ holon corresponding to
such a site. Finally note that a similar description is obtained
for the $s,4$ pseudoparticle provided that one replaces these
phase factors by one and the rotated-electron doubly occupied and
empty sites by spin-down and spin-up rotated-electron singly
occupied sites respectively.


\newpage
\begin{figure}
\begin{center}
\epsfxsize=10cm \epsfbox{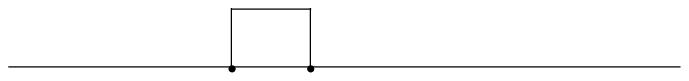}
\end{center}
\caption{Graphical representation of a charge $h_{j,\,x}
\leftrightarrow h_{j,\,x'}$ site pair or of a spin $l_{j,\,x}
\leftrightarrow l_{j,\,x'}$ site pair. In the particular case of
$x=1$ and $x'=2$ the figure represents the $h_{j,\,1}
\leftrightarrow h_{j,\,2}$ site pair of a $c,1$ pseudoparticle or
of the $l_{j,\,1} \leftrightarrow l_{j,\,2}$ site pair of a $s,1$
pseudoparticle.} \label{fig1}
\end{figure}

\begin{figure}
\begin{center}
\epsfxsize=10cm \epsfbox{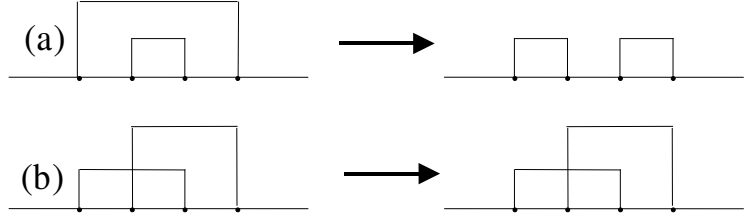}
\end{center}
\caption{Graphical representation of charge sequences of a zero
$\eta$-spin energy eigenstate with no Yang holons and two
rotated-electron doubly occupied sites and two rotated-electron
empty sites only. The $h_{j,\,g} \leftrightarrow h_{j,\,g+\nu}$
site pairs are represented as in Fig. 1. We note that the relative
height of different site pairs has no physical meaning. The goal
of using site pairs of different height is just to clearly define
the two sites of each pair. In figures (a) and (b) two possible
rotated-electron doubly occupied site/empty site pair distribution
configurations are represented. The figures show how these
distribution configurations change as a result of a cyclic
permutation which transforms the first site of the charge sequence
onto its last site. The distribution configuration (a) transforms
onto itself whereas the distribution configuration (b) transforms
onto two $c,1$ pseudoparticles. Thus according to property 6-III
the distribution configuration (a) describes a $c,2$
pseudoparticle. Note that the distribution configuration (b)
describes two $c,1$ pseudoparticles. The distribution occupancies
of the figure describe alternatively $s,1$ and $s,2$
pseudoparticles. In this case the spin sequences correspond to a
zero spin energy eigenstate with no HL spinons and with two
spin-down singly occupied sites and two spin-up singly occupied
sites only.} \label{fig2}
\end{figure}

\begin{figure}
\begin{center}
\epsfxsize=10cm \epsfbox{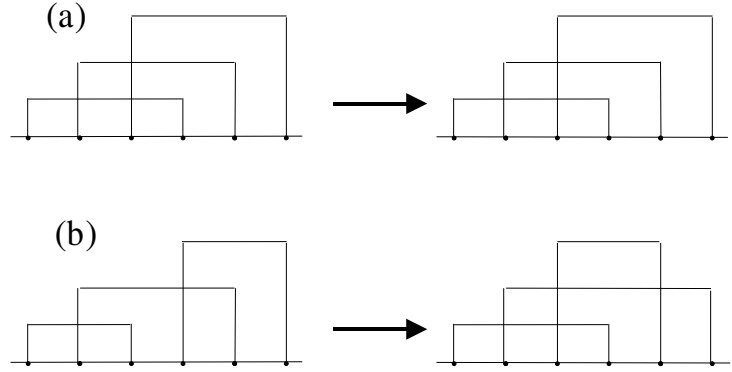}
\end{center}
\caption{Graphical representation of charge sequences of a zero
$\eta$-spin energy eigenstate with no Yang holons and three
rotated-electron doubly occupied sites and three rotated-electron
empty sites only. The $h_{j,\,g} \leftrightarrow h_{j,\,g+\nu}$
site pairs are represented as in Fig. 1. In figures (a) and (b) it
is shown how two possible rotated-electron doubly occupied
site-empty site pair distribution configurations change as a
result of a cyclic permutation which transforms the first site of
the charge sequence onto its last site. The distribution
configuration (a) transforms onto itself whereas the distribution
configuration (b) transforms onto a non-equivalent distribution
configuration. Thus according to property 6-III the distribution
configuration (a) describes a $c,3$ pseudoparticle. The figures
represent alternatively spin sequences of a zero spin energy
eigenstate with no HL spinons and with three spin-down singly
occupied sites and three spin-up singly occupied sites only.}
\label{fig3}
\end{figure}

\begin{figure}
\begin{center}
\epsfxsize=10cm \epsfbox{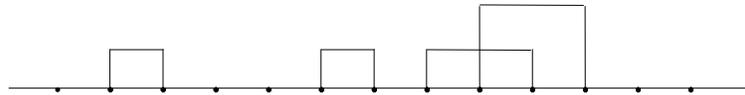}
\end{center}
\caption{Graphical representation of a domain of a charge sequence
of the effective electronic lattice including fourteen sites. The
$h_{j,\,g} \leftrightarrow h_{j,\,g+\nu}$ site pairs are
represented as in Fig. 1. The six sites with no vertical lines
correspond to Yang holons. There are two local $c,1$
pseudoparticles described by $h_{j,\,1} \leftrightarrow h_{j,\,2}$
site pairs and a local $c,2$ pseudoparticle described by a
$h_{j,\,1} \leftrightarrow h_{j,\,3}$ and a $h_{j,\,2}
\leftrightarrow h_{j,\,4}$ site pairs. Alternatively, if we
replace the Yang holons by HL spinons and the $c,\nu$
pseudoparticles by $s,\nu$ pseudoparticles the figure represents
the domain of a spin sequence with six HL spinons, two $s,1$
pseudoparticles, and a $s,2$ pseudoparticle.} \label{fig4}
\end{figure}

\begin{figure}
\begin{center}
\epsfxsize=10cm \epsfbox{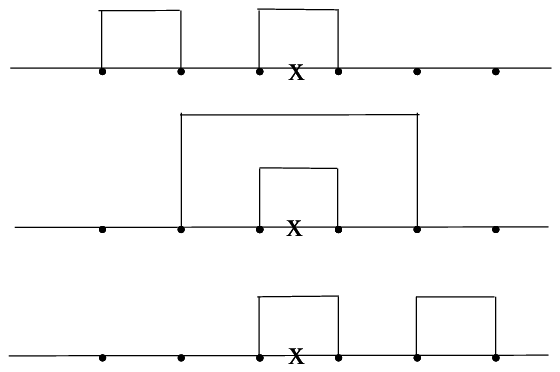}
\end{center}
\caption{An illustration of the possible positions of a local
$c,1$ pseudoparticle when it passes from the left to the
right-hand side of a steady local $c,1$ pseudoparticle. The
$h_{j,\,1} \leftrightarrow h_{j,\,g+\nu}$ site pairs are
represented as in Fig. 1. In the three rotated-electron site
distribution configurations (a)-(c) the charge-sequence position
of the steady pseudoparticle is labelled by a $X$ and remains
unchanged. These three rotated-electron site distribution
configurations correspond to three different charge sequences.
Alternatively, the figure represents the corresponding electron
site distribution configurations of two $s,1$ pseudoparticles in
three different spin sequences. } \label{fig5}
\end{figure}

\begin{figure}
\begin{center}
\epsfxsize=10cm \epsfbox{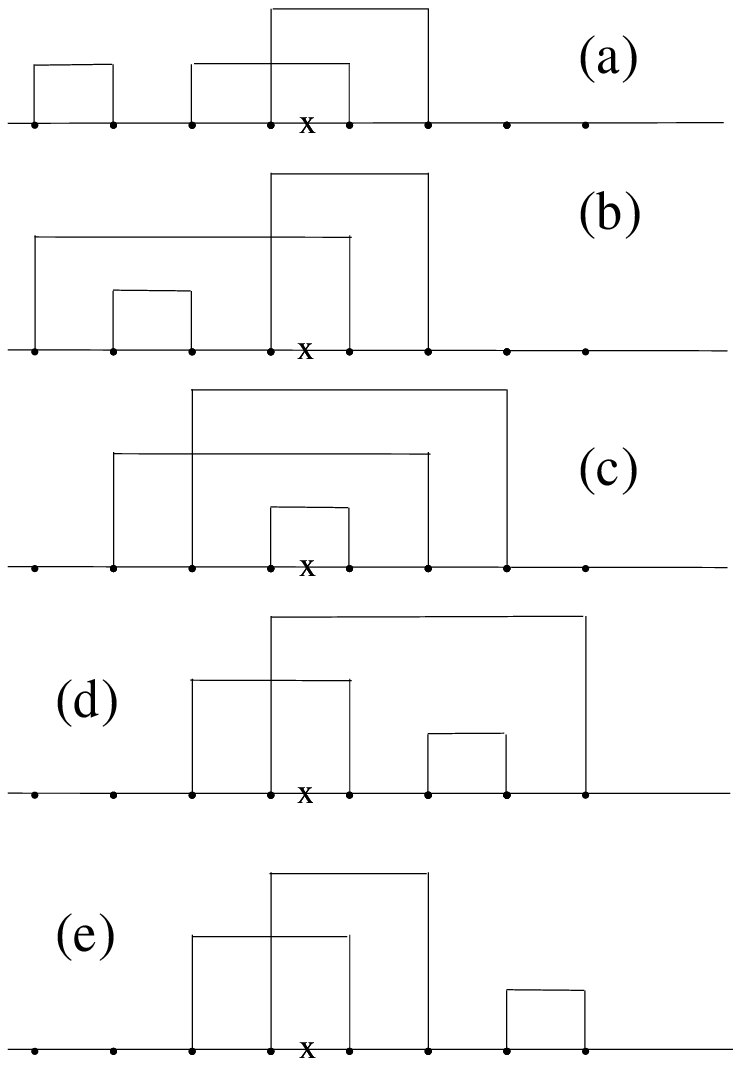}
\end{center}
\caption{An illustration of the possible positions of a local
$c,1$ pseudoparticle when it passes from the left to the
right-hand side of a steady local $c,2$ pseudoparticle. The
$h_{j,\,g} \leftrightarrow h_{j,\,g+\nu}$ site pairs are
represented as in Fig. 1. In the five rotated-electron site
distribution configurations (a)-(e) the charge-sequence position
of the steady $c,2$ pseudoparticle is labelled by a $X$ and
remains unchanged. If instead we consider a steady local $c,1$
pseudoparticle, figures (a)-(e) illustrate the possible positions
of a local $c,2$ pseudoparticle when it passes from the right to
the left-hand side of a steady local $c,1$ pseudoparticle.
However, in this second case the charge-sequence position of the
local $c,1$ pseudoparticle should remain unchanged and the $X$
point of figures (a)-(e) should be moved accordingly. In both
cases these five distribution configurations correspond to five
different charge sequences. Alternatively, the figure represents
the corresponding electron site distribution configurations of a
$s,1$ pseudoparticle and a $s,2$ pseudoparticle.} \label{fig6}
\end{figure}
\end{document}